\documentclass[12pt]{article}
\usepackage[super,compress]{cite}
\usepackage{graphicx}
\usepackage{amsmath}
\usepackage{graphicx}
\usepackage{booktabs}
\usepackage{amsfonts}
\usepackage{hyperref}
\usepackage{lmodern}
\usepackage{amssymb}
\usepackage{tikz}
\usetikzlibrary{patterns}
\usetikzlibrary{snakes}
\usepackage{yfonts}

\def\CC {{\cal C}}
\def\CD {{\cal D}}

\def\CF {{\cal F}}
\def\CG {{\cal G}}
\def\CH {{\cal H}}

\def\CK {{\cal K}}

\def\CM {{\cal M}}

\def\CR {{\cal R}}
\def\CS {{\cal S}}
\def\CT {{\cal T}}

\def\CV {{\cal V}}

\newcommand{\be}{\begin{equation}}
\newcommand{\ee}{\end{equation}}

\newcommand{\bea}{\begin{eqnarray}}
\newcommand{\eea}{\end{eqnarray}}
\newcommand{\beqa}{\begin{eqnarray}}
\newcommand{\eeqa}{\end{eqnarray}}
\newcommand{\nn}{\nonumber}

\begin{document}

\setlength{\baselineskip}{7mm}
\begin{titlepage}
 \begin{flushright}
{\tt NRCPS-HE-69-2021}
\end{flushright}

\begin{center}

{\Large ~\\{\it   Maximally Chaotic Dynamical Systems \\  
  and \\
Fundamental Interactions\footnote{
Based on lectures at the International Bogolyubov Conference "Problems of Theoretical and Mathematical Physics"  at the Steklov Mathematical Institute, Moscow-Dubna,  September 9-13, 2019 
and seminars at the Niels Bohr Institute, at the  CERN Theory Department and A. Alikhanian National Laboratory in Yerevan.  \\
\url{http://www.mathnet.ru/php/presentation.phtml?option_lang=eng&presentid=24992}  }  }}
   
\vspace{1cm}

{\sl George Savvidy}

\bigskip
 {\it Institute of Nuclear and Particle Physics, NCSR Demokritos, GR-15310 Athens, Greece}\\
{\it  A.I. Alikhanyan National Science Laboratory, Yerevan, 0036, Armenia}\\
{\it Institut f\"ur Theoretische Physik,Universit\"at Leipzig, D-04109 Leipzig, Germany }

\bigskip

\end{center}

\centerline{{\bf Abstract}}
We give a general review on the application of Ergodic theory to the investigation of  dynamics of the Yang-Mills gauge fields and of the gravitational systems, as well as its application in the Monte Carlo method and fluid dynamics. In ergodic theory the maximally chaotic  dynamical systems (MCDS) can be defined as dynamical systems that have nonzero Kolmogorov entropy. The hyperbolic dynamical systems that fulfil the Anosov C-condition belong to the MCDS insofar as they have exponential instability of their phase trajectories and positive Kolmogorov entropy.  It follows that the C-condition defines a rich class of MCDS that span over an open set in the space of all dynamical systems. The large class of Anosov-Kolmogorov MCDS is realised on Riemannian manifolds of negative sectional curvatures and on high-dimensional  tori. The interest in MCDS  is rooted in the attempts to understand the relaxation phenomena, the foundations of the statistical mechanics, the appearance of turbulence in fluid dynamics, the non-linear dynamics of Yang-Mills field and gravitating N-body systems as well as black hole thermodynamics.  Our aim is to investigate classical- and quantum-mechanical properties of MCDS and their role in the theory of fundamental interactions. 

\end{titlepage}

\section{\it Introduction}	

It seems natural to define a maximally chaotic dynamical system (MCDS) as a system that has a nonzero Kolmogorov entropy \cite{kolmo,kolmo1,kolmo2}.  A large class of hyperbolic MCDS was constructed by Anosov \cite{anosov}.   These are systems that fulfil the C-condition, a condition that is sufficient for a system to be a MCDS. The Anosov hyperbolicity C-condition is about the {\it local and uniform exponential instability of phase trajectories}\footnote{ The C-condition, which may sound similar to the existence of positive Lyapunov characteristic exponents of a dynamical system, is much stronger and is a sufficient condition for a system to be chaotic, to be a MCDS. A positive largest Lyapunov exponent does not in general indicate chaos \cite{perron,leonov1}. A negative largest Lyapunov exponent does not in general indicate stability \cite{perron,leonov1}.} that leads to the mixing of all orders and to a positive Kolmogorov entropy.  The uniqueness of the Anosov C-condition lies in the fact that it defines a rich class of MCDS that span over an open set in the space of all dynamical systems, meaning that in an arbitrary small  neighbourhood of a given MCDS the dynamical systems are homeomorphic \cite{anosov}.  

Examples of MCDS were discovered and discussed in the earlier investigations by  Lobachevsky, Artin, Hadamard, Hedlund, Hopf, Birkhoff   and others \cite{Lobachevsky,Artin,Hadamard, hedlund, hopf1,hopf,hopf2,Hopf,anosov1,Gibbs,Birkhoff,Koopman,krilov,sinai3,kornfeld,arnoldavez}, as well as in more recent  investigations \cite{rokhlin1,leonov,rokhlin,rokhlin2,smale,sinai2,sinai4,margulis,bowen0,bowen,bowen1,gines,Gutzwiller,Savvidy:2020mco}. Here we shall introduce and discuss the classical- and quantum-mechanical properties of MCDS,  the application of the ergodic  theory to the investigation of non-Abelian  gauge fields \cite{Baseyan,Natalia,Asatrian:1982ga,SavvidyKsystem,Savvidy:1982jk,Savvidy:1984gi,Hoppe,YMquantmech,Akutagawa:2020qbj,Wigner,Mehta,Dyson,Chirikov,Ermann:2021wej,Shchur,Nicolai,Banks:1996vh,Acharyya:2016fcn,Balachandran:2014iya,Pavel:2021mxn}  and gravitational systems \cite{body,Chandrasekhar,garry,Lang,Binney,Heggie}, fluid dynamics and stability of the atmosphere \cite{Milnor,turbul,Arakelian:1989,Arakelian:1988gm,Lukatzki:1981,Smolentsev,Yoshida,Dowker:1990tb} as well as application in Monte Carlo method \cite{metropolis,neuman,neuman1, sobol,lnbook,niki,yer1986a,konstantin,Savvidy:2015jva,Savvidy:2015ida,Gorlich:2016fbs,Savvidy:2018ygo,hepforge,CLHEP,Geant4,root,CMSrunII,CMS,PYTHIA,GSL,yer1986b,mixmaxGalois,pierr,Demchik:2010fd,falcion}. 
 
In recent years the quantum-mechanical concept of maximally chaotic systems was developed in a series of publications \cite{Larkin,Sekino,Shenker:2013pqa,Maldacena:2015waa,Gur-Ari:2015rcq,Cotler:2016fpe,Arefeva:1998,Arefeva:1998,Arefeva:1999,Arefeva:1999frh,Arefeva:2013uta,Hanada:2017xrv,Anous:2017mwr}.  The thermodynamics  of black holes exhibits the extraordinary property of fast relaxation, in the sense that an arbitrary perturbation to a black hole "scrambles" as fast as possible over the horizon, making it indistinguishable from a thermal distribution \cite{Wheeler}.  It has been conjectured that black holes are the fastest scramblers in nature \cite{Sekino,Shenker:2013pqa,Maldacena:2015waa}.  The influence of chaos on time-dependent double commutator of two observables can develop no faster than exponentially with the Lyapunov exponent $ 2\pi T t  $ that grows  linearly in temperature and time.  The linear growth is saturated in gravitational and dual to the gravity systems \cite{Maldacena:2015waa,Gur-Ari:2015rcq,Cotler:2016fpe}.  

We are interested  in analysing the behaviour of the out-of-time-order correlation functions (\ref{outoftime}), (\ref{fourpointfunct1}), (\ref{fourpointfunct2}) and the double commutators (\ref{commutatorL}) in the case of well defined MCDS investigating the "influence and remnants" of the classical chaos on the quantum-mechanical behaviour of the quantised  MCDS.  Considering MCDS in their quantum-mechanical regime  would help to identify the traces of classical chaos in quantum-mechanical regime and clarify the natural meaning of the quantum chaos.  In particular, we shall consider the non-Abelian gauge field theory and the Artin system that is defined on a surface of constant negative curvature, a finite-area patch on $AdS_2$ \cite{Artin,Poincare,Poincare1,Fuchs,maass,roeleke,Gelfand,selberg1,selberg2,Faddeev,Faddeev1,Takhtajan:2020hwl, hejhal2,hejhal,hejhal1,Ford,winkler,bump,Collet,Pollicot,moore,dolgopyat,chernov,Poghosyan:2018efd,Babujian:2018xoy}.\\ 
\\
 The review is organised as follows.\\
{\it Elements of Ergodic Theory. }  In the second section we shall define the Kolmogorov entropy and discuss the  classification of the dynamical systems (DS) with respect to their statistical/chaotic properties \cite{Halmos,kornfeld,arnoldavez}.   These are ergodic, mixing, n-fold mixing, and finally, the K-systems, which have mixing of all orders and the nonzero Kolmogorov entropy.  This consideration defines the hierarchy of DS by their increasing chaotic/stochastic  properties, with MCDS on the "top" of this hierarchy list.  

{\it Maximally Chaotic Dynamical Systems.} The basic question is: Do MCDS exist?  The Anosov hyperbolicity C-condition defines a rich class of MCDS that span over an open set in the space of all dynamical systems and is about the local and uniform exponential instability of the phase trajectories. It appears that a large class of MCDS can be realised on the {\it Riemannian manifolds of negative sectional curvatures and on high-dimensional tori}.  We shall consider the general properties of the MCDS in the third section. The MCDS have very strong instability of their phase trajectories and, in fact, the instability is as strong as it can be in principle \cite{anosov,anosov1}. The distance between infinitesimally close trajectories increases exponentially and on a closed phase space of the dynamical system this leads to the uniform distribution of phase  trajectories over the whole phase space and exposes extended statistical  properties \cite{anosov}.  The MCDS are structurally stable \cite{anosov},  spanning an open set in the space of all dynamical systems, which means that in an arbitrary small neighbourhood of a given MCDS the dynamical systems are homeomorphic \cite{anosov}. The other important property of the MCDS is that they have  a countable set of periodic trajectories.   The set of points on the periodic trajectories is everywhere dense in the phase space of MCDS. The periodic trajectories and non-periodic trajectories are filling out the phase space of an MCDS in a way very similar to the rational and irrational numbers on a Euclidean space \cite{anosov,bowen0,bowen,bowen1,Savvidy:2015ida,Gorlich:2016fbs}. 

{\it Anosov C-condition and Geodesic Flows.} The hyperbolic geodesic flow on Riemannian manifolds of negative sectional curvatures will be considered in the fourth section \cite{anosov,hedlund,hopf1,hopf,Hopf,Savvidy:2020mco}.  It was proven by Anosov that the geodesic flow on a closed Riemannian manifold of negative sectional curvatures fulfils the C-condition and  therefore defines a large class of MCDS.  This result provides a powerful tool for the investigation of the Hamiltonian systems \cite{Savvidy:2020mco}.  If the time evolution of a classical Hamiltonian system  under investigation can be reformulated as a geodesic flow on a Riemannian manifold and if all its sectional curvatures are negative, then it represents a MCDS system.  The MCDS  approach the equilibrium state with exponential rate that depends on the Kolmogorov-Sinai entropy.  The larger the entropy is, the faster a physical system moves toward its equilibrium \cite{krilov,turbul,kornfeld,arnoldavez,Savvidy:1982jk,body,Savvidy:2018ygo,Poghosyan:2018efd,Babujian:2018xoy}. 
 
{\it Non-Abelian Gauge Field Theory Dynamics.} In the fifth section we shall consider the classical and quantum mechanics of the Yang-Mills  fields \cite{Baseyan,Natalia,Asatrian:1982ga,SavvidyKsystem,Savvidy:1982jk,Savvidy:1984gi,Hoppe,Chirikov,Shchur,Nicolai,Banks:1996vh,Hoppe,Acharyya:2016fcn,Balachandran:2014iya,Pavel:2021mxn,Ambjorn:2000dx,Ambjorn:2000bf,Maldacena:2015waa,Gur-Ari:2015rcq,Arefeva:1998,Arefeva:1999,Arefeva:1999frh,Arefeva:2013uta}. In the case of space homogeneous gauge fields the Yang-Mills equations reduce to a classical mechanical system, so called Yang-Mills classical mechanics (YMCM), and represent the bosonic part of the matrix models \cite{Nicolai,Banks:1996vh,Hoppe,Acharyya:2016fcn,Balachandran:2014iya,Pavel:2021mxn,Ambjorn:2000dx,Ambjorn:2000bf}. It has finite degrees of freedom \cite{Baseyan,Natalia,Asatrian:1982ga,SavvidyKsystem}, isotropic and homogeneous energy momentum tensor (\ref{energymomentum})  and relativistic equation of state (\ref{equationofstate})\footnote{The apparent inhomogeneity of the energy momentum tensor (\ref{momentumdensity}) in Electrodynamics due to the term $-E_{i} E_{j} -H_{i} H_{j} $ is a critical barrier for a successful vector field driven inflation  \cite{Golovnev:2008cf,Golovnev:2008hv}. In Yang Mills theory the energy momentum tensor is perfectly homogeneous (\ref{energymomentum}) and opens a room of possibilities for a vector field driven inflation  \cite{Savvidy:2021ahq}. I would like to thank Prof. Viatcheslav  Mukhanov for the discussion of this point. }
 \cite{Savvidy:2021ahq}. By using the energy and momentum conservation integrals the system can be reduced to a system of a lower dimension, and the fundamental question is if the residual system is integrable and has additional hidden conserved integrals \cite{kolmo2} or is non-integrable and chaotic. The evolution of the YMCM can be formulated as a geodesic flow on a Riemannian manifold equipped with the Maupertuis's metric. The investigation of sectional curvatures demonstrates that in the vicinity of  the equipotential surface it is negative and causes the exponential instability of phase trajectories. The numerical integration also confirms this conclusion. The natural question that arrises here is to what extent the classical chaos influences the quantum-mechanical properties of the non-Abelian gauge fields. The corresponding quantum-mechanical system defines an important class of matrix models \cite{ Savvidy:1982jk,Savvidy:1984gi,Nicolai,Banks:1996vh,Acharyya:2016fcn,Balachandran:2014iya,Pavel:2021mxn}.  We shall discuss their spectral properties and the traces of the classical chaos in their quantum-mechanical regime.   

{\it Gravitational Systems, N-body Problem.} An interesting application of the Anosov C-systems theory was found in the investigation of the relaxation phenomena in stellar systems like globular clusters and galaxies \cite{body,garry}. Here again, one can use the  Maupertuis's metric in order to reformulate the evolution of the N-body system in Newtonian  gravity as a geodesic flow on a Riemannian manifold. Investigation of sectional curvatures allows to estimate  the exponential divergency of the phase trajectories and the relaxation time toward  the stationary distribution of the star velocities in elliptic galaxies  and globular clusters \cite{body}. This relaxation time is by few orders of magnitude shorter than the Chandrasekhar binary relaxation time \cite{Chandrasekhar,Lang,Binney,Heggie}. The  difference is rooted in the fact that in this approach one can take into account the long-range interaction between stars through their collective contribution into the sectional curvature. The Hubble Deep Field and Hubble eXtreme Deep Field images revealed a large number of distant young galaxies seemingly in a non-equilibrium state, while the stars in the nearby older galaxies  show a more regular distribution of velocities and shapes, the result of the collective relaxation phenomenon. 

{\it Artin Hyperbolic System.} Of special interest are the MCDS that are defined on a patch of the hyperbolic Lobachevsky plane of constant negative curvature.  An example of such a system was defined in a brilliant article published in 1924 by the mathematician Emil Artin \cite{Artin}. The dynamical system is defined on the fundamental region of the Lobachevsky plane that is obtained by the identification of points congruent with respect to the modular group $SL(2,Z)$, a discrete subgroup of the Lobachevsky plane isometries $SL(2,R)$ \cite{Poincare,Poincare1,Fuchs, Ford}. The fundamental region in this case is a hyperbolic triangle, a non-compact region of a finite area. The geodesic trajectories are bounded to propagate on the fundamental hyperbolic triangle and are exponentially unstable.  In the classical regime the exponential divergency of the geodesic trajectories resulted into the universal exponential decay of the classical correlation functions \cite{Poghosyan:2018efd,Collet,Pollicot,moore,dolgopyat,chernov}. The  Artin symbolic dynamics, the differential geometry and group-theoretical methods of Gelfand and Fomin \cite{Gelfand} are used to investigate the exponential decay rate of the classical correlation functions in the seventh section \cite{Poghosyan:2018efd}.   

{\it Quantum Mechanics of Artin System.}  In the eighth section we shall describe the quantisation of the Artin system and review the derivation of the  Maass wave functions  corresponding to a continuous spectrum \cite{maass}.  There is a great interest in considering quantisation of the hyperbolic dynamical systems and investigation of their quantum-mechanical properties. This subject is very closely related  to  the investigation of quantum mechanics of classically chaotic systems in gravity \cite{Maldacena:2015waa, Cotler:2016fpe,Gur-Ari:2015rcq}.  In the eighth section we shall study the behaviour of the correlation functions of the Artin hyperbolic system in its quantum-mechanical regime.  In order to investigate the behaviour of the correlation functions in the quantum-mechanical regime it is necessary  to know the spectrum of the system and the corresponding wave functions.  In the case of the modular group the energy spectrum has a continuous part originating from asymptotically free motion inside an infinitely long channel extended in the vertical direction of the fundamental region. It also has infinitely many discrete energy states  corresponding to the motion at the "bottom"  of the fundamental triangle \cite{maass,roeleke,selberg1,selberg2,bump, Faddeev,Faddeev1,hejhal2,winkler,hejhal,hejhal1,Kordyukov}. The spectral problem has a deep number-theoretical origin  and was partially solved in a series of pioneering articles \cite{maass,roeleke,selberg1,selberg2}.   It was solved partially because the discrete spectrum and the corresponding wave functions are still not known analytically \cite{Takhtajan:2020hwl}.   The general properties of the discrete spectrum were derived by using Selberg trace formula \cite{selberg1,selberg2,bump, Faddeev,Faddeev1,hejhal2}.  Numerical calculations of the discrete energy levels were performed for many energy states   \cite{winkler,hejhal,hejhal1}.  

{\it Out-of-time-order Correlation Functions.} Having in hand the explicit expression of the wave functions one can analyse the quantum-mechanical  behaviour of the correlation functions in order to investigate the traces of the classical chaos in the quantum-mechanical regime \cite{Babujian:2018xoy}.  In the ninth section we shall consider the correlation functions of the Liouiville-like  operators and shall demonstrate  that all two- and four-point correlation functions  decay exponentially with time, with the exponents that depend on temperature. Alternatively,  the double commutator of the Liouiville-like operators  separated in time  grows exponentially \cite{Babujian:2018xoy}. This growth is reminiscent of the local exponential divergency of trajectories when it was considered in the classical regime.  The results are presented in the Fig.\ref{timeevolutionofcommutator}.  The double commutator increases exponentially in time  
with the exponent  $1/\chi(\beta)$.  The ratio with respect to the maximal growth exponent is presented in Fig.\ref{timeevolutionofcommutator}. The temperature dependence of Artin exponent $1/\chi(\beta)$ relative to the maximum growth $1/\beta$ is shown by blue dots in the far right figure. At high temperatures the Artin-Lyapunov exponent $1/\chi(\beta)$ is less than the maximal exponent $1/\beta$, but at low temperatures this is not any more true and we observe a breaking of the saturation regime \cite{Maldacena:2015waa,Gur-Ari:2015rcq,Cotler:2016fpe}.  In order to confirm this result it seems important to investigate the behaviour of correlation functions and double commutators (\ref{corr1})-(\ref{commuta}) for alternative observables by using more powerful computer code and hardware than it was available to us. 
 
{\it Artin-Maass Resonances and Riemann Zeta Function Zeros.} In the tenth section we shall demonstrate that the Riemann zeta-function zeros \cite{Riemann} define the position and the widths of the resonances of the quantised Artin hyperbolic system  \cite{Savvidy:2018ffh}.  A possible relation of the zeta-function zeros and quantum-mechanical spectrum was discussed in the past,  the P\'olya-Hilbert conjecture proposed the idea of finding an eigenvalue  problem with the spectrum containing the zeros of the Riemann zeta-function \cite{Gutzwiller}.  The quantum-mechanical resonances have more complicated pole structure compared to a pure discrete spectrum and can be adequately described in terms of the scattering S-matrix theory.  We shall use the S-matrix approach to analyse the scattering phenomenon in a quantised Artin system. As it was discussed  above, the Artin dynamical system is defined on the fundamental region of the modular group on the Lobachevsky plane. It has a finite area and an infinite extension in the vertical direction that corresponds to a cusp.   In the quantum-mechanical regime the system can be associated with the narrow infinitely long waveguide stretched out to infinity along the vertical axis  and a cavity resonator attached to it at the bottom.  That suggests a physical interpretation of the Maass automorphic wave function in the form of an incoming plane wave of a given energy entering the resonator and  bouncing back to infinity.  As  the energy of the incoming wave comes close to the eigenmodes of the cavity a pronounced resonance behaviour shows up in the scattering amplitude.  The condition on the absence of incoming waves allows to find the position of these singularities \cite{Savvidy:2018ffh}. The  poles of the S-matrix are located in the complex energy plane  $E=E_n - i {\Gamma_n \over 2}$, where $E_n$ is the energy and $\Gamma_n$ is the width of the n'th resonance and can be expressed in terms of zeros $u_n$ of the Riemann zeta function $\zeta(\frac{1}{2} - i u_n) =0,~ n=1,2,....$ as
$$
E_n = {u^2_n \over 4} + {3\over 16},~~~~ \Gamma_n   = u_n. 
$$
It is an intriguing relation between the quantum-mechanical spectrum of MCDS and zeros of the Riemann zeta-function that can help to "translate and dualise" the  problem of distribution of the Riemann zeta-function zeros into its physical context \cite{maass,roeleke,selberg1,selberg2,bump, Faddeev, Faddeev1, hejhal2, winkler, hejhal, hejhal1}.   

{\it C-cascades. Entropy and Periodic Trajectories.} In the eleventh section we shall turn our attention to the investigation of the second class of the MCDS systems that is defined on high-dimensional tori \cite{anosov}.  In order that the automorphisms of a torus fulfil  the C-condition it is necessary and sufficient that the evolution operator $T$ has no eigenvalues on a unit circle and has the determinant  equal to one. Therefore  $T$ is an automorphism of the torus onto itself. All trajectories with rational coordinates, and only they, are periodic trajectories of the automorphisms of a torus.  The entropy of the C-system on a torus is equal to the logarithmic sum of all eigenvalues  that lie outside of the unit circle \cite{anosov,smale,sinai2,margulis,bowen0,bowen,bowen1}. 

{\it MIXMAX Random Number Generator.} It was suggested in 1986 \cite{yer1986a} to use the MCDS defined on a torus to generate high-quality pseudorandom numbers for the Monte-Carlo method   \cite{metropolis,neuman,neuman1,sobol,yer1986a,Demchik:2010fd,falcion}.   
Usually  pseudorandom numbers are generated by deterministic recursive rules
\cite{yer1986a,metropolis,neuman,neuman1,sobol}. Such rules produce pseudorandom numbers, and it is a great challenge to design pseudorandom number generators that produce high-quality sequences. 
Although numerous RNGs introduced in the last decades fulfil most of the requirements and are frequently used in simulations, each of them has some weak properties that influence the results \cite{pierr}, and they are less suitable for demanding MC simulations. We shall describe the details of the computer implementation of the torus automorphisms, the computation of the periods of generated random sequences used in Monte Carlo simulations. In a typical computer implementation of the torus automorphisms the initial vector will have rational components.  If the denominator is taken to be a prime number, then the recursion is realised on extended Galois field $GF[p^N]$   and  allows to find the period of the trajectories in terms of $p$ and the properties of the characteristic polynomial  of the evolution operator $T$ \cite{konstantin}.  We present the derivation of the explicit formulas for the Kolmogorov entropy in case of torus automorphisms and the properties of the periodic trajectories and their density distribution as a function of Kolmogorov entropy \cite{Gorlich:2016fbs}. 

The efficient implementation of the   MIXMAX MCDS generators  for Monte Carlo simulations was developed  in \cite{konstantin,hepforge}.  The MIXMAX generators demonstrated excellent statistical properties, high performance and superior high-quality output and became a multidisciplinary usable product. The main characteristics of the generators are: a) MIXMAX is an original and genuine 64-bit generator, is one of the fastest generators producing a 64-bit pseudorandom number in approximately 4 nanoseconds,
b) has very large Kolmogorov entropy of 0.9 per/bit,
c) long periods of order of $10^{120}$ - $10^{5000}$,
d) a new skipping algorithm generates seeds and guarantees that streams are not overlapping. The MIXMAX generators were integrated into the concurrent and distributed MC toolkit Geant4 \cite{Geant4}, the foundation library CLHEP \cite{CLHEP} and data analysis framework ROOT \cite{root}. These software tools have wide applications in High Energy Physics at CERN, in CMS experiment \cite{CMSrunII,CMS}, at SLAC, FNAL and KEK National Laboratories and are part of the CERN's active Technology Transfer policy. The generator is available in the PYTHIA event generator \cite{PYTHIA}. The MIXMAX code can be downloaded from the GSL-GNU Scientific Library \cite{GSL} and boost C++ Libraries \cite{boost}.

{\it Turning C-cascade into C-flow.}  In the thirteenth section we shall consider continuous in time dynamical systems constructed so that the discrete time slices will coincide with a given discrete authomorphism. In \cite{anosov} Anosov demonstrated how any C-cascade on a torus can be embedded into a continuous  C-flow. Here we shall describe the Anosov construction that allows to embed a discrete time evolution on a torus into an evolution that is continuous in time. The embedding was defined by a specific identification of the phase space coordinates  and  by construction of the corresponding  C-flow on a smooth Riemannian manifold of higher dimension. In Anosov construction the C-flow was not defined as a geodesic flow.  We were interested in formulating and analysing the alternative system that is defined as a geodesic flow on the same Riemannian manifold. The calculation of the corresponding sectional curvatures  demonstrated that  the geodesic flow has different dynamics  and  hyperbolic components \cite{Savvidy:2015ida}.   

{\it Infinite Dimensional Limit of C-cascade.}  In the fourteenth section we define the infinite-dimensional limit of the MCDS that are defined on N-dimensional tori. As we have just discussed, these hyperbolic systems found successful application in computer algorithms that generate high-quality pseudorandom numbers for advanced Monte Carlo simulations. The chaotic properties of these systems are increasing with $N$  because  the corresponding Kolmogorov-Sinai entropy grows  linearly with  $N$.  We are interested in considering the systems in the limit $N \rightarrow \infty$.  The limiting system defines the hyperbolic evolution of the continuous functions that is very similar to the evolution of a velocity function describing the hydrodynamic flow of fluids. We compare the chaotic properties of the limiting system with those of the hydrodynamic flow of incompressible ideal fluid on a torus investigated by Arnold. This maximally chaotic system can find application in the Monte Carlo method, statistical physics and digital signal processing.

{\it  Fluid Dynamics and Stability of the Atmosphere.} It is interesting to know if similar systems were investigated in the past. The solutions of the partial differential equation describing the evolution $t \rightarrow g_t(x)$ of the hydrodynamical flow of incompressible ideal fluid filled in a two-dimensional torus $x \in \CT^2$ can be considered as a continuous area preserving diffeomorphims $SDiff(\CT^2)$ of a torus $\CT^2 \rightarrow \CT^2$.  In the Arnold approach \cite{turbul} the ideal fluid flow is described by the geodesics $g_t(x) \in G$ on the  diffeomerphism group $G=SDiff(\CT^2)$ with the velocity $v_t(x)= \dot{g}_t(x) g^{-1}_t(x)$ belonging  to the corresponding algebra \textfrak{g}=sdiff$(\CT^2)$  of divergence-free vector fields. The Riemannian metric on the group $G$ is induced from the metric on a torus \cite{turbul}, and the stability of the geodesic flows on  the group $G$  can be analysed by investigating the behaviour of the corresponding sectional curvatures $K(v,\delta v)$ \cite{turbul,anosov}. 

It was found that the flow that is defined by a parallel velocity field on $\CT^2$ is unstable because the sectional curvatures  are negative and the flow is exponentially unstable. This shows that it is not possible to reasonably predict the weather beyond a certain period if one assumes that the Earth has a torus topology and its atmosphere is a two-dimensional incompressible fluid. A similar  stability analysis was performed  for the hydrodynamical flow on a two-dimensional sphere  $\CS^2$ in \cite{Arakelian:1989,Smolentsev,Smolentsev,Yoshida,Dowker:1990tb} as it is important to use the more realistic assumption that the surface of the Earth is a sphere.  In all these  cases  the flow is exponentially unstable in some directions and is stable in some other directions, resulting in the limitation of predictability of the hydrodynamical flow and  leading to the principal difficulties of a long-term weather forecasting.  
 
\section{ \it Hierarchy of DS and Kolmogorov Entropy}

In the ergodic theory the dynamical systems  are classified by the increase of their statistical-chaotic properties. {\it Ergodic systems} are defined as follows \cite{kornfeld,arnoldavez}.  Let $x=(q,p) $ be a point of the phase space $x \in M $ of the Hamiltonian systems.  The canonical coordinates are denoted as $q =(q_1,...,q_d)$ and $p=(p_1,...,p_d)$ are the conjugate momenta.  The phase space $M$ is equipped with a positive Liouville measure $d\mu(x)= \rho(q,p) dq_1...dq_d dp_1...dp_d$, which is invariant under the Hamiltonian flow.  The operator $T^t x = x_t$ defines the time evolution of the trajectory starting from the initial point  $x $ of the phase space. The ergodicity of the DS takes place if for almost all $x \in M$ \cite{Birkhoff} \footnote{In the following we shall be writing $dx$ instead of $d\mu(x)$ in order to simplify the expressions.}:   
\be
\lim_{t \rightarrow \infty} {1\over t} \int_{0}^{t} dt f(T^t x)  = \int_{M} f(x) dx ,
\ee
where $f(x)$ is a function/observable defined on the phase space $M $. The time average is equal to the space average almost everywhere, that is, for all $x$ except for a set of measure zero.  It follows then that  
\be
\lim_{t \rightarrow \infty} {1\over t} \int_{0}^{t} dt   f(T^t x) g(x) dx  = \int_{M} f(x) dx \int_{M} g(x) dx .
\ee
Considering the function $f = \chi_A$ on the phase space that is equal to one on a set $A \subset M$
and to zero outside, and similarly the  function $g = \chi_B$ on a set $B \subset M$,  one can get 
\be
\lim_{t \rightarrow \infty} {1\over t}~ \int_{0}^{t} dt~ \mu[ T^t A \cap  B ]  = \mu[A] \mu[B].
\ee
The measures of the points of the set $A$ that fall into the set $B$ {\it are on  average proportional} to the measures of these sets. The systems with stronger chaotic properties have been defined by Gibbs \cite{Gibbs,kornfeld}. The mixing takes place if for any two sets:
\be\label{mix1}
\lim_{t \rightarrow \infty} \mu[ T^t A \cap  B ]  = \mu[A] \mu[B],
\ee
that is, a part of the set $A$ that falls into the set $B$ {\it is  proportional} to their measures. Alternatively, 
\be\label{mix2}
\lim_{t \rightarrow \infty} \int  f(T^t x) g(x) dx  = \int_{M} f(x) dx \int_{M} g(x) dx,
\ee
which  means  that the two-point correlation function tends to zero: 
\be\label{mix2}
\CD_{t}(f,g) =   
   \lim_{t \rightarrow \infty}  \langle f(T^t x) g(x) \rangle - \langle f(x)\rangle \langle g(x) \rangle =0,
\ee
and is known in field theory language as the factorisation property of the two-point correlation functions. 
The n-fold mixing takes place if for any $n$  sets 
\be\label{mixn}
\lim_{t_n,...,t_1 \rightarrow \infty} \mu[ T^{t_n} A_n \cap ....T^{t_2} A_2 \cap T^{t_1} A_1 \cap  B ]  
= \mu[A_n]...   \mu[A_2] \mu[A_1] \mu[B]
\ee
or, alternatively, 
\be\label{mixnn}
\CD_{t}(f_n,...,f_1, g) =   
   \lim_{t_n,...,t_1 \rightarrow \infty}  \langle f_n(T^{t_n}x).....f_1(T^{t_1}x) g(x) \rangle - \langle f_n(x)\rangle .....\langle f_1(x)\rangle \langle g(x) \rangle =0.
\ee
A  class of dynamical systems that have even stronger chaotic properties was introduced by Kolmogorov in \cite{kolmo,kolmo1}.  These are the DS's which have a non-zero  entropy, so called  quasi-regular DS's, or simply  K-systems.  In order to define the Kolmogorov entropy let us consider a discrete time evolution operator $T^n x = x_n, ~n=0,1,2,..$.
Let $\alpha = \{A_i\}_{i \in I}$ ( $I$ is finite or countable)  be a  measurable partition of 
the phase space $M$ into the nonintersecting subsets $A_i$ that cover the whole phase space $M$, that is, 
\be
\mu(M \setminus \bigcup_{i \in I} A_i)=0,~~~~\mu(  A_i \bigcap   A_j)=0, i \neq j~,
\ee
and define the entropy of the partition $\alpha$ as 
\be
h(\alpha) = - \sum_{i \in I} \mu(A_i) \ln \mu(A_i).
\ee
If two partitions $\alpha_1$ and $\alpha_2$ differ by a set of 
measure zero, then their entropies are equal.  The {\it refinement partition} $\alpha$ 
\be
\alpha = \alpha_1 \vee \alpha_2  \vee ... \vee \alpha_k
\ee
of the 
collection of partitions $\alpha_1,..., \alpha_k$ 
is defined as the intersection of all their composing sets  $A_i$:
\be
\alpha = \big\{ \bigcap_{i \in I} A_i~ \vert ~A_i \in \alpha_i~ for ~all~ i  \big\}.
\ee
The entropy of the partition $\alpha$ with respect to the automorphisms T 
is defined as a limit \cite{kolmo,kolmo1,sinai3,rokhlin1,rokhlin,rokhlin2}:
\be
h(\alpha, T)= \lim_{n \rightarrow \infty} {h(\alpha \vee T \alpha \vee ...\vee T^{n-1} \alpha) \over n},~~~~
n=1,2,...
\ee
This number is equal to the entropy of the refinement 
$
\beta = \alpha \vee T \alpha \vee ...\vee T^{n-1} \alpha 
$
that was generated  during the iteration of the partition $\alpha$ 
by the  automorphism $T$. Finally the entropy of the 
automorphism $T$ is defined as a supremum: 
\be\label{supremum}
h(T) = \sup_{\{ \alpha \}} h(\alpha,T),
\ee
where the supremum is taken over all partitions $\{ \alpha \}$ of  $M$. 
It was proven that the K-systems have mixing of all orders: K-mixing $\supset$ infinite mixing, $\supset$,..n-fold mixing,..$\supset$ mixing $\supset$ ergodicity \cite{kolmo,kolmo1,sinai3,rokhlin1,rokhlin,rokhlin2}.
The calculation of the entropy for a given dynamical system seems extremely 
difficult. The theorem proven by Kolmogorov \cite{kolmo,kolmo1} tells that 
if one finds the so called "generating  partition" $\beta$, then
\be
h(T) = h(\beta,T), 
\ee
meaning  that the supremum in  (\ref{supremum}) is  reached on a generating partition 
$\beta$. In some cases the construction of the generating partition $\beta$
allows an explicit  calculation of the entropy of a given dynamical system \cite{sinai4,gines}. 

In summary, the above consideration allows to define the hierarchy of DS's with their increasing chaotic properties and with the maximally chaotic K-systems on the "top" of this hierarchy list.  The main question now is: Do the maximally chaotic systems exist and  how to decide to which ergodicity class belongs a DS under consideration? The Anosov C-condition  \cite{anosov} is a powerful tool defining the criteria under which a DS belongs to a maximally chaotic class.  The hyperbolic C-systems introduced by Anosov represent a large class of  K-systems and are structurally stable under small perturbations. We shall consider the C-systems in the next section. 

\section{\it   Anosov Hyperbolic  C-systems } 

\begin{figure}
\centering
\includegraphics[width=7cm]{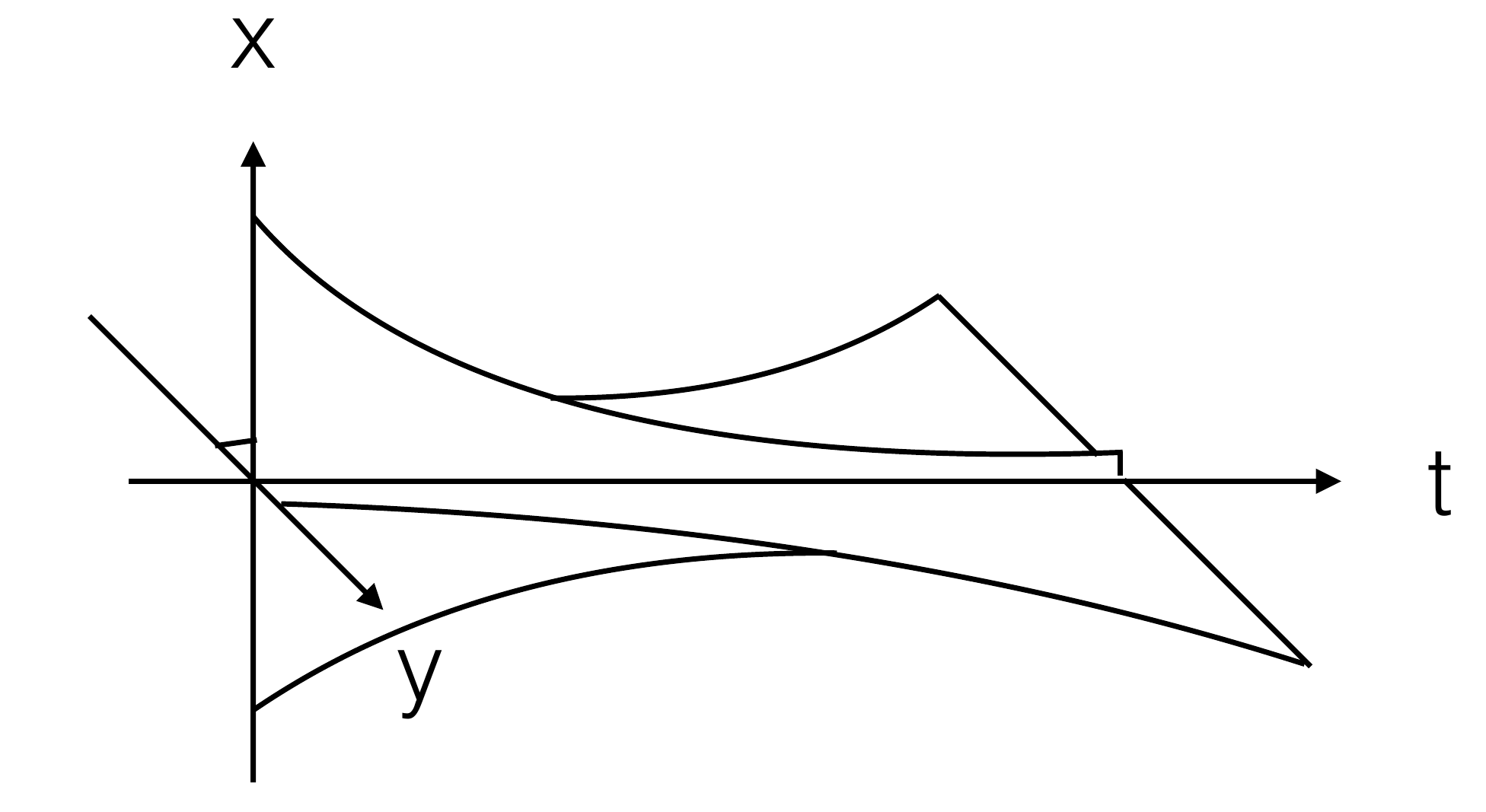}
\centering
\caption{The integral curves in the case of saddle point 
$\dot{x} =-x,~~~\dot{y}=y$ are exponentially contracting and expanding  near the solution $x=y=0$.  
A similar behaviour takes place in the neighbourhood of almost all trajectories 
of the C-system \cite{anosov} as one can get convinced inspecting the solutions (\ref{grows}) and (\ref{decay}) of the Jacobi variation equation (\ref{deviationequations1}) with negative sectional curvatures  (\ref{anosovinequality1}). } 
\label{fig1}
\end{figure}

In the fundamental work on geodesic flows on closed
Riemannian manifolds of negative curvature \cite{anosov} Anosov
pointed out  that the basic property of the geodesic flow on such manifolds 
is the {\it uniform  instability of the phase trajectories}, 
which in physical terms means that {\it in  the 
neighbourhood  of almost every fixed trajectory the trajectories 
behave similarly to the trajectories in the neighbourhood of a saddle point} (see Fig. \ref{fig1}).
In other words, the hyperbolic instability of the 
dynamical system  $ T^t $ which is defined by the equations \footnote{It is 
understood that the phase space manifold   $M$ is equipped by the invariant   
Liouville  measure \cite{anosov}.} 
\be\label{hyperbolic}
\dot{x} = f(x),~~~x=(x_1,...,x_{d})
\ee
takes place for almost all  solutions $\delta x \equiv \omega$ of the deviation equation  
\be
 \dot{\omega} = {\partial f \over \partial x } \bigg\vert_{x(t)=T^t x} \omega
\ee
in the neighbourhood of the phase trajectory  $x(t)=T^t x$, where $x\in M$. 

The exponential instability of geodesics on Riemannian manifolds of 
constant negative curvature 
has been studied by many authors, beginning with Lobachevsky \cite{Lobachevsky} and Hadamard \cite{Hadamard}
and especially  by  Artin \cite{Artin},  Hedlund  \cite{hedlund}, and Hopf \cite{hopf}.
The concept of exponential instability of a
dynamical system {\it trajectories}  appears to be extremely rich and Anosov suggested 
to elevate it into a fundamental property of a new class of dynamical systems
which he called C-systems\footnote{The letter C is used because of the "C condition"  \cite{anosov}}. 
The brilliant idea to consider dynamical systems 
which have  {\it local and homogeneous  hyperbolic instability of the phase trajectories }
is appealing to the intuition and  has deep  physical content \footnote{ The C-condition, which may sound similar to the existence of positive Lyapunov characteristic exponents of a dynamical system, is much stronger and is a sufficient condition for a system to be chaotic, to be a MCDS. A positive largest Lyapunov exponent does not in general indicate chaos \cite{perron,leonov1}. A negative largest Lyapunov exponent does not in general indicate stability \cite{perron,leonov1}.}.   
The richness of the concept 
is expressed by the fact  that  the C-systems 
occupy a nonzero volume  in the space of all dynamical 
systems \cite{anosov}\footnote{This is in a contrast 
with the integrable systems, where under arbitrary small perturbation $\delta f(x)$ of (\ref{hyperbolic}) the integrability will be partially destroyed, as it follows  from the KAM theory \cite{kolmo2}.} and have a non-zero Kolmogorov entropy.  
Anosov provided an extended list  of   MCDS  \cite{anosov}. 
The important examples of the MCDS  are:    {\it i) the geodesic flow on the Riemannian manifolds of  variable negative sectional curvatures and   ii)   C-cascades - the iterations of the hyperbolic automorphisms of tori}.   

In the forthcoming   sections we shall consider these maximally chaotic systems in details and the application of the MCDS theory to the investigation of the Yang-Mills dynamics, the N-body problem in gravity, fluid dynamics and in the Monte Carlo method.  We shall consider the quantum-mechanical properties of the maximally chaotic dynamical systems as  well and in particular  the DS's which are defined on the closed surfaces of constant negative curvature imbedded into the Lobachevsky hyperbolic plane  \cite{Artin,Gutzwiller}.

\section{\it  Geodesic Flow on  Manifolds of Negative Sectional Curvatures}

Let us consider the stability of the geodesic flow on a  Riemannian manifold $Q$ with local  coordinates $q^{\alpha} \in Q$ where $\alpha=1,2,....,3 N$.
The functions $q^{\alpha}(s) \in Q$ define a one-parameter  integral curve $\gamma(s)$  
on a Riemannian  manifold $Q$ and the corresponding  velocity vector
\be
u^{\alpha}= {d q^{\alpha} \over d s},~~~~~~\alpha=1,2,....,3 N~.
\ee
The proper time parameter $s$ along the $\gamma(s)$ is equal to  the length, while   
the Riemannian metric on $Q$ is defined as 
$$ds^2 = g_{\alpha\beta} d q^{\alpha} d q^{\beta},$$
and therefore 
\be\label{measure1}
g_{\alpha\beta}u^{\alpha} u^{\beta}=1.
\ee
A one-parameter family of deformations assumed to 
form a congruence of world lines 
$
q^{\alpha}(s) \rightarrow q^{\alpha}(s,\upsilon).
$
In order to characterise the infinitesimal deformation of the curve $\gamma(s)$ it is
convenient to define a separation vector 
$
\delta q^{\alpha} = {\partial q^{\alpha} \over \partial \upsilon} d \upsilon,
$
where $\delta q^{\alpha}$ is a separation of points having equal distance from some arbitrary
initial points along two neighbouring curves. 

The  resulting phase space manifold $(q(s),u(s)) \in M$ has a bundle structure 
with the base $q \in Q $ and the  spheres 
$S^{3N-1}$ of unit tangent vectors $u^{\alpha}$ (\ref{measure1}) as fibers.
The integral curve $\gamma(s)$ fulfils the geodesic equation 
\be\label{geodesicequation}
{d^2 q^{\alpha} \over d s^{2}}+ \Gamma^{\alpha}_{\beta\gamma} {d  q^{\beta} \over d s }
{d  q^{\gamma} \over d s }=0
\ee
and the {\it relative acceleration} depends only on the Riemann curvature:
\bea\label{deviationequations1}
{D^2 \delta q^{\alpha} \over ds^2}
 = - R^{\alpha}_{\beta\gamma\sigma} u^{\beta} \delta q^{\gamma} u^{\sigma} .
\eea
The above form of the Jacobi equation is difficult  to analyse, first 
of all because it
is written in terms of covariant derivatives $D   u^{\alpha}  = d u^{\alpha}  +
\Gamma^{\alpha}_{\beta\gamma} u^{\beta} d q^{\gamma} $.  And secondly because it is written in terms of separation of points on  trajectories  instead of distance between trajectories.  Following Anosov it is convenient to represent the  Jacobi equation in terms of simple derivatives. The norm of the deviation $\delta q$  has the form
$
 \vert \delta q  \vert^2   \equiv g_{\alpha\beta} \delta q^{\alpha} \delta q^{\beta}
$
and  its  second derivative is
\bea
{d^2\over ds^2} \vert \delta q  \vert^2  &=& 
2 g_{\alpha\beta}   \delta q^{\alpha} 
{D^2  \delta q^{\beta}\over ds^2} + 2 g_{\alpha\beta}  {D  \delta q^{\alpha}\over ds} 
{D  \delta q^{\beta}\over ds}  .
 \eea
Using (\ref{deviationequations1}) we shall get the Anosov form of the Jacobi equation 
\bea
 {d^2\over ds^2} \vert \delta q  \vert^2 &=& 
 -2   R_{\alpha \beta\gamma\lambda}  \delta q^{\alpha} u^{\beta} \delta q^{\gamma} u^{\lambda} +2 \vert  {D  \delta q \over ds} \vert^2=-2 K(q,u,\delta q) ~\vert u \wedge \delta q \vert^2 +2 \vert  \delta u  \vert^2 ~~~~~~~
\eea
where $K(q,u,\delta q)$ is the sectional curvature  in the two-dimensional directions 
defined by the velocity vector 
$u^{\alpha}$ and the deviation vector $\delta q^{\beta}:$
\be\label{sectional0}
 K(q,u, \delta q)  = { R_{\alpha\beta\gamma\sigma} \delta q^{\alpha} 
 u^{\beta} \delta q^{\gamma}  u^{\sigma} \over  \vert u \wedge \delta q  \vert^2 } . 
 \ee
One can decompose the deviation vector $\delta q$ into 
 longitudinal  and transversal components  
 $
 \delta q^{\alpha} = \delta q^{\alpha}_{\perp} + \delta q^{\alpha}_{\parallel},
 $ 
where $\delta q_{\parallel}$  describes a translation 
along the geodesic trajectories and has no physical interest, the transversal 
component $\delta q_{\perp}$ describes a physically relevant distance between original and infinitesimally close  trajectories $\vert u \delta q_{\perp} \vert =0$.  
Such decomposition allows to derive the Jacobi equation 
only in terms of transversal deviation\footnote{The area spanned 
by the bivector is simplifies  $\vert u \wedge \delta q_{\perp} \vert^2= \vert u \vert^2  \vert   \delta q_{\perp}  \vert^2 - \vert u \delta q_{\perp} \vert^2 = \vert \delta q_{\perp} \vert^2$, because 
$ \vert u \vert^2  =1$ and $\vert u \delta q_{\perp} \vert =0$.}:
\bea\label{transversaldev}
{d^2\over ds^2} \vert \delta q_{\perp}  \vert^2 &=& 
-2 K(q,u,\delta q_{\perp}) ~\vert \delta q_{\perp} \vert^2 
+2 \vert   \delta u_{\perp}  \vert^2. 
\eea
Because the last term is positive function the following inequality takes place for
{\it relative acceleration}: 
\bea\label{anosovinequality}
{d^2\over ds^2} \vert \delta q_{\perp}  \vert^2 \geq
-2 K(q,u,\delta q_{\perp}) ~\vert \delta q_{\perp} \vert^2. 
\eea
If the sectional curvature is negative and uniformly bound from above by a constant $\kappa$:
\be
K(q,u,\delta q_{\perp}) \leq - \kappa < 0, ~~~\text{where}~~~~~  \kappa = \min_{(q, u,\delta q_{\perp})} \vert K(q,u,\delta q_{\perp}) \vert 
\ee  
then 
\bea\label{anosovinequality1}
{d^2\over ds^2} \vert \delta q_{\perp}  \vert^2 \geq
2 \kappa ~\vert \delta q_{\perp} \vert^2.
\eea
The phase space of solutions of the second-order differential equation is divided into two separate sets  $X_q$ and $Y_q$\footnote{It follows from the variation equation (\ref{anosovinequality1})  and the boundary condition imposed on the deviation $\delta q_{\perp}$ and its first derivative ${d \over ds} \vert \delta q_{\perp}  \vert^2$  that for all $s$ the 
$
{d^2\over ds^2} \vert \delta q_{\perp}  \vert^2  >  0,
$
therefore  $\vert \delta q_{\perp}  \vert^2$ is a convex function and its graph is convex downward. Thus the variation 
equation   has no conjugate points because if $\delta q(s_1)=0$, $\delta q(s_2)=0$ and  $s_1 \neq s_2$
then  $\delta q(s) \equiv 0$ for $s_1 \leq s \leq s_2 $. The sets $X_q$ and $Y_q$ are defined as follows. The set $X_q$ consists of the vectors  $(\delta q(s), {d \delta q(s) \over d s}  )  \rightarrow 0$ when $s \rightarrow + \infty$ and the set $Y_q$ of the vectors $(\delta q(t), {d \delta q(s) \over d s}  )  \rightarrow 0$ when $s \rightarrow - \infty$. If $(\delta q(s), {d \delta q(s) \over d s}  )_{s=0} \in X_q$ then the first derivative is negative 
${d\over ds} \vert \delta q_{\perp}  \vert^2  < 0$  for all $s$.  As well if $(\delta q(s), {d \delta q(s) \over d s}  )_{s=0} \in Y_q$ then 
${d\over ds} \vert \delta q_{\perp}  \vert^2  > 0$  for all $s$.}
. The set $Y_q$ consists of the solutions with positive first derivative 
$$
{d \over ds} \vert \delta q_{\perp}(0)  \vert^2 > 0    
$$ 
and exponentially grows with $s \rightarrow + \infty$
\be\label{grows}
\vert \delta q_{\perp}(s) \vert \geq {1\over 2} \vert \delta q_{\perp}(0) \vert e^{\sqrt{2\kappa} s},
\ee
while the set $X_q$ consists of the solutions  with negative first derivative
$$
  {d \over ds} \vert \delta q_{\perp}(0)  \vert^2  < 0 
  $$ 
and decay exponentially with $s \rightarrow + \infty$
\be\label{decay}
\vert \delta q_{\perp}(s) \vert \leq {1\over 2} \vert \delta q_{\perp}(0) \vert e^{-\sqrt{2\kappa} s}.
\ee
This proves that the geodesic flow on closed Riemannian manifold of negative curvature fulfils the C-conditions, is therefore maximally chaotic and tends to equilibrium with exponential rate. We shall define a relaxation time as \cite{yer1986a,body}
\be\label{relaxationtime}
\tau = {1 \over \sqrt{2\kappa}}~,
\ee
 which is inversely proportional to the Kolmogorov entropy. 
 
 The above consideration demonstrate that the geodesic flow on closed Riemannian manifold of negative sectional curvature fulfils the C-condition and  defines a large class of maximally chaotic systems with nonzero Kolmogorov entropy.  This result provides a powerful tool for the investigation of the Hamiltonian systems.  If the time evolution of a classical Hamiltonian system  under investigation can be reformulated as the geodesic flow on the Riemannian manifold and if all sectional curvatures are negative, then it represents a MCDS. In physical terms this means that the phase space of a DS does not have invariant tori of an integrable system and its trajectories cover the whole phase space  \cite{kolmo2}.  In the next sections we shall apply this approach to the investigation of Yang-Mills Classical Mechanics \cite{Baseyan,Natalia,SavvidyKsystem,Savvidy:1982jk,Savvidy:1984gi}, to the N-body problem in Newtonian gravity \cite{body} and Monte Carlo method \cite{yer1986a}. The MCDS have a tendency to approach the equilibrium state with exponential rate depending on the entropy (\ref{relaxationtime}).  That will allow to calculate the relaxation  time of stars in galaxies \cite{body} and the quality of Monte Carlo generators \cite{yer1986a}.  The larger the entropy is, the faster a physical system tends to its equilibrium \cite{krilov,turbul,kornfeld,arnoldavez,Savvidy:2018ygo,Poghosyan:2018efd,Babujian:2018xoy}. 
 
\section{\it  Classical and Quantum Mechanics of Yang-Mills field}

It is crucially important to find and analyse the classical solutions of the Yang-Mills equations without the external sources in Minkowski space, which may prove to be useful for the construction and study of the Yang-Mills theory vacuum and asymptotic states \cite{Savvidy:1977as}. Searching the solutions of the classical Yang-Mills equations for which the Poynting vector vanishes, that is there is no energy flux,   it was found that space-homogeneous gauge fields satisfy the above condition \cite{Baseyan,Natalia,Asatrian:1982ga,SavvidyKsystem,Savvidy:1982jk,Savvidy:1984gi}. 

For space-homogeneous gauge fields $\partial_i A^a_k =0$, $i,k=1,2,3$ the Yang-Mills equation reduces to a classical mechanical system  with the Hamiltonian of the form \cite{Baseyan,Natalia,Asatrian:1982ga,SavvidyKsystem,Savvidy:1982jk,Savvidy:1984gi}
\be\label{YMclassical}
H= \sum_{i} {1\over 2} Tr \dot{A}_{i} \dot{A}_{i} + {g^2\over 4} \sum_{i,j}Tr [A_i,A_j]^2,
\ee
where the gauge field $A^a_{i}(t)$ depends only on time,  $i=1,2,3$, the index $a=1,...,N^2-1$ for $SU(N)$ group and in the Hamiltonian gauge $A_0=0$ the Gaussian constraint has the form
\be\label{hamiltonian} 
\CG=[\dot{A}_{i}, A_i]=0.
\ee
It is natural to call this system the Yang-Mills Classical Mechanics (YMCM). It is a mechanical system with $3 \cdot(N^2-1)$ degrees of freedom.  It is important to investigate classical equations of motion of this class of non-Abelian gauge fields, the properties of the separate solutions and of the system as a whole. In particular its integrability versus chaotic properties of the system. The YMCM has a number of conserved integrals: the space and isospin angular momenta  
\be\label{momenta}
m_i = \epsilon_{ijk} A^{a}_{j} \dot{A}^{a}_{k}, ~~~n^a = f^{abc} A^{a}_{i} \dot{A}^{a}_{i},~~~i=1,2,3~~~~a=1,...,N^2-1
\ee
in total $3 + (N^2-1)$ integrals, plus the energy integral (\ref{YMclassical}) (the  $n^a =0$ due to the constraint (\ref{hamiltonian})).  The original "white colour" solution found in \cite{Baseyan} has the form 
\be
A^{a}_{i} = \delta^{a}_{i} f(t),
\ee
and the corresponding chromoelectric and chromomagnetic fields  are:
\be
E^{a}_{i} = \delta^{a}_{i} \dot{f}(t),~~~~H^{a}_{i} = g \delta^{a}_{i} f^2(t).
\ee
The energy density is:
\be\label{energydensity}
\epsilon =T_{00}= {3\over 2} ( \dot{f}^2+ g^2 f^4 )=\mu^4,
\ee
where $\mu^4$ is a constant of dimension $mass^4$.  Unusual property of this solution is that the chromoelectric and chromomagnetic fields are parallel to each other and therefore the energy flux, the Pyonting vector, vanishes 
\be\label{Pyontingvector}
T_{0i}=S_i = \epsilon_{ijk}  E^{a}_{j} H^{a}_{k} =0.
\ee
The other important property is that the space components of $T_{\mu\nu}$ are diagonal:
\be\label{momentumdensity}
T_{ij} =  {1\over 2}\delta_{ij}  (E^{a}_{k} E^{a}_{k} +H^{a}_{k} H^{a}_{k})-E^{a}_{i} E^{a}_{j} -H^{a}_{i} H^{a}_{j} ={1\over 2}\delta_{ij} \Big( \dot{f}^2+ g^2 \dot{f}^4 \Big)=\delta_{ij} p
\ee 
and the full energy momentum tensor has the form of a relativistic matter: 
\be
\label{energymomentum}
T_{\mu\nu} = 
   \begin{pmatrix} 
      \epsilon & 0 & 0 & 0  \\
      0 & p & 0 & 0  \\
      0 & 0 & p & 0  \\
      0 & 0 & 0 & p   
   \end{pmatrix}.
\ee
Indeed it follows from relations (\ref{energydensity}), (\ref{Pyontingvector}) and  (\ref{momentumdensity})  that the Yang Mills equation of state is equivalent to a homogeneous  relativistic matter\footnote{The apparent inhomogeneity of the energy momentum tensor (\ref{momentumdensity}) in Electrodynamics due to the term $-E_{i} E_{j} -H_{i} H_{j} $ is a critical barrier for a successful vector field driven inflation  \cite{Golovnev:2008cf,Golovnev:2008hv}. In Yang Mills theory the energy momentum tensor is perfectly homogeneous (\ref{energymomentum}) and opens a room of possibilities for a vector field driven inflation  \cite{Savvidy:2021ahq}. I would like to thank Prof. Viatcheslav  Mukhanov for the discussion of this point. }
\be\label{equationofstate}
p = {1\over 3} \epsilon. 
\ee
The space homogeneous time-dependent vacuum solutions of the Yang Mills equations were considered in the context of the cosmological models and inflation in \cite{Savvidy:2021ahq,Maleknejad:2011sq,Elizalde:2012yk,Adshead:2012qe,Pasechnik:2013sga,Pasechnik:2016sbh,Pasechnik:2016sbh}.

The question is if the YMCM system is exactly integrable and has additional integrals of motions or it is a chaotic system. If the number of conserved integrals coincide with the number of degrees of freedom, then the system is exactly integrable and the phase trajectories lie on high-dimensional  tori, if there are less integrals, then the trajectories lie on a manifold of a larger dimension, and if there is no conserved integrals at all, then the trajectories are distributed over the full phase-space \cite{kolmo2}. 
\begin{figure}
\centering
\includegraphics[width=5cm]{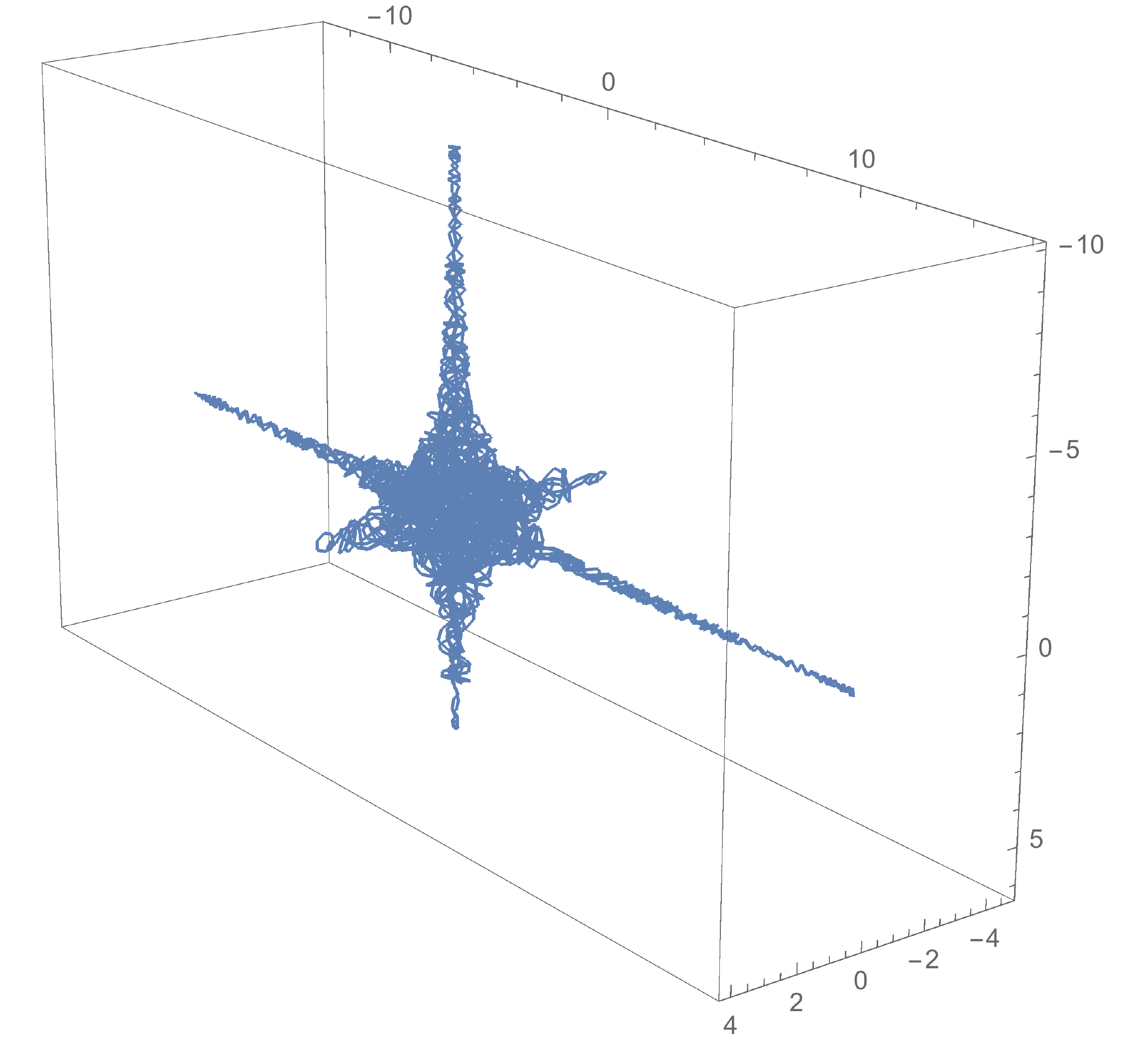}
\centering
\caption{A single trajectory of the YMCM  system integrated over a large time interval. The trajectory scatters on the equipotential surface  $ x^2_1 x^2_2 +x^2_2 x^2_3+x^2_3 x^2_1  = 1 $, densely filling the interior  region and making visible the six channels  of the equipotential surface on which a trajectory scatters.} 
\label{fig17}
\end{figure}

Let us consider in details the case of the $SU(2)$ gauge group by introducing the angular variables   which allow to separate   the angular  motion from the oscillations by using the substitution 
\be 
A = O_1 E O^T_2,
\ee
where $E=(x(t),y(t),z(t))$ is a diagonal matrix and $O_1,O_2$ are orthogonal matrices parametrised by  Euler angular variables. In this variables the Hamiltonian (\ref{hamiltonian}) will take the form \cite{Asatrian:1982ga}
\be\label{YMCM}
H_{FS}= {1\over 2}  (\dot{x}^2 + \dot{y}^2 + \dot{z}^2 )+  {g^2\over 2}( x^2y^2 +y^2z^2 +z^2x^2 ) + T_{YM},
\ee
where $T_{YM}$ is the rotational kinetic energy of the the  Yang-Mills "top" spinning in space and iso-space. The question is whether the YMCM system (\ref{YMCM}) is an integrable system or not \cite{Natalia,Asatrian:1982ga,SavvidyKsystem,Savvidy:1982jk}.  The general behaviour of the colour amplitudes  $(x(t),y(t),z(t))$  in time is characterised by rapid oscillations, decrease in some colour amplitudes and growth in others, colour "beats" \cite{Natalia} Fig.\ref{fig17}. The strong instability of the trajectories with respect to small variations of the initial conditions in the phase space led  to the conclusion that the system is stochastic and non-integrable. The search of the conserved integrals of the form $F(p_x,p_y,p_z,x,y,z)$ fulfilling the equation $\{ F, H_{FS}\} =0$ also confirms their absence, except the Hamiltonian itself. 
The evolution of the YMCM can be formulated as the geodesic flow on a Riemannian manifold with the Maupertuis's metric (see the details in the next section). The investigation of the sectional curvature \cite{Savvidy:1982jk} demonstrated that it is negative on the equipotential surface and generates exponential instability of the trajectories Fig.\ref{fig17}.  The solutions of an YMCM system in an arbitrary coordinate system \cite{Baseyan} (after a Lorentz boost) are nonlinear plane waves $A^a_{\mu}(k\cdot x)$ with a nonzero square of the wave vector $k^2=\mu^2$ chaotically oscillating in space-time.

The natural question which arrises here is to what extent the classical chaos influences 
the quantum-mechanical properties of the gauge fields. The significance of the answer to this question consists in the following. In field theory, e.g. in QED, the electromagnetic field is represented in the form of a set of harmonic oscillators whose quantum-mechanical properties (as of an integrable system) are well known, and the interaction between them is taken into account by perturbation theory. Such an approach excellently describes the experimental situation. In QCD the state of things is quite different. The properties of the YMCM as of a MCDS, cannot be established to any finite order of the perturbation theory. Therefore to understand QCD it seems important to investigate the quantum-mechanical properties of the systems which in the classical limit are maximally chaotic.

The natural question arising now is what quantum-mechanical properties does the system with the Hamiltonian (\ref{YMclassical})  possess if in the classical limit $\hbar \rightarrow 0$ it is maximally chaotic. What is the structure of the energy spectrum and of the wave functions of quantised gauge system, if in the classical limit it is chaotic. The Schr\"odinger equation for the gauge field theory in the $A_0 =0$ gauge has the following form \cite{Savvidy:1984gi,Maldacena:2015waa,Gur-Ari:2015rcq,Arefeva:1998,Arefeva:1999,Arefeva:1999frh,Arefeva:2013uta}:
\be\label{fulleq}
{1\over 2} \int d^3 x [ - {\delta^2 \over \delta A^a_i \delta A^a_i } + H^a_i H^a_i ] \Psi[A]=  E \Psi[A],
\ee
where $H^a_i = {1\over 2} \epsilon_{ijk} G^a_{jk}(A)$ and the constraint equations are:
\be\label{fulleq1}
[\delta^{ab} \partial_i + g f^{acb} A^c_i] {\delta \over \delta A^b_i  } \Psi[A]=0.
\ee
In case of space-homogeneous fields the equations will reduce to the Yang-Mills quantum-mechanical system (YMQM) with finite degrees of freedom which defines a special class of quantum-mechanical matrix  models \cite{ Savvidy:1982jk,Savvidy:1984gi,YMquantmech,Akutagawa:2020qbj}.   At zero angular momentum  $\hat{m}_i = 0$ (\ref{momenta}) the YMQM Schr\"odinger equation takes the form   (equations (20) and (21) in  \cite{Savvidy:1984gi})
\be
\Big\{-{1\over 2} D^{-1} \partial_i D \partial_i + {g^2\over 2}(x^2_1 x^2_2 +x^2_2 x^2_3+x^2_3 x^2_1) \Big\}\Psi = E \Psi ,
\ee
where $D(x) = \vert (x^2_1 - x^2_2)(x^2_2 - x^2_3)(x^2_3 - x^2_1)\vert$. Using the substitution 
\be
\Psi(x) = {1\over \sqrt{D(x)}}~ \Phi(x)
\ee
and the fact that the $D(x)$ is a harmonic function $ \partial^2_i D(x)=0$ the equation can be reduced to the form 
\beqa
 -{1\over 2} \partial^2_i  \Phi +  {1\over 2} \sum_{i<j} \Big( {1\over (x_i-x_j)^2} +{1\over (x_i+x_j)^2}   + g^2  x^2_i x^2_j\Big) \Phi=   E \Phi .
\eeqa
The analytical investigation of this Schr\"odinger equation is a challenging problem because the equation cannot be solved by separation of variables as far as all canonical symmetries are already extracted and the residual system possess no continuous symmetries. Nevertheless some important properties of the energy spectrum can be  established by calculating the volume of the classical phase space defined by the condition $H(p,q) \leq E$. It follows that classical phase space volume is finite and the energy spectrum of YMQM system is discrete  \cite{ Savvidy:1982jk,Savvidy:1984gi,YMquantmech}. Typically the classically chaotic systems have no degeneracy of the energy spectrum, the energy levels "repulse" from each other similar to the distribution of the eigenvalues of the matrices with randomly distributed elements \cite{Wigner,Mehta,Dyson}. The resent investigations revile that the YMQM successfully describes the QCD glueball spectrum \cite{Balachandran:2014iya,Pavel:2021mxn}.

In the next section we shall consider the $N$-body problem in classical Newtonian gravity 
analysing the geodesic flow on a Riemannian manifold equipped  with the Maupertuis  metric. 

\section{ N-body Problem. Collective Relaxation of Stellar Systems}

The important application of the theory of MCDS and geodesic flow of manifolds of negative sectional curvatures have found in astrophysics and cosmology. The $N$-body problem in Newtonian gravity can be formulated as a geodesic flow on Riemannian manifold with the conformal Maupertuis  metric  \cite{body}
\be\label{metricM}
ds^2 = (E-U) d \rho^2 = W \sum^{3N}_{\alpha=1} (d q^{\alpha})^2,~~~~~
U=-G \sum_{a < b}{M_a M_b \over \vert \vec{r}_a - \vec{r}_a \vert },
\ee
where $W=E-U$ and $\{q^{\alpha}\}$ are the coordinates of the $N$ stars: 
\be
\{q^{\alpha}\} = \{  M^{1/2}_1 \vec{r}_1,.....,M^{1/2}_N \vec{r}_N      \},~~~~\alpha =1,...,3N.
\ee
The equation of the geodesics (\ref{geodesicequation}) on Riemannian manifold  with the metric 
$g_{\alpha\beta} = W \delta_{\alpha\beta}$ in (\ref{metricM}) has the form 
\be
{d^2 q^{\alpha} \over ds^2 } + {1\over 2W}\Big(   2 {\partial W \over \partial q^{\gamma}}     
{d q^{\gamma} \over ds} {d q^{\alpha} \over ds}- g^{\alpha\gamma}  {\partial W \over \partial q^{\gamma}}    g_{\beta\delta} {d q^{\beta} \over ds} {d q^{\delta} \over ds}     \Big) =0
\ee
and coincides  with the classical N-body equations  when the proper time interval $ds$ is replaced by the time interval $dt$  of the form $ds = \sqrt{2} W dt$. The Riemann curvature  in (\ref{deviationequations1})
for the Maupertuis  metric  has the form 
\bea
R_{\alpha\beta\gamma\delta} &=& {1\over 2W} [ W_{\beta\gamma} g_{\alpha\delta}- 
W_{\alpha\gamma} g_{\beta\delta} - W_{\beta\delta } g_{\alpha\gamma} +W_{\alpha\delta } g_{\beta\gamma}] -\nn\\
&-& {3\over 4 W^2} [ W_{\beta} W_{\gamma} g_{\alpha\delta} - 
W_{\alpha} W_{\gamma} g_{\beta\delta} - W_{\beta} W_{\delta } g_{\alpha\gamma} +W_{\alpha} W_{\delta } g_{\beta\gamma}] + \nn\\
&+&{1\over 4 W^2} [ g_{\beta \gamma} g_{\alpha\delta} - 
g_{\alpha \gamma} g_{\beta\delta} ] W_{\sigma}W^{\sigma},
\eea
where $W_{\alpha} = \partial W/ \partial q^{\alpha}$ , $W_{\alpha\beta} = \partial W/ \partial q^{\alpha}\partial q^{\beta}$ and the scalar curvature is 
\bea\label{scalarcurve}
R = 3N(3N -1) \Big[- {\triangle W \over 3N W^2} -({1\over 4} - {1\over 2N}){(\nabla W)^2 \over W^3} \Big]
\eea
and $\triangle W  = \partial^2 W/ \partial q^{\alpha} \partial q^{\alpha}$,~  $\nabla W  = (\partial W/ \partial q^{\alpha})( \partial W/ \partial q^{\alpha})$.
Let us now calculate the sectional curvature (\ref{sectional0}):
\bea
R_{\alpha\beta\gamma\sigma} \delta q^{\alpha} 
 u^{\beta} \delta q^{\gamma}  u^{\sigma}&=& {1\over 2W} [~ 2 \vert u W^{''} \delta q \vert \vert u \delta q \vert - 
\vert  \delta q  W^{''}  \delta q \vert  \vert u u \vert - \vert  u  W^{''} u \vert  \vert \delta q \delta q \vert~] -\nn\\
&-& {3\over 4 W^2} [~ 2 \vert u W^{'}\vert  \vert  W^{'} \delta q \vert \vert u \delta q \vert - 
\vert  \delta q  W^{'}  \vert   \vert   W^{'}  \delta q \vert   \vert u u \vert  -  \vert  u  W^{'}  \vert   \vert    W^{'}  u \vert   \vert \delta q \delta q \vert ~ ] + \nn\\
&+&{1\over 4 W^2} [ ~ \vert u \delta q \vert^2- 
\vert u u \vert  \vert \delta q \delta q \vert ~ ] ~\vert W^{'}W^{'} \vert. 
\eea
For the normal deviation $q_{\perp}$   we have $\vert u \delta q_{\perp} \vert =0 $ and taking into account that the velocity is normalised to unity $\vert u u \vert  =1$   we have 
\bea\label{deviations}
R_{\alpha\beta\gamma\sigma} \delta q^{\alpha}_{\perp} 
 u^{\beta} \delta q^{\gamma}_{\perp}  u^{\sigma}&= &   \Big(~ {3\over 4 W^2} \vert  u  W^{'}  \vert^2     ~   
-{1\over 4 W^2} \vert W^{'}W^{'} \vert  -{1\over 2W}  \vert u W^{''} u \vert ~\Big)  \vert \delta q_{\perp}   \vert^2   \nn\\
&-& {1\over 2W}   ~    
\vert  \delta q_{\perp}  W^{''}  \delta q_{\perp} \vert   +{3\over 4 W^2} ~   
\vert  \delta q_{\perp}  W^{'}  \vert^2 .
\eea
A remarkable regularity in the light distribution in elliptical galaxies suggests that they have reached some form of natural equilibrium and therefore one can conjecture that the average value of the velocities and deviations can be taken in the form:
\be
\overline{u^{\alpha} u^{\beta}} = {1\over 3N} g^{\alpha\beta} \vert u u \vert,~~~~~\overline{\delta q_{\perp}^{\alpha} \delta q_{\perp}^{\beta}} = {1\over 3N} g^{\alpha\beta}  \vert \delta q_{\perp}   \vert^2~,
\ee
That reduces the equation (\ref{deviations}) to the form which contains  only the scalar curvature (\ref{scalarcurve}):
\bea\label{sectional2}
R_{\alpha\beta\gamma\sigma} \delta q^{\alpha}_{\perp} 
 u^{\beta} \delta q^{\gamma}_{\perp}  u^{\sigma}= \Big[-{1\over 3N} {\triangle W \over  W^2} -({1\over 4} - {1\over 2N}){(\nabla W)^2 \over W^3}\Big]  \vert \delta q_{\perp}   \vert^2 .
\eea
Thus the sectional curvature is proportional to the scalar curvature in this case. 
As the number of stars in  galaxies is very large, $N \gg 1$,  we shall get that the dominant term in sectional curvature (\ref{sectional2}) is negative: 
\be
K(q,u, \delta q)  = { R_{\alpha\beta\gamma\sigma} \delta q^{\alpha} 
 u^{\beta} \delta q^{\gamma}  u^{\sigma} \over  \vert \delta q_{\perp}   \vert^2}=   - {1\over 4}  {(\nabla W)^2 \over W^3}  < 0.
\ee
Finally the deviation equation (\ref{anosovinequality}) will take the form 
\bea\label{anosovinequality3}
{d^2\over dt^2} \vert \delta q_{\perp}  \vert^2 \geq
  {(\nabla W)^2 \over W}  ~\vert \delta q_{\perp} \vert^2,
\eea
where we used the relation $ds =   \sqrt{2} W dt$. The solution of this equation was  considered in section four and has the form (\ref{anosovinequality1}) with $2 \kappa =   (\nabla W)^2 / W$. Thus the relaxation time (\ref{relaxationtime}) can be defined as 
\be\label{relaxationtime1}
\tau = \sqrt{{W \over (\nabla W)^2 }},
\ee
where $W= T= N {M  <v^2> \over 2}$ is total kinetic energy of the stars and $\nabla W$ is a sum of the forces acting on the stars.  Each term in the sum can be approximated by a force ${G M \over d^2}$ acting on a star by a nearby star at a distance $d$, where  $d$ is the mean distance between stars. Thus we can get 
\be\label{relaxationtime}
\tau  \propto \Big({N M <v^2> \over 2} ~{d^4 \over  N M (G M)^2 } \Big)^{1/2}= {v \over 2 G M n^{2/3} }
\ee
Comparing collective relaxation time with the Chandrasekhar relaxation time $\tau_b$ which is due to the binary encounters of stars  \cite{Chandrasekhar} one can get 
\be
{\tau_b \over \tau} = { v^3 \over G^2 M^2 n \log N } {2 G M n^{2/3} \over v}= {v^2 \over G M n^{1/3}} {2 \over  \log N} ~\propto~ {d \over r_*}
\ee
where $r_* = {2 G M \over v^2}$ is a radius of effective binary scattering of stars. As far as the astrophysical observations revile that $d \gg r_*$ we will get that the collective relaxation time $\tau$ is much shorter than the binary relaxation time $\tau \ll \tau_b$.  These time scales together with the dynamical time scale $\tau_d ={D\over v}= D^{3/2} (G N M)^{-1/2}$, the scale corresponding to a time interval for a star to move around a gravitating system of a characteristic size $D$, are in the following relation 
\be\label{ratios}
\tau \approx  {D\over d} \tau_d, ~~~~\tau_b \approx  {D \over r_*} \tau_d
\ee
reflecting the appearance and the correspondence of time ($\tau_d, \tau, \tau_b$) and length ($D,d,r_*$) scales in the gravitational systems\footnote{ The above consideration was instigated during a private presentation of the collective relaxation mechanism to Prof. Viktor Ambartsumian. At the end of the presentation he made a remark that there should be some sort of correspondence between time and length scales in the extended gravitational systems.  After returning back to the office I calculated the ratios (\ref{ratios}) and found that indeed there is a direct correspondence between time scales  ($\tau_d, \tau, \tau_b$) and length scales ($D,d,r_*$)!} .  One can estimate the relaxation time (\ref{relaxationtime}) for the elliptic galaxies and globular clusters\footnote{In 1986 a seminar was organised at the ITEP in Moscow to present the results on the collective relaxation mechanism.  After the seminar Prof. Lev Okun suggested that one can arrange a meeting with Prof. Vladimir Arnold for further discussions of the collective relaxation. The meeting was organised at the Moscow State University and then at his home. Instead of discussing the N-body problem - it seemed that he had already been acquainted with the results on the collective relaxation mechanism - Arnold in a very clear physical terms explained the direct and inverse two-dimensional Radon transformation and presented his book "Catastrophe Theory" where he discussed caustics, a wave front propagation and classification of bifurcations \cite{Arnols-caus}.  At the end of the discussion Arnold suggested that the results should be presented also to Prof. Yakov Zeldovich. The meeting was scheduled at the Moscow State University where he had a lecture on that day. After the lecture he felt uncomfortable to proceed with the discussion at the University and drove his Volga car to the Sternberg Astronomical Institute. During the drive he told that in the last lecture he had presented to the students  the Pauli exclusion principle and then added that together with George Gamow they had attempted to "explain" it by repulsive force, but it came out to be impossible. (In Pauli's " General Principles of Quantum Mechanics" the author discussed  the attempts to explain the exclusion principle by a singular interaction force between two particles and remarked that in such attempts the difficulty lies in keeping the antisymmetric functions still regular, a  constraint that is difficult to fulfil.  A mathematically flawless realisation of the program was found by  Jaff\'e \cite{Jaffe}. Pauli stressed that the singularities were such that they barely could be realised in reality.)   In  Sternberg Institute Zeldovich walked around but then suggested to drive to the Kapitza Institute of Physical Problems where he had recently become the head of the theoretical department. The discussion took place in the Landau office that had beautiful armchairs and sofa with a blackboard in front, at the upper left corner of which  was the phrase written by chalk and signed by Dirac:  " It is more important to have beauty in one's equations than to have them fit experiment."  The question that was raised by Zeldovich during the presentation was about a possible overestimation of the number of star phase trajectories.  It was an unexpected question.   Arnold asked me to let him know how the meeting with Zeldovich went. I told him about the concern of Zeldovich regarding the statistics of the particle distribution in the phase space.  He responded that he already had a conversation with Zeldovich and the question has been settled, there were no overestimation of the number of star phase trajectories.  The question was about the statistical distribution of $N$ particles/stars in the phase space: Should the particles be considered as identical or distinguishable with exclusion or without exclusion principle?  The discussion of  similar questions can be  found in Lynden-Bell article  \cite{Lynden-Bell}).   Maybe the question echoed the previous conversation in the car of the exclusion principle. }
\be
\tau \approx 10^8 yr \Big({v \over 10 km/s}\Big) \Big( {n \over 1 pc^{-3}}\Big)^{-2/3} \Big( { M \over M_{\odot}  }\Big)^{-1},
\ee
where $v$ is the mean velocity, $n$ is the density and $M$ is the mean mass of stars  \cite{body}. This time is by few orders of magnitude shorter than the binary relaxation time\footnote{In 1990 I sent the article \cite{body} by a surface mail to Prof. Subrahmanyan Chandrasekhar and then visited him in Chicago University in 1991. He had the article on his desk, and we went through the derivation of the collective relaxation time. He asked me if a possible direct encounters of stars had been taken into consideration in this derivation. The first term in the sectional curvature (\ref{sectional2}) contains the Laplacian of the gravitational potential and as a consequence has a sum of delta function terms that correspond  to the direct encounters of stars. In a system with a large number of stars this term is suppressed by the factor $1/N$, and it can be safely discarded.  It seems that the observational data are also supporting the idea that direct encounters are rare. At the end of the discussion he asked me if I am working also in the field of particle  physics. I responded that Yang-Mills theory is my first love. Then  Chandrasekhar told that he divided theories into two categories: God-made and Man-made: Electrodynamics and General Relativity are God-made theories, and Yang-Mills theory is a Man-made theory! It seems that it was a reflection of his deep knowledge of these fields and of their compelling  beauty!} \cite{Chandrasekhar,garry,Lang}.  

It is interesting to observe that the Hubble Deep Field and the Hubble eXtreme Deep Field images revealed a large number of distant young galaxies seemingly in a non-equilibrium state, while the stars in the nearby older galaxies  show a more regular distribution of velocities and shapes, reflecting the  collective relaxation mechanism of stars. 

\section{\it  Artin Hyperbolic System}

In order to understand better the interrelation between classically chaotic systems and their quantum mechanical counterparts it seems natural to consider DS that are defined on closed surfaces of constant negative curvature \cite{Lobachevsky,Artin,Hadamard,hedlund,hopf,anosov}. These DS systems have been studded for a long time by mathematicians  \cite{Poincare,Poincare1,Fuchs}
and have deep roots in number theory, differential geometry, group theory, the theory of Fuchsian groups and automorphic forms  
\cite{maass,roeleke,selberg1,selberg2,bump,Faddeev,Faddeev1,hejhal2,winkler,hejhal,hejhal1,Takhtajan:2020hwl}.

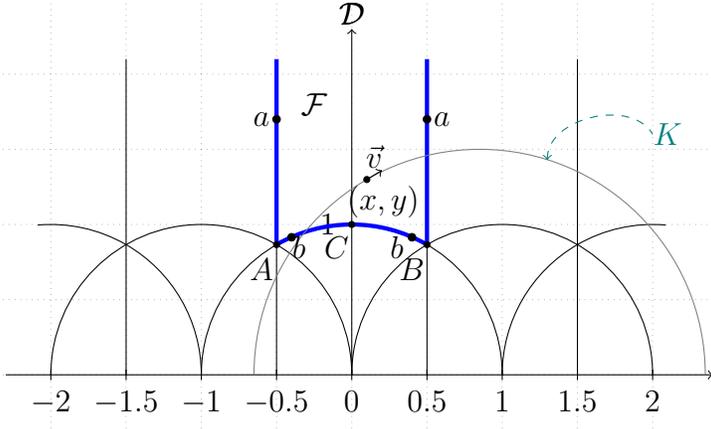
\begin{figure}
\centering
\begin{tikzpicture}[scale=2]
\hspace*{-3cm} 
\clip (-4.3,-0.4) rectangle (4.5,2.8);
\draw[step=.5cm,style=help lines,dotted] (-3.4,-1.4) grid (3.4,2.6);
\draw[,->] (-2.3,0) -- (2.4,0); \draw[->] (0,0) -- (0,2.3);
\foreach \x in {-2,-1.5,-1,-0.5,0,0.5,1,1.5,2}
\draw (\x cm,1pt) -- (\x cm,-1pt) node[anchor=north] {$\x$};
\foreach \y in {1}
\draw (1pt,\y cm) -- (-1pt,\y cm) node[anchor=east] {$\y$};
\draw (0,0) arc (0:180:1cm);
\draw (1,0) arc (0:180:1cm);
\draw (2,0) arc (0:180:1cm);
\draw (-1,0) arc (0:95:1cm);
\draw (1,0) arc (0:-95:-1cm);
\draw (1.5,0)--(1.5,2.1);
\draw (0.5,0)--(0.5,0.86602540378);
\draw (-0.5,0)--(-0.5,0.86602540378);
\draw (-1.5,0)--(-1.5,2.1);
\draw[ ultra thick, blue] (0.5,0.86602540378)--(0.5,2.1);
\draw[ ultra thick, blue] (-0.5,0.86602540378)--(-0.5,2.1);
\draw[ultra thick, blue] (0.5,0.86602540378) arc (60:120:1cm);
\draw (-0.25,1.8) node{ $\CF$};
\draw (0,2.4) node{ $\CD$};
\draw [thin,gray](2.35,0) arc (0:180:1.5cm);
\draw[fill] (0.1,1.29903810568) circle [radius=0.02];
\draw [->,] ((0.1,1.29903810568) to ((0.2,1.36);
\draw (0.21,1.15) node{ $(x,y)$};
\draw (0.15,1.45) node{ $\vec{v}$};
\draw (0,2.4) node{ $\CD$};
\draw (-0.6,0.7) node{ $A$};
\draw (0.4,0.7) node{ $B$};
\draw (-0.1,0.85) node{ $C$};
\draw[fill] (-0.5,0.86602540378) circle [radius=0.02];
\draw[fill] (0,1) circle [radius=0.02];
\draw[fill] (0.5,0.86602540378) circle [radius=0.02];
\draw[fill] (-0.5,1.7) circle [radius=0.025];
\draw[fill] (0.5,1.7) circle [radius=0.025];
\draw (-0.6,1.7) node{ $a$};
\draw (0.6,1.7) node{ $a$};
\draw[fill] ( -0.4,0.915) circle [radius=0.025];
\draw[fill] ( 0.4,0.915) circle [radius=0.025];
\draw (-0.35,0.85) node{ $b$};
\draw (0.3,0.85) node{ $b$};
\draw [<-,thin,  teal,dashed] (1.3,1.43) to [out=90,in=120] (2,1.6) ;
\draw  [thin,  teal] (2.1,1.6) node{ $K$};
\end{tikzpicture}
\caption{The  fundamental region $\CF$ is the hyperbolic triangle $ABD$,  the vertex D is at infinity of the $y$ axis. The edges of the triangle are the arc $AB$,  the rays $AD$   and $BD$. The points on the edges $AD$ and $BD$ and the points of the arks $AC$ with $CB$ 
should be identified by the transformations $w=z+1$ and $w = -1/z$ in order to form the Artin {\it  surface} $\bar{\CF}$
by  "gluing" the opposite sides of the modular triangle together Fig.\ref{scattering}.  The modular transformations (\ref{modular}) of the fundamental  region $\CF$ create a regular tessellation of the  whole Lobachevsky plane by congruent hyperbolic triangles. 
 K is the geodesic trajectory passing through the point ($x,y$) of $\CF$ in the $\vec{v}$ direction.}
\label{fig5} 
\end{figure}

Let us consider the DS's which are defined on closed surfaces on the Lobachevsky plane of constant negative curvature.  An example of such system has been defined in a brilliant article published in 1924 by the mathematician Emil Artin  \cite{Artin} (see also \cite{Gutzwiller,Gutzwiller:1971fy,Gutzwiller:1980}). The dynamical system is defined on the fundamental region of the Lobachevsky plane which is obtained by the identification of points congruent with respect to the modular group $SL(2,Z)$, a discrete subgroup of the Lobachevsky plane isometries $SL(2,R)$. The fundamental region in this case is a hyperbolic triangle Fig.\ref{fig5}. The geodesic trajectories  are bounded to propagate on the fundamental hyperbolic triangle.  The area  of the fundamental region is finite  and gets a topology of sphere by  "gluing" the opposite sides of the triangle as it is shown in Fig.\ref{fig5} and Fig.\ref{fig11}.   The  Artin symbolic dynamics, the differential geometry and group-theoretical methods of Gelfand and Fomin can be used to investigate the decay rate of the classical and quantum mechanical correlation functions.  The following three sections are devoted to the Artin system and are based on the results published in the articles \cite{Poghosyan:2018efd,Babujian:2018xoy,Savvidy:2018ffh}.

Let us start with the Poincare model of
the Lobachevsky plane, i.e. the upper half of the complex plane:
$H$=$\{z \in \mathbb{C}$, $\Im z >0\}$ supplied with the metric (we
set $z=x+i y$) 
\bea 
ds^2=\frac{dx^2+dy^2}{ y^2}\, 
\label{metric_hp}
\eea 
with the Ricci scalar $R=-2 $. Isometries 
of this space are given by $SL(2,\mathbb{R})$ transformations. 
The $SL(2,\mathbb{R})$ matrix ($a$,$b$,$c$,$d$ are real and $ad-bc=1$ )
\[
g=\left(
\begin{array}{cc}
a&b\\c&d
\end{array}
\right)
\] 
acts on a point $z$ by linear fractional substitutions
$
z\rightarrow \frac{az+b}{cz+d}~.
$
Note also that $g$ and $-g$ give the same transformation, hence 
the effective group is $SL(2,\mathbb{R})/\mathbb{Z}_2$.
We'll be interested in the space of orbits of a discrete subgroup
$\Gamma \subset SL(2,\mathbb{R})$ in $H$. Our main example will be 
the modular group $\Gamma=SL(2,\mathbb{Z})$. A standard choice of the 
fundamental region $\CF$ of $SL(2,\mathbb{Z})$ is displayed in Fig.\ref{fig5}.
The fundamental region $\CF$ of the  modular group consisting of those  points between the lines
$x=-\frac{1}{2}$ and $x=+\frac{1}{2}$ which lie outside the unit circle in  Fig.\ref{fig5}.
The modular triangle $\CF$ has two 
equal angles  $\alpha = \beta = \frac{\pi }{3}$ and the third one is equal to zero, $\gamma=0$, 
thus $\alpha + \beta + \gamma = 2 \pi /3 < \pi$.
The area  of the fundamental region is finite and equals to $\frac{\pi }{3}$ and gets a topology of sphere
by  "gluing" the opposite sides of the triangle. The invariant area element on the Lobachevsky plane is proportional to the square root of the determinant  of the metric (\ref{metric_hp}):
\be\label{me}
d \mu(z)= {dx dy \over y^2} \,.
\ee
Thus
$
\text{area}(\CF)=\int _{-\frac{1}{2}}^{\frac{1}{2}} dx
\int _{\sqrt{1-x^2}}^{\infty }\frac{dy}{y^2}  =\frac{\pi }{3}\,.
$
Following the Artin construction let us consider the model of the Lobachevsky plane realised 
in the upper half-plane $y>0$ of the complex plane  $z=x+iy \in \CC$
with  the Poincar\'e metric which is given by the line element (\ref{metric_hp}).

The Lobachevsky plane is a surface of a constant negative curvature, because its  curvature is equal to $R=g^{ik}R_{ik}= -2$  and it is twice the Gaussian curvature $K=-1$.
This  metric has two well known properties: 1) it is invariant with respect to all linear substitutions which form the group $g \in G$ of isometries of the Lobachevsky plane\footnote{$G$ is a subgroup of all M\"obius transformations. }:
 \beqa\label{real_frac_trans}
w= g \cdot z \equiv \left(
\begin{array}{cc}
a  & b   \\
c   & d  
\end{array} \right) \cdot z  \equiv \frac{a z +b}{c z +d},  
\eeqa
where $a, b,  c, d $ are {\it real coefficients of the matrix} $g $ and the determinant 
of $g$ is positive, $ a d -  b c  > 0 $.   The geodesic lines are either semi-circles orthogonal to the real axis  or rays perpendicular to the real axis. The equation for the geodesic lines on a curved surface has the form  (\ref{geodesicequation}),  where  the  Christoffer symbols  are evaluated for the metric (\ref{metric_hp}). The geodesic equations  take the form 
\beqa
&&\frac{d^2 x}{d t^2}-\frac{2}{ y} \frac{d x}{d t}\frac{d y}{d t}=0\,, ~~~~~
 \frac{d^2 y}{d t^2}+\frac{1}{ y}\left(\frac{d x}{d t}\right)^2-
\frac{1}{ y}\left(\frac{d y}{d t}\right)^2=0, \nn
\eeqa
and they have two solutions:
\beqa\label{traject}
& x(t)-x_0=r \tanh \left(t \right),~~~~  y(t)=\frac{r}{\cosh\left(t \right)}~~~&\leftarrow
\text{orthogonal semi-circles}~ \nn\\
& x(t)=x_0,~~~~~~~~~~~~~~~~~~~~~~  y(t)=e^t ~~~ &\leftarrow
\text{perpendicular rays }~.
\eeqa
Here $x_0 \in (-\infty, +\infty), t \in (-\infty, +\infty)$ and $r \in (0, \infty)$.
By substituting each of the above solutions into the metric (\ref{metric_hp}) one 
can get convinced that a point on the geodesics curve moves with a unit velocity (\ref{measure1})
\be\label{unit}
{ds \over dt } =1.
\ee 
In order to construct a closed surface $\CF$ on the  Lobachevsky plane, one can identify all points in the upper half of the plane which are related to each other by the substitution (\ref{real_frac_trans})  with the integer coefficients and a unit determinant. These transformations form  a modular group $ SL(2,\mathbb{Z})$. The two points $z$ and $w$  are identified if:
\begin{equation}\label{modular}
w =\frac{mz+n}{pz+q},~~~~
d=\left(
\begin{array}{cc}
m & n \\
p & q \\
\end{array} \right), ~~~~d \in SL(2,\mathbb{Z}) 
\end{equation}
with  integers $m$, $n$, $p$, $q$   constrained by the condition $mq-pn=1$.  The  $SL(2,\mathbb{Z})$  is 
the discrete subgroup of the isometry transformations $SL(2,\mathbb{R})$ of (\ref{real_frac_trans})\footnote{The modular group $SL(2,\mathbb{Z})$ serves as an example of the Fuchsian group \cite{Poincare,Fuchs}. Recall that Fuchsian groups are discrete subgroups of the group  of all isometry transformations $SL(2,\mathbb{R})$ of 
(\ref{real_frac_trans}). The Fuchsian group allows to tessellate the hyperbolic plane with regular polygons as faces, one of which can play the role of the fundamental region.}. 
The identification creates a regular tessellation of the  Lobachevsky plane by congruent hyperbolic triangles in Fig. \ref{fig5}.   The Lobachevsky plane is covered by the infinite-order triangular tiling. 
One of these triangles can be chosen as a fundamental region. That fundamental region $\CF$ of the above modular group (\ref{modular}) is the
well known "modular triangle", consisting of those  points between the lines
$x=-\frac{1}{2}$ and $x=+\frac{1}{2}$ which lie outside the unit circle in  Fig. \ref{fig5}.
The area  of the fundamental region is finite and equals to $\frac{\pi }{3}$. 
Inside the modular triangle $\CF$ there is exactly one representative
among all equivalent points of the Lobachevsky plane with the exception 
of the points on the triangle edges which are opposite to each other. 
These points can be identified in order to form a {\it closed surface} $\bar{\CF}$
by  "gluing" the opposite edges of the modular triangle together.
In  Fig. \ref{fig5} one can see the pairs of points on the sides of the triangle  
which are identified. Now one can consider the behaviour of the geodesic trajectories defined on the  surface $\bar{\CF}$ of constant negative curvature. 
Let us consider an arbitrary point $(x,y) \in \CF$ and the 
velocity vector $\vec{v} = (\cos \theta, \sin \theta)$. These are the coordinates of the 
phase space $(x,y,\theta) \in \CM$, and they uniquely determine  the geodesic trajectory  as 
the orthogonal circle $K$ in the whole Lobachevsky plane.  As this trajectory "hits" the edges 
 of the fundamental region $\CF$ 
and goes outside of it,  one should apply the modular transformation (\ref{modular}) to that parts of the circle $K$ which lie  outside of $\CF$ in order to return them back to the $\CF$. 
That algorithm will define the whole trajectory on $\bar{\CF}$ for $t \in (-\infty, +\infty)$.

Let us describe the time evolution of the physical observables $\{f(x,y, \theta)\}$ which are defined on the phase space  $(x,y,\theta) \in \CM$, where $z =x+iy \in \bar{\CF}$ and $\theta \in S^1$ is a
direction of a unit velocity vector.   For that one should know a  time evolution of geodesics on the phase space $\CM$. The simplest motion on  the ray  $CD$ in Fig.\ref{fig5},  
  is given by  the solution (\ref{traject})
$x(t)=0,~~  y(t)=e^{t} $ and can be represented as a group transformation (\ref{real_frac_trans}):
\beqa\label{transformation1}
  z_1(t) =g_{1}(t) \cdot i  = \left(
\begin{array}{cc}
e^{t/2} & 0 \\
0 & e^{-t/2} \\
\end{array} \right) \cdot i  = i e^{t}~,~~~  t \in (-\infty, +\infty).
\eeqa
The other motion on the circle of a unit radius, the arc $ACB$ in Fig.\ref{fig5}, is given by the transformation 
\beqa\label{transformation2}
  z_2(t) =g_{2}(t) \cdot i  = \left(
\begin{array}{cc}
\cosh(t/2) & \sinh(t/2) \\
\sinh(t/2) & \cosh(t/2) \\
\end{array} \right) \cdot i  = {i \cosh(t/2) +  \sinh(t/2) \over i \sinh(t/2) + \cosh(t/2)} .
\eeqa
Because the isometry group $SL(2,\mathbb{R})$ acts transitively on the Lobachevsky plane, any  geodesic 
can be  mapped into any other  geodesic through the action of the  group element $g \in SL(2,\mathbb{R}) $ (\ref{real_frac_trans}), thus 
the  generic trajectory can be represented  in the following form: 
\be\label{transformation3}
z(t)= g g_{1}(t) \cdot i = \begin{pmatrix}
 a e^{t/2}  & b e^{-t/2} \\
  c e^{t/2} & d e^{-t/2}
 \end{pmatrix} \cdot  i , ~~~z(t)= { i a e^{ t} + b  \over i c e^{ t} + d }~.
\ee
This provides a convenient description of the time evolution of the geodesic flow on the whole Lobachevsky plane with a unit velocity vector (\ref{unit}). In order to project this motion into  the
closed surface  $\bar{\CF}$ one should identify the group elements $g \in SL(2,\mathbb{R})$ which are 
 connected by the modular transformations $SL(2,\mathbb{Z})$. For that one can use  
a parametrisation of the group elements  $g \in SL(2,\mathbb{R})$ defined in \cite{Gelfand}. Any element $g$ can be defined by the parameters   $(\tau, \omega_2)$
\be\label{phasespace}
(\tau, \omega_2), ~~~\tau \in \CF ,~~~~\omega_2 = {e^{i \theta} \over  \sqrt{ y}} ,~~~
0 \leq \theta \leq 2 \pi,
\ee
where $\tau=x+iy$ are the coordinates in the fundamental region $\CF$ and the 
angle $\theta $ defines the direction of the unit velocity vector $\vec{v} = (\cos \theta, \sin \theta)$  at the point $(x,y)$ (see Fig.\ref{fig5}).   The functions $\{f(x,y, \theta)\}$ on the phase space $  \CM$ can be written 
as depending on  ($ \tau, \omega_2$)  and the invariance of  the functions  
with respect to the modular transformations  (\ref{modular}) takes the form\footnote{This defines the automorphic functions, the generalisation of the trigonometric, hyperbolic, elliptic  and other periodic functions \cite{Poincare,Ford}.}
\be\label{periodic}
f( \tau',\omega'_2) = f\Big( { m \tau+ n \over p\tau +q } ,(p \tau +q) \omega_2 \Big) = f (\tau,\omega_2).
\ee
The evolution of the  function $\{f(\tau, u)\}$, where  $ u= e^{-2i \theta }$,  under the geodesic flow $g_1(t)$  (\ref{transformation1}) is defined   by the mapping 
 \be\label{evolu1}
 \tau' = 
 {\tau \cosh( t/2 )  + u ~\overline{\tau} ~\sinh( t/2 )  \over \cosh( t/2 )   + u ~\sinh( t/2 ) },~~~~~
u' = 
{u \cosh( t/2 )  + \sinh( t/2 )   \over  \cosh( t/2 )    + u \sinh( t/2 ) },
\ee 
The evolution of the  observables under the geodesic flow $g_2(t)$ (\ref{transformation2}) has a  
similar form, except of an additional factor $i$ in front of the variable  $u$.
These expressions allow to define the transformation of the functions  $\{ f(x,y,u)\}$
under the time evolution as $f(x,y,u) \rightarrow  f(x',y',u')$ where
\be\label{coor1}
x'=x'(x,y,u,t),~~~y'=y'(x,y,u,t),~~~u' = u'(u, t). 
\ee
By using the Koopman \cite{Koopman}  theorem this transformation of functions can be 
expressed as an action of a one-parameter 
group of  unitary operators $U_{t}$:
\be\label{geoflow}
U_{t} f(g ) = f(g g_t ).
\ee
Let us calculate  transformations which are induced by $g_1(t)$  and $g_2(t)$  
 in (\ref{transformation1})-(\ref{transformation2}).  The time evolution   is given by the equations (\ref{evolu1}) and (\ref{coor1}):
\beqa\label{evolu}
&U_1(t) f(\tau,u) = f\Big({\tau \cosh( t/2 )  + u ~\overline{\tau} ~\sinh( t/2 )  \over \cosh( t/2 )   + u ~\sinh( t/2 ) } ,~ { u \cosh( t/2 )  + \sinh( t/2 )   \over  \cosh( t/2 )    + u \sinh( t/2 ) }\Big),\nn\\
&U_2(t) f(\tau,u) = f\Big(
{\tau \cosh( t/2 )  + i u ~\overline{\tau} ~\sinh( t/2 )  \over \cosh( t/2 )   + i u ~\sinh( t/2 ) },~{ u \cosh( t/2 )   -i \sinh( t/2 )   \over  \cosh( t/2 )    + i  u \sinh( t/2 ) } \Big).
\eeqa
A one-parameter family of unitary operators $U_t$ can be represented as an exponent of the 
self-adjoint  operator $U_t=\exp(i  H  t)$, thus we have 
\begin{equation}
U_{t} f(g )= e^{i  H  t} f(g ) = f(g g_t )
\end{equation}
and by differentiating it over the time $t$ at $t=0$ we shall get 
$
   Hf =-i \frac{ \mathrm d}{ \mathrm dt}U_t f \vert_{t=0}, 
$
that allows to calculate the operators $H$ corresponding to the  $U_1(t)$ and $U_2(t)$. 
Differentiating over time in (\ref{evolu})  we shall get for $H_1$ and $H_2 $:
    \begin{eqnarray}\label{koopman}
   2  H_1  =\frac{y }{u}\left(\frac{\partial }{\partial x}+i \frac{\partial }{\partial y}\right)-i \frac{\partial }{\partial u} - u y \left( \frac{\partial }{\partial x}-i\frac{\partial }{\partial y}\right) 
+i u^2 \frac{\partial }{\partial u}  \nn\\
 2 i H_2  =\frac{y }{u}\left(\frac{\partial }{\partial x}+i \frac{\partial }{\partial y}\right)-i \frac{\partial }{\partial u} + u y \left( \frac{\partial }{\partial x}-i\frac{\partial }{\partial y}\right)
-i u^2 \frac{\partial }{\partial u} . 
    \end{eqnarray}
Introducing annihilation and creation operators  $ H_{-}=H_1 -iH_2 $ and  $ H_{+}=H_1 +iH_2 $
 yields 
\begin{eqnarray}
H_{+}=
\frac{y }{u}\left(\frac{\partial }{\partial x}+i \frac{\partial }{\partial y}\right)-i \frac{\partial }{\partial u},~~~ 
H_{-}=- u y \left( \frac{\partial }{\partial x}-i\frac{\partial }{\partial y}\right)
+i u^2 \frac{\partial }{\partial u} 
\end{eqnarray}
and by calculating the commutator $[H_+, H_-]$ we shall get 
$
H_{0}=u \frac {\partial }{\partial u}$
and  their  $sl(2,R)$ algebra is:
\begin{eqnarray}
\left[H_+,H_-\right]=2 H_0, ~~
\left[H_0,H_+\right]=-H_+,~~
\left[H_0,H_-\right]=H_-.
\end{eqnarray}
We can also calculate the expression for the invariant Casimir operator: 
\be
H= {1\over 2} (H_+H_- + H_-H_+) - H^2_0 = -y^2(\partial^2_x + \partial^2_y) + 2 i y\partial_x  u\partial_u   = 
-y^2(\partial^2_x + \partial^2_y) -  y\partial_x  \partial_{\theta}.  
\ee
Consider a class of functions which fulfil the following two equations: 
\begin{eqnarray}
H_0 f(x,y,u) = -{N \over 2}  f(x,y,u) ,  ~~~~~
    H_{-} f(x,y,u)=0 ,  
    \end{eqnarray}
where $N$ is an integer number. The first equation has  the solution $f_N(x,y,u)= ({ 1  \over  u y})^{N/2} \psi(x,y)=\omega^N_2 \psi(x,y)$ and by substituting it into the second one we shall get 
\be
N \psi(\tau,\overline{ \tau }) +(\overline{ \tau }-\tau)\frac{\partial \psi(\tau,\overline{ \tau })}{\partial \tau } =0.
\ee
By taking  $\psi(\tau,\overline{ \tau }) = (\overline{ \tau }-\tau)^N \Phi(\overline{ \tau },\tau) $ we shall get 
the equation $\frac{\partial \Phi}{\partial \tau } =0 $, that is, $\Phi$
is a anti-holomorphic function and  $f(\omega_2,\tau,\overline{ \tau })$ takes the 
form\footnote{The factors  $(2 i)^N$ have been  absorbed  by the redefinition of $\Phi$.}
 \begin{eqnarray}\label{classfunc}
     f(\omega_2,\tau,\overline{ \tau })=\omega^N_2 (\overline{\tau}-\tau)^N \Phi
     (\overline{\tau}) = {1 \over \overline{\omega_2}^N }\Phi
     (\overline{\tau}).
     \end{eqnarray}
The invariance under the action of the modular transformation (\ref{periodic}) will take 
the form
\be
\Phi ({\frac{m \overline{\tau} + n  }{p \overline{\tau} + q  }}) =  \Phi(\overline{\tau}) 
(p \overline{\tau} + q )^N 
\ee
and $\Phi(\overline{\tau})$  is a theta  function of weight $N$ \cite{Poincare,Ford}. 
The invariant integration measure on the group $G$ is given by the formula 
 \cite{Hopf,Gelfand}
\begin{equation}
d \mu = \frac{dx dy}{y^2} d\theta
\end{equation}
and the invariant product of functions on the  phase space $(x,y,\theta) \in \CM$  will be given by the integral 
\beqa
(f_1 ,f_2)
&=& \int_{0}^{2\pi} d \theta \int_{\CF}f_1(\theta,x,y) \overline{f_2(\theta,x,y) }
{d x d y \over y^2}.
\eeqa
It was demonstrated that the functions on the phase space are of the form (\ref{classfunc}),
where $\tau = x + i y$ and ${(\tau-\overline{\tau})\over 2 i }= y$, thus  the expression for 
the scalar product  will takes the following form: 
\beqa
(f_1 ,f_2)
&=& \int_{0}^{2\pi} d \theta \int_{\CF} \Phi_1 (\overline{\tau})  ~  \overline{\Phi_2 (\overline{\tau})}  {1 \over   \vert \omega_2 \vert^{2N} }
  {d x d y \over y^2}=  2\pi  \int_{\CF} \Phi_1 (\overline{\tau})  ~  \overline{\Phi_2 (\overline{\tau})}  y^{N-2}
  d x d y ,
\eeqa
where $N \geq 2$. This expression for the scalar product allows to calculate the two-point correlation functions under the geodesic flow  (\ref{transformation1})-(\ref{transformation3}).

A  correlation function can be defined as an integral over a pair of  functions 
in which the first one is stationary and the second one evolves with the 
geodesic flow:
\beqa
\CD_{t}(f_1,f_2) &=&   \int_{\CM}f_1(g) \overline{f_2(g g_t) } d \mu .
\eeqa
By using (\ref{evolu1}) and (\ref{coor1})  one can represent the integral in the following form    \cite{Savvidy:1982jk}:
\beqa
\CD_{t}(f_1,f_2) &=&    
= \int_{0}^{2\pi}\int_{\CF}f_1[x,y,\theta] ~\overline{f_2[x'(x,y,\theta,t), y'(x,y,\theta,t), \theta'(\theta,t)] }
{d x d y \over y^2}d \theta.
\eeqa
From   (\ref{geoflow}), (\ref{evolu})  and (\ref{evolu1}), (\ref{evolu}) it follows that
\beqa
&f_1(\omega_2,\tau,\overline{ \tau }) = {1 \over  \overline{\omega_2}^N  } \Phi_1 (\overline{\tau}),~~~\\
&\overline{f_2(\omega'_2,\tau',\overline{ \tau' })} ={1 \over   \omega_2^N  (\cosh( t /2)  +  e^{-2i\theta} \sinh( t /2) )^N }  
\overline{ \Phi_2 \Big( {\overline{\tau} \cosh( t/2 )  + e^{-2i\theta} ~\tau ~\sinh( t/2 ) 
 \over \cosh( t/2 )   + e^{-2i\theta} ~\sinh( t/2 )  }\Big)}.\nn
\eeqa
Therefore the correlation function takes the following form:
\beqa
\CD_t(f_1,f_2) 
&=&   \int_{0}^{2\pi} d \theta \int_{\CF}    \Phi_1 (\overline{\tau})  \overline{\Phi_2 (\overline{\tau'})}  ~ {y^{N-2 } d x d y \over    (\cosh( t/2 )   +  e^{-2i\theta} ~\sinh( t/2 )  )^N }  .  \nn
\eeqa
The upper bound on the correlations functions is 
\beqa
 \vert \CD_t(f_1,f_2) \vert   \leq   
    \int_{0}^{2\pi} d \theta \int_{\CF}  \Big\vert    \Phi_1 (\overline{\tau})  \overline{\Phi_2 (\overline{\tau'})}  \Big\vert ~\Big\vert    {y^{N-2 } d x d y \over    (\cosh( t/2 )   +  e^{-2i\theta} ~\sinh( t/2 )  )^N }  \Big\vert ~ \nn
\eeqa
and  in the limit $t \rightarrow + \infty$ the correlation function exponentially decays:
\beqa
\vert \CD_t(f_1,f_2) \vert  & \leq &     ~ M_{\Phi_1 \Phi_2}(\epsilon)~  e^{-{N\over 2}\vert  t \vert }~~.
\eeqa
If the surface curvature is $K$ and the metric has the form  $ds^2 = {dx^2 +dy^2 \over K  y^2}$,
 then in the last formula the exponential factor will be \cite{Poghosyan:2018efd,Babujian:2018xoy,Savvidy:2018ffh}. 

\beqa
\vert \CD_t(f_1,f_2) \vert  & \leq &     ~ M_{\Phi_1 \Phi_2}(\epsilon)~  e^{-{N\over 2} K \vert t \vert}~
\eeqa
and the characteristic time decay   (\ref{relaxationtime}), (\ref{relaxationtime1})  \cite{yer1986a,Savvidy:2018ygo}  will take the form 
\be
\tau = {2\over N K }.
\ee
The decay time  of the correlation functions 
is shorter when the surface has a larger negative curvature or, in other words, when the divergency  
of the trajectories is stronger. 

 The behaviour of the phase trajectories and of the correlations functions discussed in the previous sections emphasise the fact that a local exponential divergency of the phase trajectories is translated into the exponential decay of the correlation functions at a universal rate expressible in terms of the entropy $h(T )$. This observation also provides a qualitative understanding of why  the correlation functions decay exponentially: under the action of the hyperbolic evolution $T^t$ on the observable $f(T^t x)$ the initial oscillating frequencies of $f(x)$ are stretched apart toward the high frequency modes, while the frequencies of the  observable $g(x)$ remain stationary.  As a result the overlapping integral of the function $f(T^t x)$ with the $g(x)$ falls exponentially.
 
The earlier investigation of the correlation functions of  Anosov geodesic flows was performed in 
\cite{Collet,Pollicot,moore,dolgopyat,chernov} by using alternative approaches.  In our analyses we have used the time evolution equations and the properties of the automorphic functions  \cite{Poghosyan:2018efd}. 

\section{\it  Quantum Mechanics of Artin System} 
 
In the previous section we analysed the behaviour of the classical correlation functions defined on the phase space of the Artin system, which represent a finite-area patch on $AdS_2$ and demonstrated their exponential decay. In this section we shall investigate the of quantum-mechanical behaviour of the correlation functions of quantised Artin system \cite{Babujian:2018xoy,Savvidy:2018ffh}.  There is a great interest in considering quantisation of the hyperbolic dynamical systems and investigation of their quantum-mechanical properties \cite{Savvidy:1982jk,Savvidy:1984gi,Gutzwiller,Gutzwiller:1971fy,Gutzwiller:1980}.  This subject is very closely related  with  the investigation of quantum mechanical properties of classically non-integrable systems.  

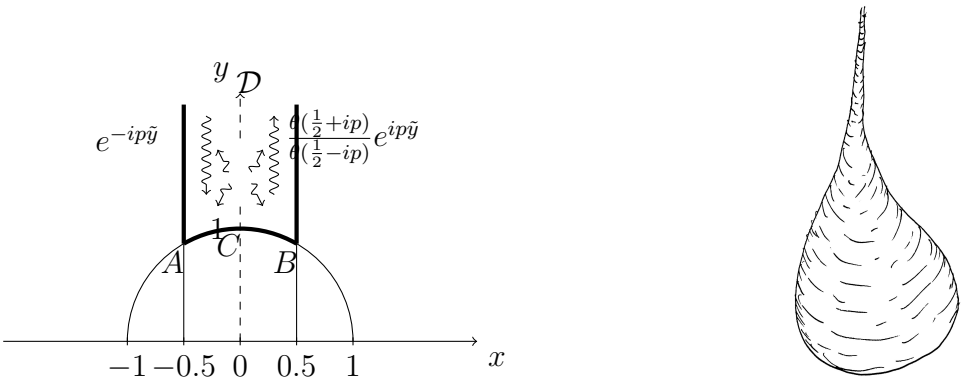
\begin{figure}[htbp]
	 \hspace*{-4cm} 
	\begin{tikzpicture}[scale=1.5]
	\clip (-4.3,-0.4) rectangle (4.5,2.8);
	\draw[,->] (-2.1,0) -- (2.1,0)  node[anchor=north west] {$x$ };
	\draw[dashed] (0,0) -- (0,1.2);
	\draw[dashed,->]  (0,1.8)--(0,2.2)  node[anchor=south east] {$y$};
	\foreach \x in {-1,-0.5 ,0,0.5,1}
	\draw (\x cm,1pt) -- (\x cm,-1pt) node[anchor=north] {$\x$};
	\foreach \y in {1}
	\draw (1pt,\y cm) -- (-1pt,\y cm) node[anchor=east] {$\y$};
	\draw (1,0) arc (0:180:1cm);
	\draw (0.5,0)--(0.5,0.86602540378);
	\draw (-0.5,0)--(-0.5,0.86602540378);
	\draw[ ultra thick, black] (0.5,0.86602540378)--(0.5,2.1);
	\draw[ ultra thick, black] (-0.5,0.86602540378)--(-0.5,2.1);
	\draw[ultra thick, black] (0.5,0.86602540378) arc (60:120:1cm);
	\draw (-0.6,0.7) node{ $A$};
	\draw (0.4,0.7) node{ $B$};
	\draw (-0.1,0.85) node{ $C$};
	\draw [->,snake=snake,
	segment amplitude=.6mm,
	segment length=1.5mm,
	line after snake=1mm] (-0.3,2) -- (-0.3,1.3);
	\draw [->,snake=snake,
	segment amplitude=.6mm,
	segment length=1.5mm,
	line after snake=1mm] (0.3,1.3)-- (0.3,2);
	%%%%%%%%%%%%%%%%%%%%%%%%%%%%%%%%%%%%%%%%%%%%%%%%%%%%
	\draw [->,snake=snake,
	segment amplitude=.6mm,
	segment length=1.5mm,
	line after snake=1mm] (0.1,1.5) -- (0.2,1.7);
	%%%%%%%%%%%%%%%%%%%%%%%%%%%%%%%%%%%%%%%%%%%%%%%%%%%
	\draw [->,snake=snake,
	segment amplitude=.6mm,
	segment length=1.5mm,
	line after snake=1mm] (-0.1,1.5) -- (-0.2,1.7);
	%%%%%%%%%%%%%%%%%%%%%%%%%%%%%%%%%%%%%%%%%%%%%%%%%%%
	%%%%%%%%%%%%%%%%%%%%%%%%%%%%%%%%%%%%%%%%%%%%%%%%%%%
	\draw [->,snake=snake,
	segment amplitude=.6mm,
	segment length=1.5mm,
	line after snake=1mm] (-0.1,1.4) -- (-0.2,1.2);
	%%%%%%%%%%%%%%%%%%%%%%%%%%%%%%%%%%%%%%%%%%%%%%%%%%%
	%%%%%%%%%%%%%%%%%%%%%%%%%%%%%%%%%%%%%%%%%%%%%%%%%%%
	\draw [->,snake=snake,
	segment amplitude=.6mm,
	segment length=1.5mm,
	line after snake=1mm] (0.1,1.4) -- (0.2,1.2);
	%%%%%%%%%%%%%%%%%%%%%%%%%%%%%%%%%%%%%%%%%%%%%%%%%%%
	\draw (1,1.8) node{ $\frac{\theta({1\over 2}+i p)}{\theta({1\over 2}-i p)}e^{i p \tilde{y}}$};
	\draw (-1,1.8) node{ $e^{-i p \tilde{y}}$};
	%%%%%%%%%%%%%%%%%%%%%%%%%%%%%%%%%%%%%%%%%%%%%%%%%%%%%%%
	\draw (0.08,2.3) node{ $\CD$};
	\end{tikzpicture}
	 \includegraphics[angle=0,width=3cm]{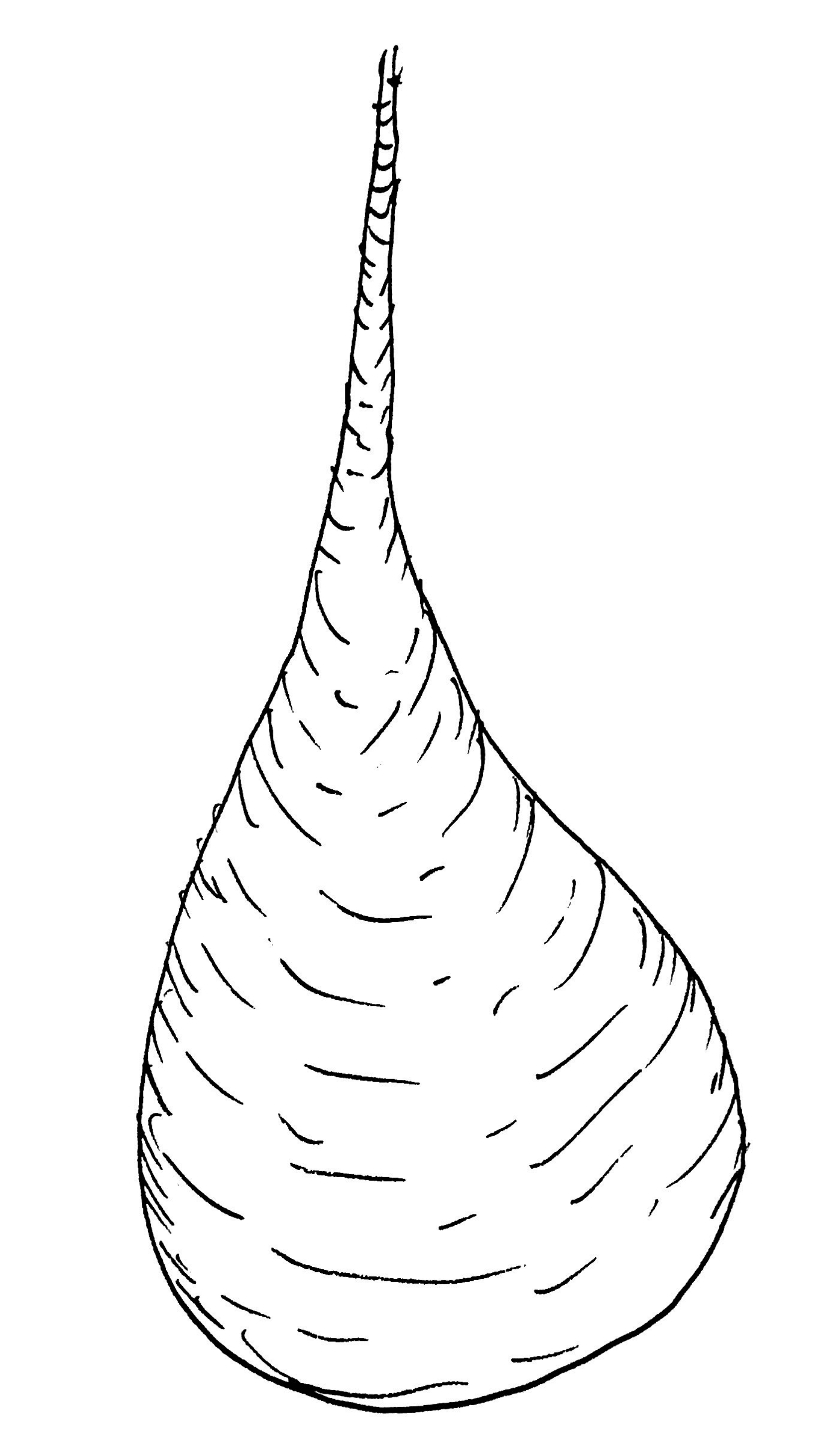}

	\caption{ The incoming and outgoing plane waves. The plane wave  $e^{-i p \tilde{y}  }$ incoming from infinity of the $y$ axis,    the vertex $\CD$,  elastically scatters on the boundary $ACB$ of the fundamental triangle  $\CF$ in Fig. \ref{fig5}, the bottom of the Artin surface $\bar{\CF}$.  The reflection amplitude $\theta(\frac{1}{2} +i p) / \theta(\frac{1}{2} -i p)$ is a pure phase and is given by the expression in front of the outgoing plane wave $e^{  i p \tilde{y}}$.
The rest of the wave function describes the standing waves in the $x$ direction between boundaries $x=\pm 1/2$ with the amplitudes, which are exponentially decreasing. On the right figure is an artistic image of the Artin surface, as far as it cannot be smoothly embedded  into $R^3$.}
	
	\label{scattering}
\end{figure}

 In classical regime the exponential divergency of  geodesic trajectories instigate the universal exponential decay of its classical correlation functions \cite{Savvidy:2018ygo,Poghosyan:2018efd}. 
In order to investigate the behaviour of the correlation functions in quantum-mechanical regime it is necessary  to know the spectrum of the system and the corresponding wave functions. The spectral problem has deep number-theoretical origin  and was partially solved in a series of pioneering publications \cite{maass,roeleke,selberg1,selberg2}.   It was solved partially because the discrete spectrum and the corresponding wave functions are not known analytically \cite{Takhtajan:2020hwl}.  The energy spectrum has  a continuous part corresponding to the  free motion along the infinite  "y -channel"  extended in the vertical direction of the fundamental region $\CF$ as well as infinitely many discrete energy states  corresponding to a bounded motion at the "bottom"  of the fundamental triangle Fig.\ref{scattering}.  The general properties of the discrete spectrum have been investigated by using Selberg trace formula \cite{selberg1,selberg2,bump, Faddeev,Faddeev1,hejhal2}.  Numerical calculation of the discrete energy levels were performed for many energy states   \cite{winkler,hejhal,hejhal1}.  

Here we shall review the quantisation of the Artin system defined on a finite-area patch on $AdS_2$ \cite{Babujian:2018xoy,Savvidy:2018ffh} and the derivation of the Maass wave function \cite{maass} corresponding to the continuous spectrum. We shall use  the Poincar\'e representation of the  Maass non-holomorphic automorphic wave function. By introducing a natural physical variable $\tilde{y}$ for the distance in the vertical direction  $\int dy/y =  \ln y = \tilde{y} $   and the corresponding momentum $p_y$ we shall 
 represent the  wave functions in the form appealing to the physical intuition: 
\bea\label{alterwave0}
&\psi_{p} (x,\tilde{y})  
= e^{-i p \tilde{y}  }+{\theta(\frac{1}{2} +i p) \over \theta(\frac{1}{2} -i p)}   \, e^{  i p \tilde{y}}  +{ 4  \over  \theta(\frac{1}{2} -i p)}   \sum_{l=1}^{\infty}\tau_{i p}(l)
K_{i p }(2 \pi  l e^{\tilde{y}} )\cos(2\pi l x).~~~~
\eea
The first two terms describe the incoming and outgoing plane waves. The plane wave  $e^{-i p \tilde{y}  }$ incoming from infinity  (the vertex $\CD$) along the $y$ axis of Fig. \ref{scattering}   elastically scatters on the boundary $ACB$ of the fundamental triangle  $\CF$ and reflects backwards $e^{  i p \tilde{y}}$.  The reflection amplitude is a pure phase and is given by the expression in front of the outgoing plane wave: 
\be\label{phase}
{\theta(\frac{1}{2} +i p) \over \theta(\frac{1}{2} -i p)} = \exp{[i\, \varphi(p)]}.
\ee
The rest of the wave function describes the standing waves $\cos(2\pi l x)$ between boundaries $x=\pm 1/2$
and exponentially decreasing amplitude $K_{i p }(2 \pi  l y )$ in $y$ and index $l$ (\ref{bessel}). The continuous energy spectrum is given by the formula  
\be
E=   p^2  + \frac{1}{4} .
\ee
It was conjectured that the wave functions of the discrete spectrum have the following form \cite{maass,roeleke,selberg1,selberg2,hejhal,hejhal1, winkler}:
\bea\label{wavedisc0}
\psi_n(z) &=&   \sum_{l=1}^{\infty} c_l(n) \,
 K_{i u_n }(2 \pi  l e^{\tilde{y}} ) 
\left\{  \begin{array}{ll} 
\cos(2\pi l x) \\
\sin(2\pi l x)   \\
\end{array} \right. , 
\eea
where the spectrum $E_n = {1\over 4} + u^2_n$ and the coefficients $c_l(n)$  are not known analytically but were computed numerically for many 
values of $n$ \cite{hejhal,hejhal1,winkler}.

Having in hand the explicit expression of the wave function one can analyse a quantum-mechanical  behaviour of the correlation functions and double commutators defined in \cite{Maldacena:2015waa}: 
\bea\label{basicopera1}
&\CD_2(\beta,t)=  \langle   A(t)   B(0) e^{-\beta H}   \rangle ,~~~~ 
 \CD_4(\beta,t)=  \langle  A(t)   B(0) A(t)   B(0)e^{-\beta H}   \rangle \\
\label{basicopera2}
&C(\beta,t) =   -\langle [A(t),B(0)]^2 e^{-\beta H} \rangle~,
\eea
and the operators $A$ and $B$ which we  chose to be of  the Louiville type \cite{Babujian:2018xoy}:
\be\label{basicopera}
A(N)=  e^{-2 N \tilde{y}},~~~N=1,2,.....
\ee
Analysing  the basic matrix elements of the Louiville-like  operators (\ref{basicopera}) we shall demonstrate   that all two- and four-point correlation functions (\ref{basicopera1}) decay exponentially in time, with the exponents which depend on temperature Fig.\ref{twopointfunc} and Fig.\ref{fourpointfunc}.  These exponents define the quantum-mechanical decay time $t_d(\beta)$. 
The square of the commutator of the Louiville-like  operators  separated in time (\ref{basicopera2}) grows exponentially Fig.\ref{timeevolutionofcommutator} \cite{Babujian:2018xoy}. This growth is reminiscent of the local exponential divergency of classical trajectories. The exponential growth Fig.\ref{timeevolutionofcommutator} of the commutator (\ref{basicopera2}) was approximated by the expression  
\be\label{exponentc12}
C(\beta,t) \sim K(\beta) \,e^{ {2\pi  \over  \chi(\beta)} t }.
\ee
It was conjectured \cite{Maldacena:2015waa} that a maximal growth of double commutator is linear in temperature  
\be\label{exponentc1t}
C(\beta,t)_{max} \sim K(\beta) \,e^{ {2\pi  \over  \beta} t }  .
\ee
It is therefore important to investigate the ratio with respect to the maximal growth exponent 
\be\label{exponentc123}
{C(\beta,t)_{max} \over  C(\beta,t)_{Artin-AdS_2}}  \sim R(\beta) \,e^{( {1  \over  \beta}  - {1  \over  \chi(\beta)} ) 2\pi  t }.
\ee
The result is presented in Fig.\ref{timeevolutionofcommutator}. The temperature dependence of  $1/\chi(\beta)$ relative to the maximum growth $1/\beta$ is shown by blue dots in the far right figure. At high temperatures the Artin Lyapunov exponent $1/\chi(\beta)$ is less than the maximal exponent $1/\beta$, but at low temperatures this is not any more true and we observe a breaking of the saturation regime \cite{Maldacena:2015waa,Gur-Ari:2015rcq,Cotler:2016fpe}.\\
\\
 {\bf{1.} }  {\it Quantisation }\\
\\
Let us consider the geodesic flow on $\CF$  described by the  action (\ref{metric_hp})
\bea
S=\int ds=\int \frac{\sqrt{\dot{x}^2+\dot{y}^2}}{ y}\,\,dt~
\eea 
and the equations of motion
\bea
 \frac{d}{dt}\,\, \frac{\dot{x}}{\,y\sqrt{\dot{x}^2+\dot{y}^2}}=0, ~~~~~~~~
 \frac{d}{dt}\,\, \frac{\dot{y}}{\,y\sqrt{\dot{x}^2+\dot{y}^2}}
+\frac{\sqrt{\dot{x}^2+\dot{y}^2}}{\,y^2}=0.
\label{EOM_generic}
\eea 
 Notice the invariance of the action and of the equations under time reparametrisations  
$t\rightarrow t(\tau)$. The presence of this local gauge symmetry 
indicates that we have a constrained dynamical system \cite{Faddeev:1969su}. 
A convenient  gauge fixing which
specifies the time parameter $t$ to be proportional  to the proper time, is archived by imposing 
the condition  
\bea
{ \dot{x}^2+\dot{y}^2 \over y^2}=2 H \,,
\label{gauge}
\eea
where $H$ is a constant. 
In this gauge the equations (\ref{EOM_generic}) will take the form \cite{Faddeev:1969su}
\bea
\frac{d}{dt}\,\,( \frac{\dot{x}}{\,y^2 })=0,~~~~~ \frac{d}{dt}\,\, (\frac{\dot{y}}{\,y^2})
+\frac{2H }{\,y}=0.
\label{EOM_generic1}
\eea 
Defining the canonical momenta as $
p_x = \frac{\dot{x}}{\, y^2},  ~ 
  p_y= 
\frac{\dot{y}}{\,y^2},
$ conjugate to the coordinates 
$(x,y)$,    
one can get the geodesic equations   (\ref{EOM_generic1})  in the Hamiltonian form: 
 \bea
 \dot{ p_x}=0,~~~~
\dot{p_y} =-
\frac{2H }{\,y} .
\label{EOM_generic2}
\eea 
The Hamiltonian will take the form
 \bea
 H = {1\over 2}y^2 (p^{2}_{x} +p^{2}_{y})
\label{hamiltonian1}
\eea 
and the corresponding equations will take the following form: 
\bea
&& \dot{x} = \frac{\partial H }{\, \partial p_x} = y^2 p_x,~\dot{y} = \frac{\partial H }{\, \partial p_y} = y^2 p_y
\\
&&\dot{p_x} = - \frac{\partial H }{\, \partial x} = 0,~\dot{p_y} = -\frac{\partial H }{\, \partial y} = - y(p^2_x +p^2_y)= -\frac{2H }{\,y},\nn
\eea
and they coincide with  (\ref{EOM_generic2}).
The advantage of the gauge (\ref{gauge}) is that the Hamiltonian (\ref{hamiltonian1}) coincides with the constraint. 
 
Now it is fairly standard to quantize this Hamiltonian system by replacing in (\ref{hamiltonian1})   
$
p_x=-i\frac{\partial}{\partial x},  
p_y=-i\frac{\partial}{\partial y}
$
and considering time independent  Schr\"odinger equation
$
H\psi = E \psi.
$
The resulting equation explicitly reads:
\bea
-y^2(\partial_x^2+\partial_y^2)\psi= E \psi.
\eea 
On the LHS one easily recognises  the  Laplace operator in Poincare metric (\ref{metric_hp})\cite{maass,roeleke,selberg1,selberg2,bump,Faddeev,Faddeev1,hejhal2}. It is easy to see that the Hamiltonian is positive semi-definite Hermitian operator: 
 \bea
-\int \psi^*(x,y )\, y^2(\partial_x^2+\partial_y^2) \, \psi(x,y ) {d x dy \over y^2}  
=  \int (\vert \partial_x \psi(x,y)\vert^2+ \vert \partial_y \psi(x,y)\vert^2) d x dy  \geq 0.~~~~
\label{ semiposti}
\eea 
It is convenient toparametrisation of the energy $E=s(1-s)$ and 
to rewrite the Schr\"odinger equation as
\bea
-y^2(\partial_x^2+\partial_y^2)~ \psi(x,y)  = s(1-s) ~\psi(x,y).
\label{Laplace_eq}
\eea 
As far as $E$ is real and semi-positive and parametrisation is symmetric with respect to $s \leftrightarrow 1-s $ it follows that the parameter $s$ should be chosen within the ranges
\be\label{srange}
s \in  [1/2 , 1  ]~~ \text{or } ~~ s=1/2 +i u ,~~~u ~\in~ [0,\infty].
\ee
One should impose the "periodic" boundary condition on the wave function  with respect to the modular group 
\bea\label{invariance}
\psi(\frac{a z+b}{c z+d})=\psi(z),~~~\left(
\begin{array}{cc}
a&b\\c&d
\end{array}
\right) \in SL(2,Z)
\eea
in order to have  the wave function which is properly defined on the fundamental 
region $\bar{\CF}$ shown in Fig. \ref{fig5} .
Taking into account that the transformation $T:z\rightarrow z+1$ 
is the element  of $SL(2,Z)$ and  imposing the periodicity condition (\ref{invariance})
$\psi(z)=\psi(z+1)$ one can get  the following Fourier expansion
$
\psi(x,y)=\sum_{n=-\infty}^\infty f_n(y)\exp(2\pi i n x).
$  
Inserting this into Eq. (\ref{Laplace_eq}), for the Fourier 
component $f_n(y)$ one can get the equation:
$$
\frac{d^2f_n(y)}{dy^2}+(s(1-s)-4\pi^2n^2)f_n(y)=0~.
$$ 
In the case $n\neq 0$ the solution which exponentially decays at 
large $y$ reads  
$
f_n(y)=\sqrt{y} K_{s-\frac{1}{2}}(2\pi n|y|)
$
and for $n=0$ one can get
$$
f_0(y)=c_0 y^s+c^{'}_0 y^{1-s}.
$$
Thus the solution can be represented in the form \cite{maass,roeleke,selberg1,selberg2,bump,Faddeev,Faddeev1,hejhal2}
\bea\label{solution3}
\psi(x,y) =  c_0 y^s+c^{'}_0 y^{1-s}  
+ \sqrt{y}\sum_{n=-\infty \atop n \neq 0}^\infty c_n  K_{s-\frac{1}{2}}(2\pi n|y|) \exp(2\pi i n x),
\eea
where the coefficients $c_0, c^{'}_0, c_n$ should be defined via periodic boundary condition (\ref{invariance}),  that is,  with respect to the second transformation $S:z\rightarrow -1/z$ ~:
$
\psi(z)=\psi(-1/z).
$
The corresponding functional equation defines the coefficients $c_0, c^{'}_0, c_n$. It is difficult to solve this equation and we will consider an alternative solution in the next section. \\ 
\\
{\bf{2.} }  {\it  Maass Wave Functions of Continuous Spectrum}\\
\\
 Another option is to take a particular solution and perform summation over all nonequivalent transformations of the $SL(2,Z)$ group \cite{Poincare,Poincare1,maass,roeleke,selberg1,selberg2,bump,Faddeev,Faddeev1,hejhal2}. Let us demonstrate this strategy by using  the solution (\ref{solution3}) when $c_0 =1, c^{'}_0=c_n=0$:
\[
\psi(z)= y^s = (\Im z)^s\, .
\] 
The $\Gamma_{\infty}$ is the
subgroup of $\Gamma = SL(2,Z)$ generating shifts $z\rightarrow z+n$, 
$n\in Z$.  Since $y^s$ is already invariant with respect to $\Gamma_{\infty}$, 
one should perform summation over the conjugacy classes 
$\Gamma_{\infty} \backslash \Gamma $.  There is  a bijection between the set of mutually 
prime pairs  $(c,d)$ with $(c,d)\neq (0,0)$ and the set of conjugacy 
classes $\Gamma_\infty\backslash \Gamma $. The fact that the integers
$(c,d)$ are mutually prime integers means that their greatest common divisor 
(gcd) is equal to one: $gcd(c,d)=1$.  As a result, it is defined by the classical Poincar\'e  series representation \cite{Poincare,Poincare1} and  
for the sum of our interest we get 
\bea
\psi_{s}(z) \equiv \frac{1}{2}\sum_{\gamma \in \Gamma_\infty\backslash \Gamma}
(\Im (\gamma z))^s   
=\frac{1}{2}\sum_{(c,d) \in \mathbb{Z}^2 \atop gcd(c,d)=1} \frac{y^s}{((c x+d)^2+c^2y^2)^s}~,
\label{sum1}
\eea   
where, as explained above, the sum on r.h.s. is taken over all 
mutually prime pairs $(c,d)$.  To evaluate the sum one should multiply  both sides of the
eq. (\ref{sum1}) by 
$
\sum_{n=1}^{\infty}\frac{1}{n^{2s}}
\equiv \zeta(2s)
$ \cite{maass}
so that the wave function will be expressed in terms of the Eisenstein series:  
\bea\label{solution4}
\zeta(2s) \, \psi_{s}(z) =\frac{1}{2} 
\sum_{(m,k) \in \mathbb{Z}^2 \atop (m,k)\neq(0,0)}
\frac{y^s}{((m x+k)^2+m^2y^2)^s} .
\eea
The evaluation of the sum can be performed explicitly and allows to represent the (\ref{solution4}) in the following form:
\bea
\zeta(2s) \, \psi_{s}(x,y) = \zeta(2s) y^s + \frac{\sqrt{\pi}\Gamma(s-\frac{1}{2})\zeta(2s-1)}
{\Gamma(s)}\,y^{1-s} +\nn\\
+\sqrt{y} \frac{4 \pi^s}{\Gamma(s)} \sum_{l=1}^{\infty}\tau_{s-\frac{1}{2}}(l)
K_{s-\frac{1}{2}}(2 \pi  ly )\cos(2\pi l x) ,
\eea 
where the modified Bessel's $K$ function is given by the expression
\be\label{bessel}
K_{i u}(y) = {1\over 2} \int^{\infty}_{-\infty} e^{- y \cosh t} e^{i u t}   dt
\ee
and 
$
\tau_{i p}(n )=\sum_{a \cdot b=n}\left(\frac{a}{b}\right)^{ip} .
$
By using Riemann's reflection relation
\bea 
\zeta (s)=\frac{\pi ^{s-\frac{1}{2}} \Gamma \left(\frac{1-s}{2}\right)}
{ \Gamma \left(\frac{s}{2}\right)}\,\zeta (1-s)
\eea
and introducing the function
\bea\label{thata1}
\theta(s)=\pi ^{-s} \zeta (2 s) \Gamma (s)
\eea  
we get an elegant expression of the eigenfunctions obtained by Maass \cite{maass}:
\bea\label{elegant}
\theta(s) \psi_{s}(z) =\theta (s)y^s+\theta(1-s)\,y^{1-s}  
+4\sqrt{y} \sum_{l=1}^{\infty}\tau_{s-\frac{1}{2}}(l)
K_{s-\frac{1}{2}}(2 \pi  ly )\cos(2\pi l x) .\qquad 
\eea 
This wave function is well defined in the complex $s$ plane and has a simple pole at $s=1$.
The physical continuous spectrum was defined in (\ref{srange}), where  $s=\frac{1}{2} +iu $, $u \in [0,\infty]$, therefore  
\be\label{eigenvalues}
E= s(1-s)= \frac{1}{4} +u^2 .
\ee
The continuous spectrum wave functions $\psi_s(z)$ are delta function normalisable \cite{maass,roeleke,selberg1,selberg2,Faddeev,bump}.
The wave function (\ref{elegant}) can be  conveniently represented also in the form 
\bea
\psi_{ \frac{1}{2} +i u}(z) = y^{ \frac{1}{2} +i u}+{\theta(\frac{1}{2} -i u) \over \theta(\frac{1}{2} +i u)}   \,y^{\frac{1}{2} -i u}  
+{4\sqrt{y} \over \theta(\frac{1}{2} +i u)}   \sum_{l=1}^{\infty}\tau_{i u}(l)
K_{i u }(2 \pi  ly )\cos(2\pi l x) , 
\eea
where 
$
K_{-i u}(y ) =K_{i u }( y),~~~~~\tau_{-i u}(l) =\tau_{i u}(l)~. 
$
The physical interpretation of the wave function becomes  more transparent if one introduce the new variables 
\be\label{newvariab}
\tilde{y} = \ln y,~~~~ p= -u,~~~~E=   p^2  + \frac{1}{4}, 
\ee
as well as  the alternative normalisation of the wave function $\psi_{p} (x,\tilde{y})  \equiv
y^{- \frac{1}{2}} \psi_{ \frac{1}{2} +i u}(z)  $
\bea\label{alterwave}
\psi_{p} (x,\tilde{y})    
 = e^{-i p \tilde{y}  }+{\theta(\frac{1}{2} +i p) \over \theta(\frac{1}{2} -i p)}   \, e^{  +i p \tilde{y}}  + { 4  \over  \theta(\frac{1}{2} -i p)}  \sum_{l=1}^{\infty}\tau_{i p}(l)
K_{i p }(2 \pi  l e^{\tilde{y}} )\cos(2\pi l x) .
\eea
The first two terms describe the incoming and outgoing plane waves. The plane wave  $e^{-i p \tilde{y}  }$ incoming along the $y$ axis elastically scatters on the boundary $ACB$ of the fundamental region $\CF$ Fig.\ref{scattering} .  The reflection amplitude is a pure phase and is given by the expression in front of the outgoing plane wave $e^{  i p \tilde{y}}$ 
\be\label{phase}
{\theta(\frac{1}{2} +i p) \over \theta(\frac{1}{2} -i p)} = \exp{[i\, \varphi(p)]}.
\ee
The rest of the wave function describes the standing waves between boundaries $x=\pm 1/2$
with exponentially decreasing  amplitudes $K_{i p }(2 \pi  l y )$.

In addition to the continuous spectrum the system (\ref{Laplace_eq}) 
may have  a discrete spectrum \cite{maass,roeleke,selberg1,selberg2,Faddeev,bump}.
The number of discrete states is infinite: $E_0=0 < E_1 < E_2 < ....\rightarrow \infty$.  The spectrum is extended to infinity -  unbounded from above -  and lacks any accumulation  point  except infinity. 
 The wave functions of the discrete spectrum have the form (\ref{wavedisc0}) \cite{maass,roeleke,selberg1,selberg2, winkler,hejhal,hejhal1}. 

\section{\it Out-of-Time-Order Correlation Functions of Artin System}

Here we are interested to analyse the behaviour of the out-of-time-order correlation functions (\ref{outoftime}), (\ref{fourpointfunct1}), (\ref{fourpointfunct2}) and the double commutators (\ref{commutatorL}) in the case of well defined Artin MCDS investigating the "influence and remnants" of the classical chaos on the quantum mechanical behaviour of the quantised system.  Considering Artin system  in its quantum mechanical regime  would help to identify the traces of classical chaos  and clarify a natural meaning of quantum chaos on a finite-area patch on $AdS_2$ \cite{Artin,Poincare,Poincare1,Fuchs,maass,roeleke,Gelfand,selberg1,selberg2,Faddeev,Faddeev1,Takhtajan:2020hwl, hejhal2,hejhal,hejhal1,Ford,winkler,bump,Collet,Pollicot,moore,dolgopyat,chernov,Poghosyan:2018efd,Babujian:2018xoy}.
 
The two-point correlation function is defined as: 
\bea\label{corr1}
&\CD_2(\beta,t)=  \langle     A(t)   B(0) e^{-\beta H}   \rangle = 
  \sum_{n}  \langle  n \vert e^{i H t } A(0) e^{-i H t } B(0) e^{-\beta H} \vert n  \rangle=\nn\\
 &=\sum_{n,m} e^{i (E_n -E_m)t - \beta E_n}   \langle  n \vert   A(0)\vert m  \rangle  \langle m\vert  B(0)  \vert n  \rangle.
\eea
The energy eigenvalues (\ref{eigenvalues}) are parametrised by $n = {1\over 2} +i u$, $E_n =\frac{1}{4} + u^2 $ and $m = {1\over 2} +i v$, $E_m =\frac{1}{4} + v^2 $, thus \cite{Babujian:2018xoy}
\bea
&\CD_2(\beta,t)=\int^{+\infty}_{0} \int^{+\infty}_{0} du \,  dv  ~ e^{i (u^2 -v^2)t - \beta( \frac{1}{4} + u^2)}  
  \\
&\int_{\CF}\psi_{ \frac{1}{2} -i u}(z)  \, A \, \psi_{ \frac{1}{2} +i v }(z) \,  d\mu(z) 
\int_{\CF}\psi_{ \frac{1}{2} -i v }(w)  \, B \, \psi_{ \frac{1}{2} +i u}(w) \,  d\mu(w)~,\nn
\eea
where the complex conjugate  function is $\psi^*_{ \frac{1}{2} +i u}(z) =\psi_{ \frac{1}{2} -i u}(z) $.
Defining the basic matrix element as
\bea\label{basicmatrix}
A_{uv} = 
\int_{\CF}  \psi_{ \frac{1}{2} -i u}(z)  \, A \, \psi_{ \frac{1}{2} +i v }(z) \,  d\mu(z)  = 
\int^{1/2}_{-1/2} dx  \int^{\infty}_{\sqrt{1-x^2}} {dy \over y^2}  \psi_{ \frac{1}{2} -i u}(z)  \, A \, \psi_{ \frac{1}{2} +i v }(z)   
\eea
for the two-point correlation function one can get 
\bea
\CD_2(\beta,t)=  \int^{+\infty}_{-\infty} e^{i (u^2 -v^2)t - \beta( \frac{1}{4} + u^2)}   
A_{uv}\,  B_{vu} \, du dv.~~~
\eea
In terms of the new variables (\ref{newvariab}) the  basic matrix element (\ref{basicmatrix}) will take the form 
\bea\label{basicmatrixelement}
 A_{p q}  
= \int^{1/2}_{-1/2} dx  \int^{\infty}_{{1\over 2}\log(1-x^2)}  dy  
  \psi^*_{p} (x,y)  \,  ( e^{- \frac{1}{2} y}   A \,   e^{  \frac{1}{2} y})  \, \psi_{ q} (x,y).~~~~~ 
 \eea
The matrix element (\ref{basicmatrix}), (\ref{basicmatrixelement}) plays a fundamental role in the investigation of the correlation functions because all correlations can be expressed through it.  One should choose also appropriate observables 
$A$ and  $B$.  The operator $y^{-2} $  seems very appropriate for two reasons. Firstly, the convergence of the integrals over the fundamental region $\CF$ will be well defined. Secondly, this operator is reminiscent of 
the exponentiated Louiville  field  since $y^{-2} = e^{-2 \tilde{y}}$ . Thus the interest is in calculating  the matrix element  (\ref{basicmatrixelement})
for the observables in the form of the  Louiville-like operators \cite{Babujian:2018xoy}:
\be\label{Louiville-like}
A(N)=  e^{-2 N y} 
\ee  
with matrix element  
\bea
&A_{p q}(N) =\int^{1/2}_{-1/2} dx  \int^{\infty}_{{1\over 2}\log(1-x^2)}  dy  ~ \psi^*_{p} (x,y)  \,   e^{- 2 N y}  \, \psi_{ q} (x,y) ,\nn\\&  ~~ N=1,2,...
 \eea
 The other interesting observable is  $A = \cos(2\pi N x), N=1,2,... $.
The evaluation of the above matrix elements is convenient to perform using a perturbation expansion in which the part of the wave function (\ref{alterwave}) containing the Bessel's functions and the contribution of the discrete spectrum is considered as a perturbation \cite{Faddeev}. These terms of the perturbative expansion are small and don't change the physical behaviour of the correlation functions.  The reason behind this fact is that  in the integration region $\Im z \gg 1, \Im w \gg 1$ of the matrix element (\ref{basicmatrix}) the Bessel's functions  decay exponentially. Therefore the contribution of these high modes is small (analogues to the so called mini-superspace approximation in the Liouville theory).   In the first approximation of the wave function (\ref{alterwave}) for the matrix element one can get \cite{Babujian:2018xoy}
 \bea
& A_{pq}(N)= {\, _2F_1 \left(  \frac{1}{2}, N+  i {p-q \over 2 }  ;    \frac{3}{2};   \frac{1}{4}          \right)  \over   2 N +i (p-q)}  
+{\, _2F_1 \left(  \frac{1}{2}, N+i {p+q \over 2 }  ;    \frac{3}{2};   \frac{1}{4}          \right)  \over   2 N+ i (p+q)} e^{-i \varphi(q)}  \nn\\
&+{\, _2F_1 \left(  \frac{1}{2}, N- i {p+q \over 2 }  ;    \frac{3}{2};   \frac{1}{4}          \right)  \over   2 N -i (p+q) } e^{i \varphi(p)} + 
{\, _2F_1 \left(  \frac{1}{2}, N-i {p-q \over 2 }  ;    \frac{3}{2};   \frac{1}{4}          \right)  \over   2 N+ i (p-q )} e^{i (\varphi(p)-\varphi(q))}, 
\eea
where the scattering phase $\varphi(p)$ was defined in (\ref{phase}). Thus  the correlation function between Louiville-like  fields  taken in the power $N$ and $M$ respectively is:
\bea\label{twopointcorrel}
\CD_2(\beta,t)=  \int^{+\infty}_{-\infty} e^{i (p^2 -q^2)t - \beta( \frac{1}{4} + p^2)}   
A_{pq}(N)\,  A_{qp}(M) \, dp dq~. 
\eea
 \begin{figure}[h]
 \centering
        \includegraphics[width=0.3\textwidth]{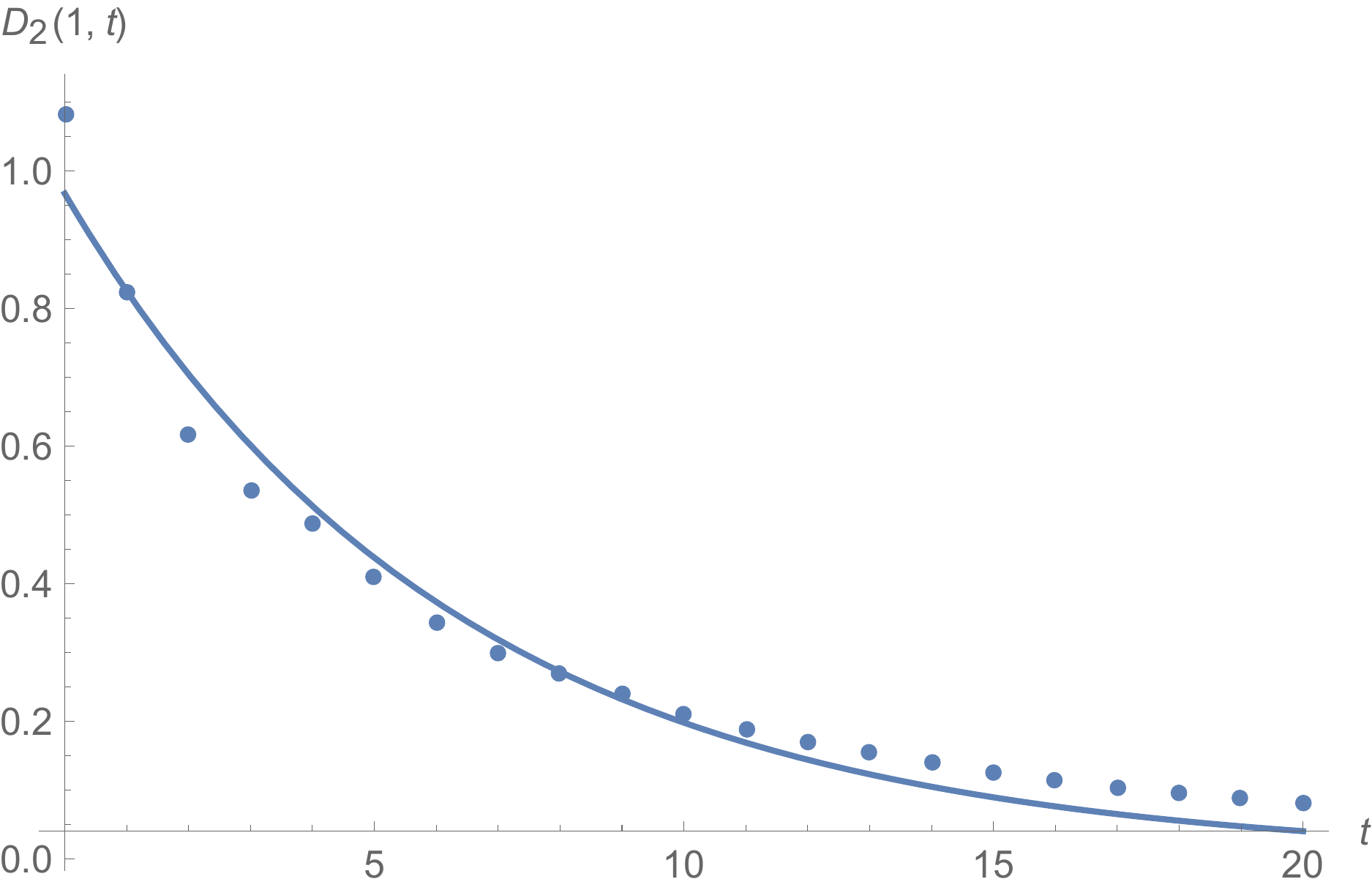}~~~~
         \includegraphics[width=0.3\textwidth]{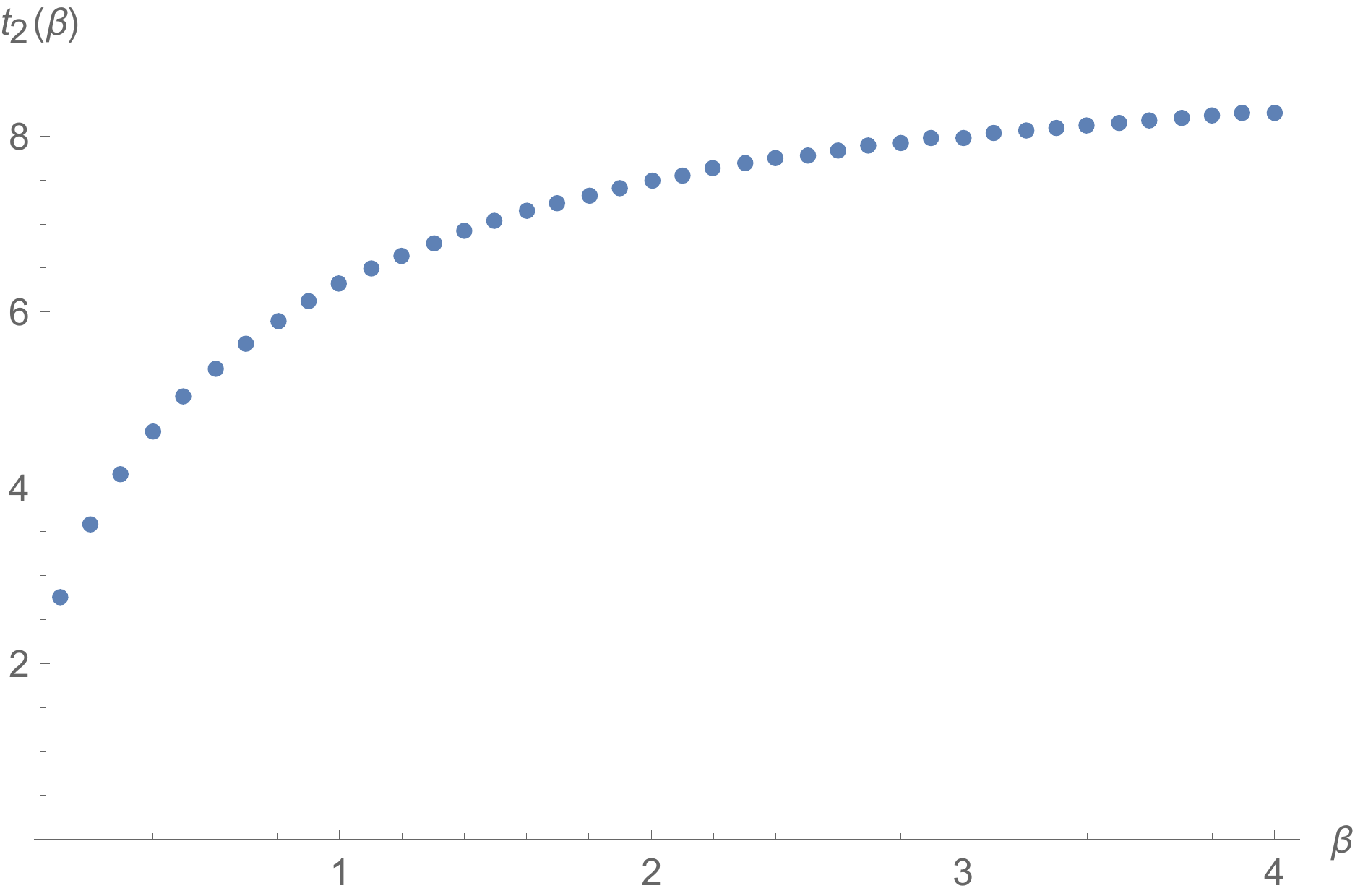}~~~~
         \includegraphics[width=0.3\textwidth]{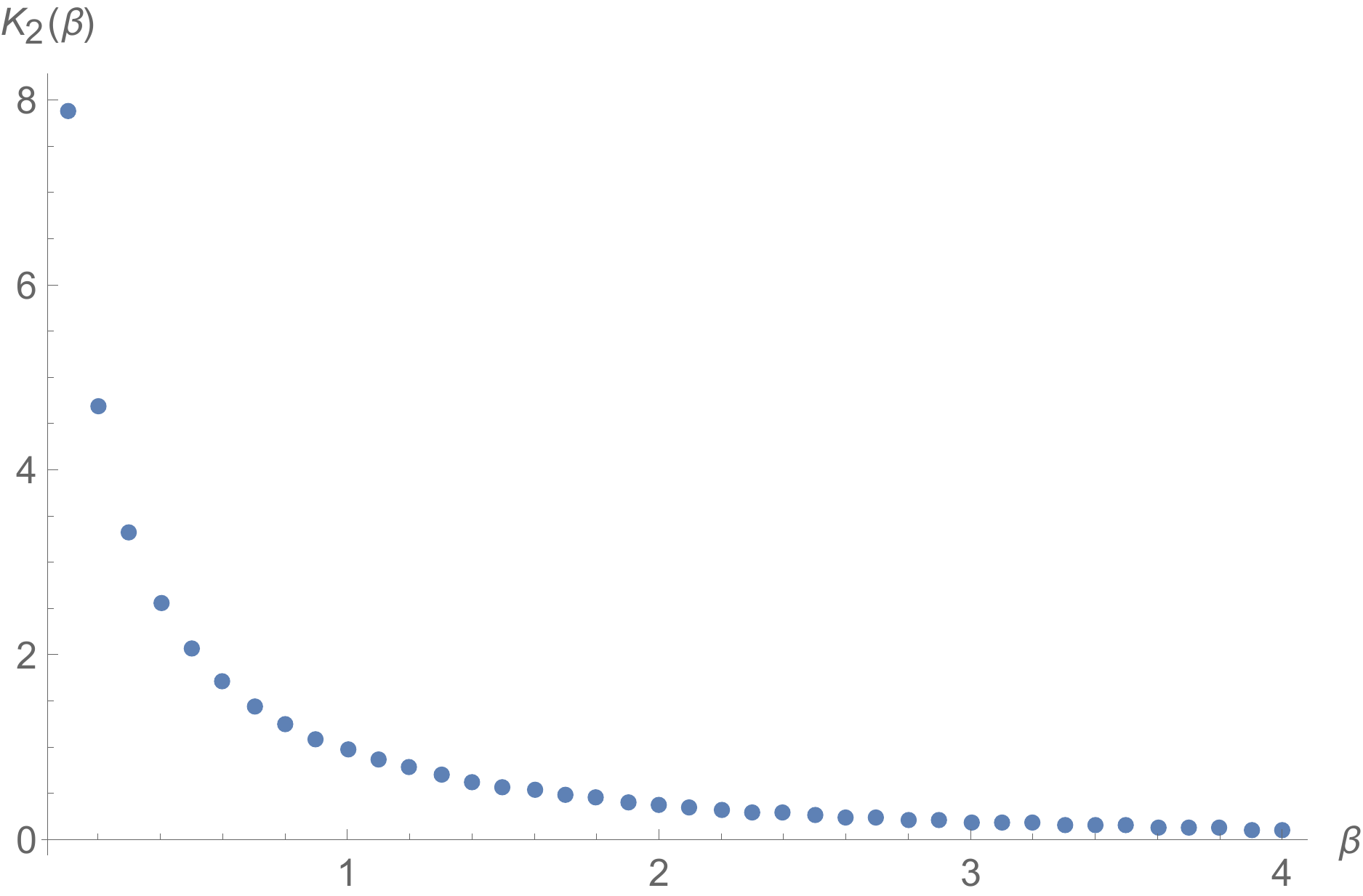}
        \caption{The exponential decay of the two-point correlation function $\CD_2(\beta,t)$ as a function of time  at   temperature $\beta =1$.  The points are fitted by the curve $K(\beta) \exp{(- t / t_2(\beta))}$. The exponent $t_2(\beta)$ has a well defined high and low temperature limits. The limiting values in dimensionless units are $t_2(0) \approx 0.276$ and $t_2(\infty) \approx 0.749$. The temperature dependence of $K_2(\beta)$ is shown on the l.h.s. graph.
}
\label{twopointfunc}
\end{figure}
This expression is very convenient for the analytical and numerical analyses.   
It is expected that the two-point correlation function decay exponentially  \cite{Maldacena:2015waa}
\be\label{twopoint}
\CD_2(\beta,t)  \sim K_2(\beta)~e^{-{t \over t_2(\beta)}}, 
\ee
where $ t_2(\beta)$ is a characteristic time scale of the quantum-mechanical system. The exponential decay of the two-point correlation function with time at different  temperatures is shown in Fig.\ref{twopointfunc}. The dependence of the exponent $ t_2(\beta)$ and of the prefactor $K_2(\beta)$ as a function of temperature are presented in Fig.\ref{twopointfunc}.   
 At high and low temperatures $ t_2(\beta)$ tends to the fixed values shown in the Fig.\ref{twopointfunc} in dimensionless units. The out-of-time-order four-point correlation function of interest was defined in \cite{Maldacena:2015waa} as follows:
\bea\label{outoftime}
&\CD_4(\beta,t)=  \langle   A(t)   B(0) A(t)   B(0)e^{-\beta H}   \rangle= 
 \sum_{n,m,l,r} e^{i (E_n -E_m+E_l - E_r)t - \beta E_n} \nn\\
 & \langle  n \vert   A(0)\vert m \rangle \langle m\vert  B(0)   \vert l \rangle
\langle l \vert   A(0)\vert r \rangle  \langle r\vert  B(0)   \vert n \rangle.   
\eea
The other important observable  is the double commutator of the Louiville-like  operators  separated in time \cite{Maldacena:2015waa} 
\be\label{commutatorL}
C(\beta,t) = -\langle [A(t),B(0)]^2 e^{-\beta H} \rangle~.
\ee 
The energy eigenvalues we shall parametrise as $n = {1\over 2} +i u$, $m = {1\over 2} +i v$,$l = {1\over 2} +i l$ and $r = {1\over 2} +i r$, thus  from (\ref{outoftime}) we shall get \cite{Babujian:2018xoy}
\bea\label{fourmatrix}
&\CD_4(\beta,t)= \int^{+\infty}_{-\infty} e^{i (u^2 -v^2 + l^2  - r^2)t - \beta( \frac{1}{4} + u^2)}   
  A_{uv}\,  B_{vl} \, A_{lr}\,  B_{ru} \,  du dv dl dr ~. 
 \eea
 \begin{figure}
 \centering
 \includegraphics[angle=0,width=5cm]{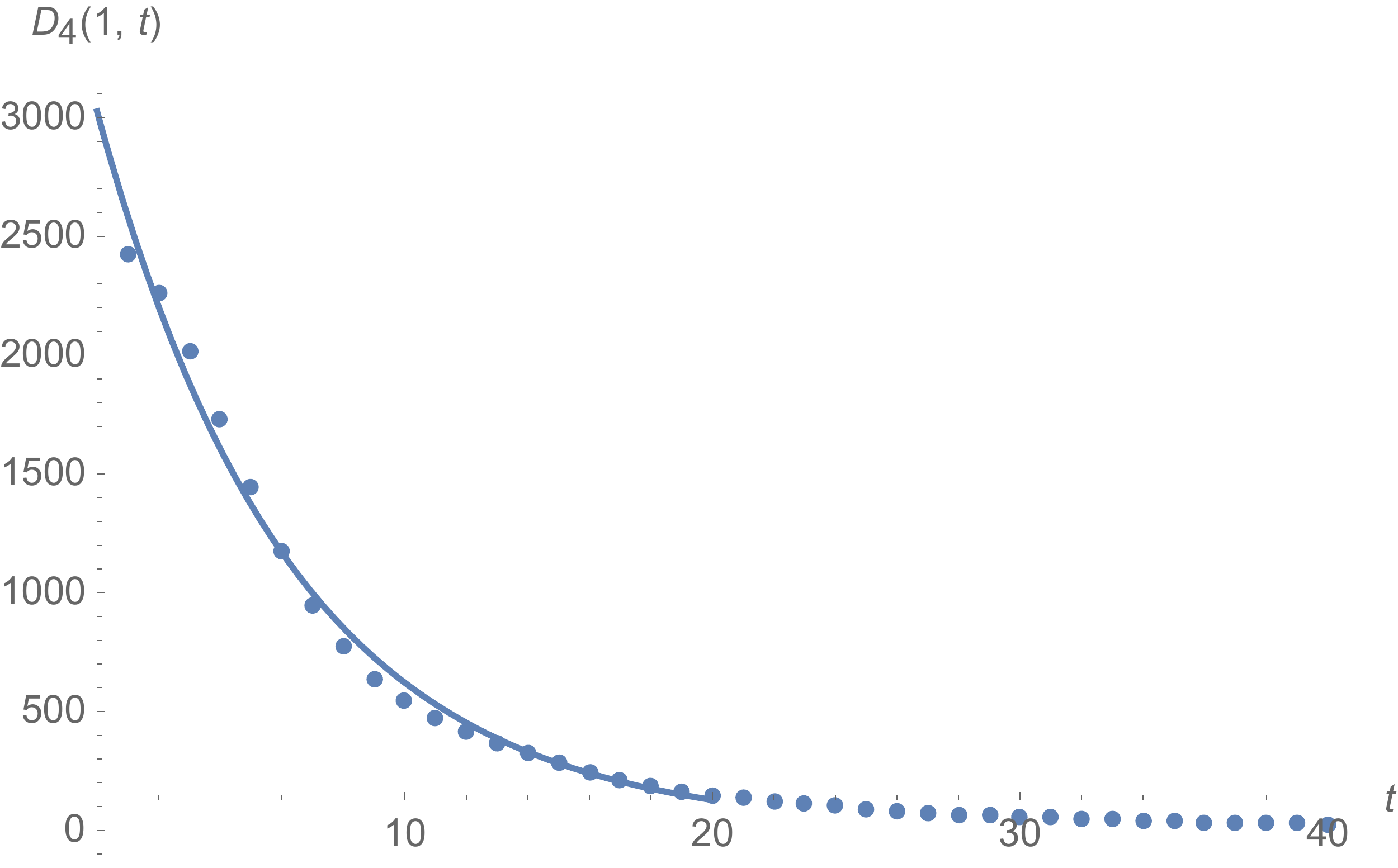}~~~~~~~~~
  \includegraphics[angle=0,width=5cm]{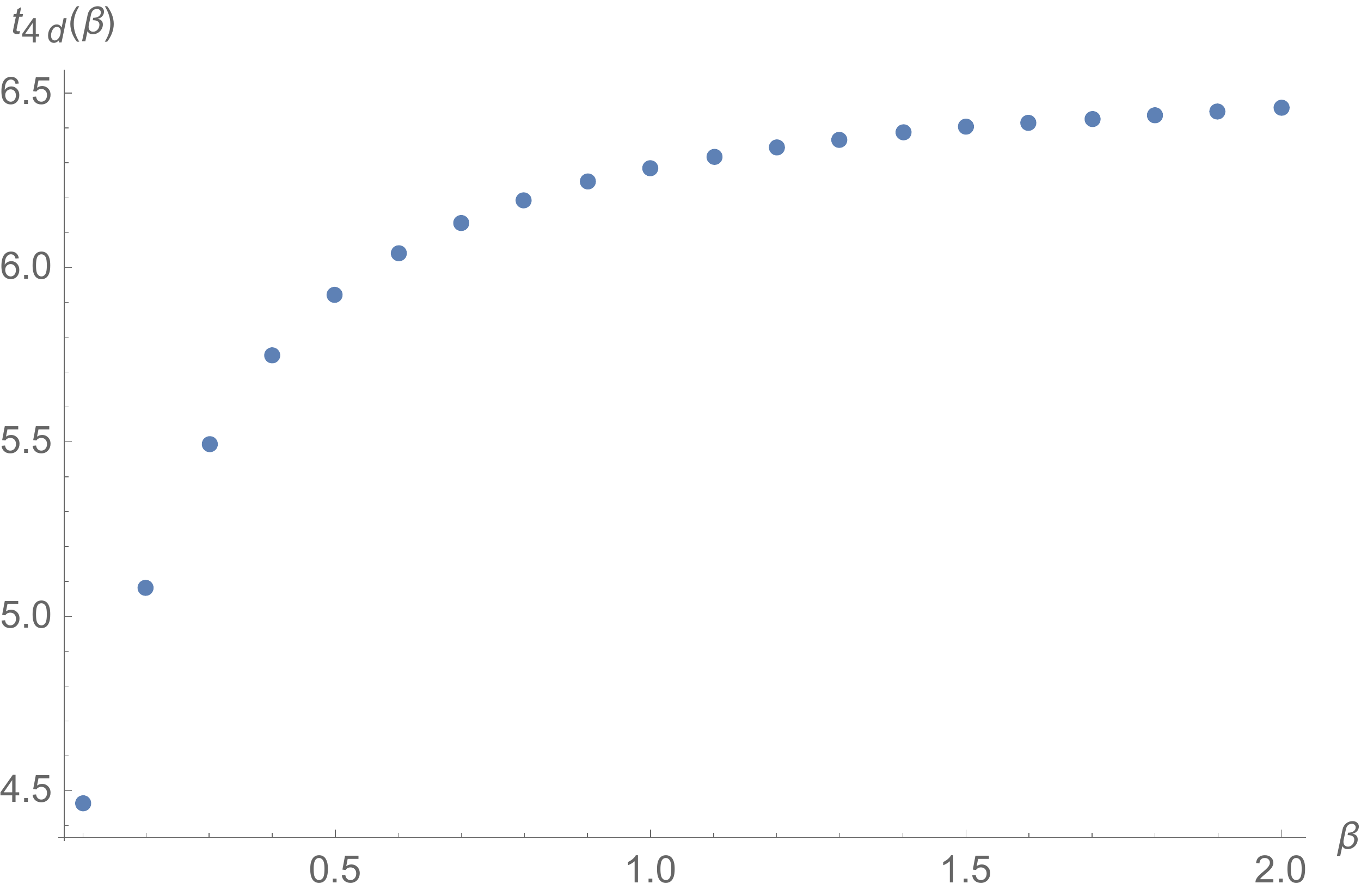}
\caption{  The exponential decay of the correlation function $\CD_4(\beta,t)$ as a function of time at $\beta =1$. The functions $\CD'_4(\beta,t),\CD''_4(\beta,t),\CD'''_4(\beta,t)$ demonstrate a similar exponential  decay $  ~\exp{(-{t \over t_{4}(\beta)} )}$. The temperature dependence of the exponent $t_{4}(\beta)$ has a well defined high and low temperature limits and is shown on the r.h.s. graph. 
The corresponding limiting values of the function $t_{4}(\beta)$ in dimensionless units are $t_{4}(0)=0,112$ and $t_{4}(\infty)=0,163$. The exponent $t_2(\beta)$ of the two-point correlation function is shown in the Fig.\ref{twopointfunc}.
}
\label{fourpointfunc}
\end{figure}
In terms of the variables (\ref{newvariab})   the four-point correlation function (\ref{fourmatrix})  will take the following form: 
\bea
&\CD_4(\beta,t)=  \int^{+\infty}_{-\infty} e^{i (p^2 -q^2 + l^2  - r^2)t - \beta( \frac{1}{4} + p^2)}   
 A_{pq}(N)\,  A_{ql}(M)\,  A_{lr}(N)\,  A_{rp}(M) \,   dp dq  dl dr .  ~~~~~~~~
\eea
As it was suggested in \cite{Maldacena:2015waa}, the most important correlation function indicating the traces of the classical chaotic dynamics in quantum regime is (\ref{commutatorL}) 
\bea\label{commutatorL1}
C(\beta,t) = \langle [A(t),B(0)]^2 e^{-\beta H} \rangle= - \CD_4(\beta,t) + \CD'_4(\beta,t) +\CD''_4(\beta,t)-\CD'''_4(\beta,t). 
\eea
For the Artin system one can get that \cite{Babujian:2018xoy}
\bea\label{fourpointfunct1}
 \CD'_4(\beta,t) +  \CD''_4(\beta,t) =   
  2 \int^{+\infty}_{-\infty} e^{- \beta( \frac{1}{4} + p^2)}  ~\cos{ ( q^2 - r^2)t }  \nn\\
 ~~~~~~~~~~~~~~~~~A_{pq}(N)\,  A_{ql}(M)\,  A_{lr}(N)\,  A_{rp}(M) \,   dp dq  dl dr   
\eea
and 
\bea\label{fourpointfunct2}
\CD_4(\beta,t) +  \CD'''_4(\beta,t)  
= 2 \int^{+\infty}_{-\infty} e^{- \beta( \frac{1}{4} + p^2)}  ~\cos{(p^2 -q^2 + l^2  - r^2)t }   \nn\\
 ~~~~~~~~~~~~~~~~~A_{pq}(N)\,  A_{ql}(M)\,  A_{lr}(N)\,  A_{rp}(M) \,   dp dq  dl dr. 
\eea
The Fig.\ref{fourpointfunc} shows the behaviour of the four-point correlation $\CD_4(\beta,t)$ as the function of the temperature and time.  All four correlation functions decay exponentially as it is shown in Fig.\ref{fourpointfunc}  
\be\label{fourpoint}
\CD_4(\beta,t)  \sim K(\beta)~e^{-{t \over t_{4}(\beta)}}.
\ee
\begin{figure}
 \centering
 \includegraphics[angle=0,width=4cm]{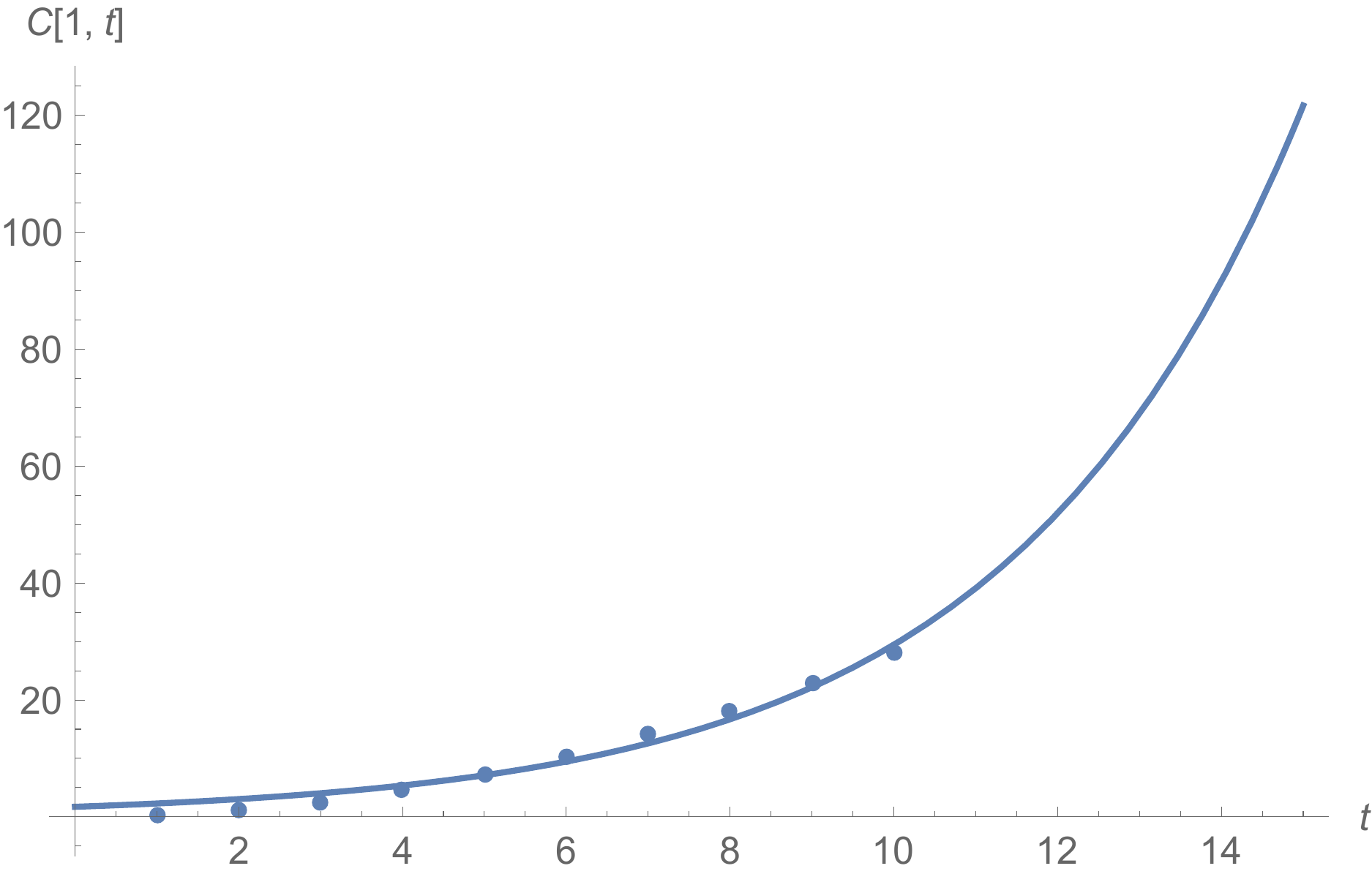}~~
  \includegraphics[angle=0,width=4cm]{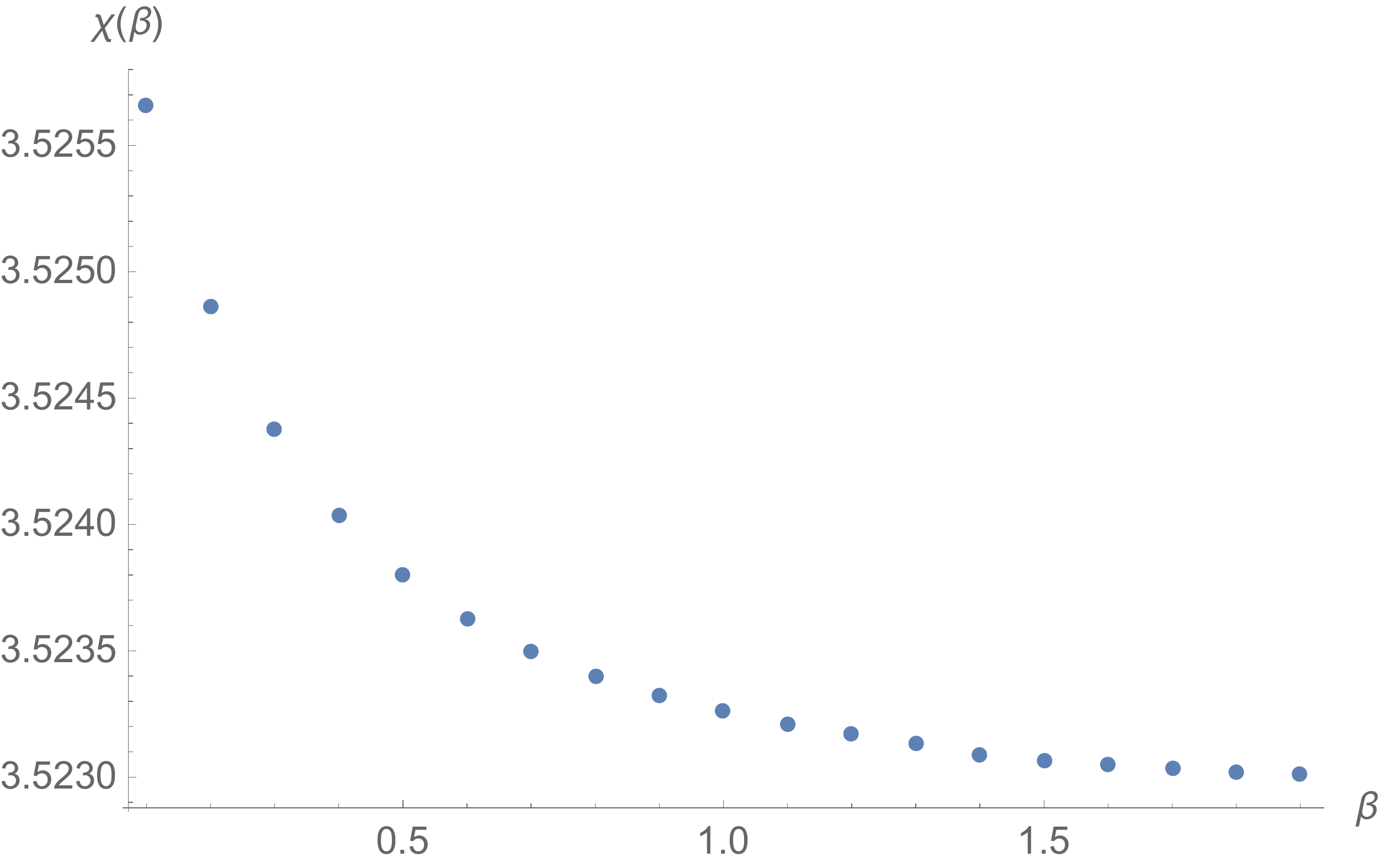}
  \includegraphics[angle=0,width=4cm]{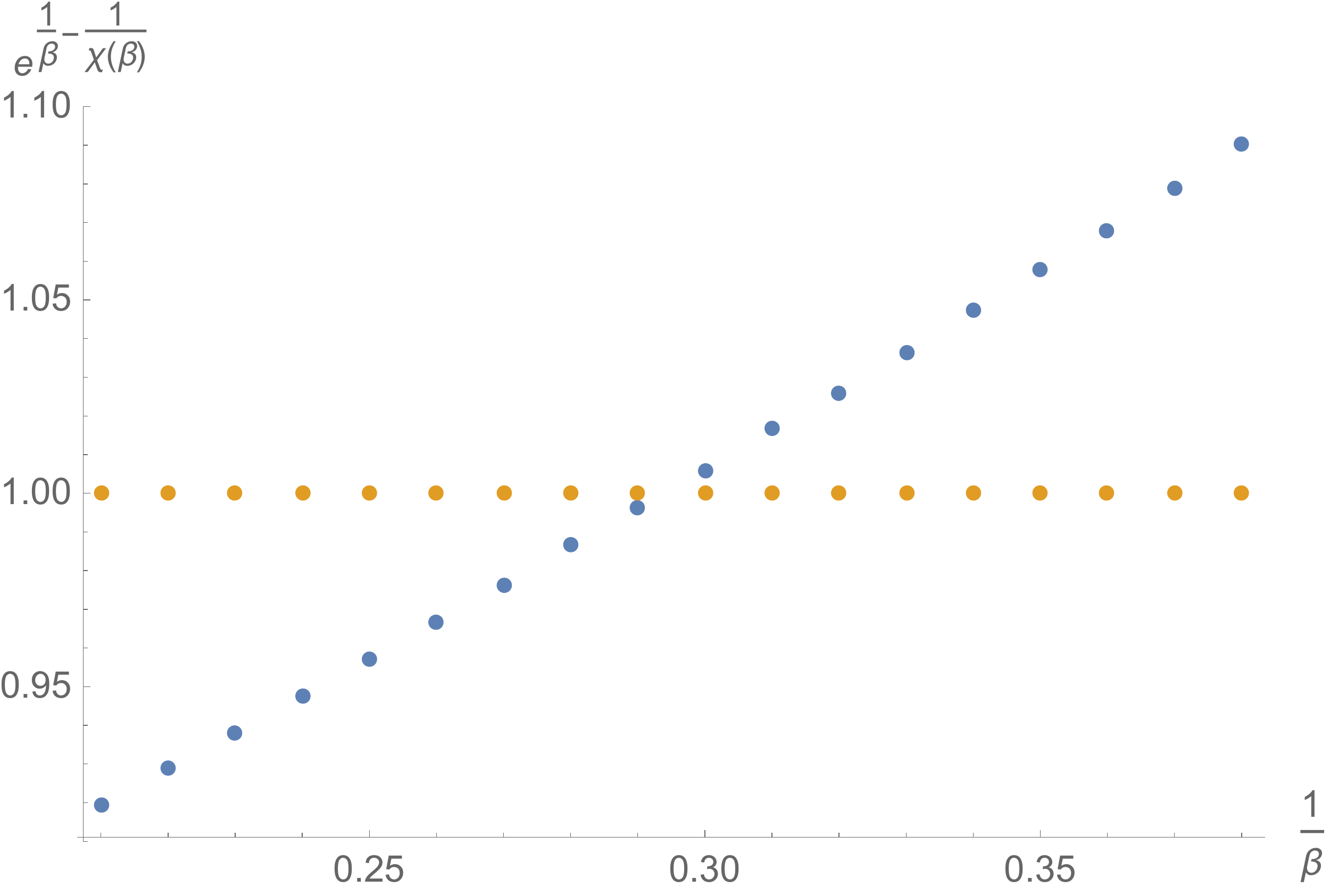}
\caption{  Time evolution of the correlation function $C(\beta,t)$ (\ref{commuta}) at temperature $\beta =1$.  For the short  time intervals the function $C(\beta,t)$ exponentially increases with time.  The exponent  $\chi(\beta)$  (\ref{exponentc122}) slowly decreases with temperature $\beta$. The temperature dependence of Artin exponent $1/\chi(\beta)$ relative to the maximum growth $1/\beta$ is shown by blue dots in the far right figure. At high temperatures the Lyapunov exponent $1/\chi(\beta)$ in Artin system is smaller than the maximal one $1/\beta$, but at low temperatures this is not any more true, we observe a breaking of the saturation regime.   
}
\label{timeevolutionofcommutator}
\end{figure}
Turning to the investigation of the double commutator  (\ref{commutatorL1}) it is convenient to represent it in the following form \cite{Babujian:2018xoy}:  
\bea\label{commuta}
&C(\beta,t) =2 \int^{+\infty}_{-\infty} e^{- \beta( \frac{1}{4} + p^2)} 
~\{ \cos{ ( q^2 - r^2)t }  - \cos{ (p^2 -q^2 + l^2  - r^2)t }  \}\nn\\
&A_{pq}(N)\,  A_{ql}(M)\,  A_{lr}(N)\,  A_{rp}(M) \,   dp dq  dl dr  ,
\eea
where  (\ref{fourpointfunct1}) and (\ref{fourpointfunct2}) have been  used.   
The results of the integration are presented in the Fig.\ref{timeevolutionofcommutator}.  For a short time intervals the  $C(\beta,t)$ increases exponentially in time  
\be\label{exponentc122}
C(\beta,t)_{Artin-AdS_2} \sim K(\beta) \,e^{ {2\pi  \over  \chi(\beta)} t }, 
\ee
with the exponent  $1/\chi(\beta)$.  The important ratio with respect to the maximal growth exponent 
\be\label{exponentc123}
{C(\beta,t)_{max} \over  C(\beta,t)_{Artin-AdS_2}}  \sim R(\beta) \,e^{( {1  \over  \beta}  - {1  \over  \chi(\beta)} ) 2\pi  t }
\ee
is presented in Fig.\ref{timeevolutionofcommutator}. The temperature dependence of Artin exponent $1/\chi(\beta)$ relative to the maximum growth $1/\beta$ is shown by blue dots in the far right figure. At high temperatures the Artin Lyapunov exponent $1/\chi(\beta)$ is less than the maximal exponent $1/\beta$, but at low temperatures this is not any more true and we observe a breaking of the saturation regime \cite{Maldacena:2015waa,Gur-Ari:2015rcq,Cotler:2016fpe}.  In order to confirm this result it seems important to investigate the behaviour of correlation functions and double commutators (\ref{corr1})-(\ref{commuta}) for alternative observables, as well as using more powerful computer codes and hardware as it was available to us.

\section{ \it Artin-Maass  Resonances and Riemann Zeta Function Zeros  }

\begin{figure}
 \centering
 \includegraphics[angle=0,width=3cm]{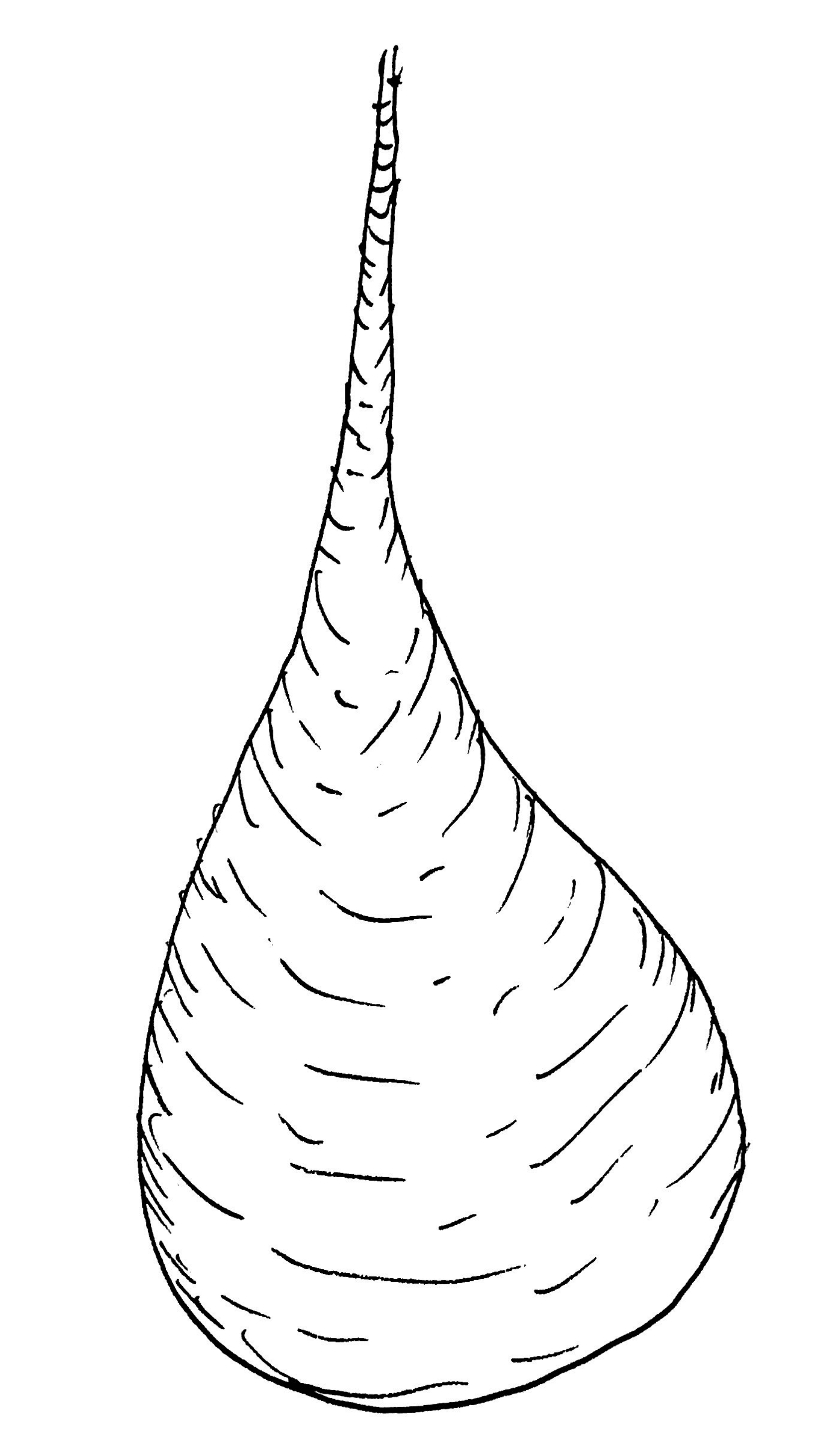}
\caption{ The Arin  system is defined on a non-compact surface $\bar{\CF}$ of constant negative curvature which has a topology of sphere with a cusp on the north pole which is stretched to infinity. The deficit angles on the vertices of the Artin surface can be computed using the formula $2\pi - \alpha$, thus $\int K \sqrt{g} d^2 \xi = (-1) {\pi \over 3} + (2\pi - 2{ \pi \over 3})+ (2\pi - 2 { \pi \over 2}) + (2\pi - 0 ) =4\pi$.}
\label{fig11}
\end{figure}

Here we shall demonstrate that the Riemann zeta function zeros define the position and the widths of the resonances of the quantised Artin system \cite{Savvidy:2018ffh}.  As it was discussed in previous sections 
the Artin  system is defined on the fundamental region of the modular group on the Lobachevsky plane. It has a finite area and an infinite extension in the vertical direction that correspond to a cusp.  In classical regime the geodesic flow on this non-compact surface of constant negative curvature represents one of the most chaotic dynamical systems, has mixing of all orders, continuous classical Koopman  spectrum (\ref{geoflow}), (\ref{koopman}) and non-zero Kolmogorov entropy.  In quantum-mechanical regime the system can be associated with the narrow infinitely long waveguide stretched out to infinity along the vertical axis  and a cavity resonator attached to it at the bottom  Fig.\ref{fig11}.  That suggests a physical interpretation of the Maass automorphic wave function in the form of an incoming plane wave of a given energy which is entering the resonator,  bouncing and scatters back to infinity.  As  the energy of the incoming wave comes close to the eigenmodes of the cavity a pronounced resonance behaviour shows up in the scattering amplitude \cite{Savvidy:2018ffh}.

We already presented above   (\ref{alterwave})  the Maass wave function \cite{maass} in terms of the  natural physical variable $\tilde{y}$, which is the distance in the vertical direction on the Lobachevsky plane  $  \ln y = \tilde{y} $,   and of the corresponding momentum $p$ \cite{Babujian:2018xoy}.  The plane wave  $e^{-i p \tilde{y}  }$ incoming from infinity $\CD$ along the $y$ axis in Fig.\ref{fig5}, Fig.\ref{scattering} and Fig.\ref{fig11}  elastically scatters on the boundary $ACB$ of the fundamental triangle  $\CF$.  The reflection amplitude is a pure phase and is given by the expression in front of the outgoing plane wave $e^{  i p \tilde{y}}$ :
\be\label{Smatrix}
S={\theta(\frac{1}{2} +i p) \over \theta(\frac{1}{2} -i p)}  =\exp{[ 2\, i\, \delta(p)]}.
\ee
The continuous energy spectrum is given by the formula   (\ref{newvariab}) \cite{Babujian:2018xoy}
\be\label{energy}
E=   p^2  + \frac{1}{4} .
\ee
As we suggested  the system can be described in terms of  an  infinitely long narrow waveguide with a cavity resonator attached to it at the bottom Fig.\ref{scattering} and Fig.\ref{fig11}).  In order to support this interpretation we will calculate the area of the Artin surface which is below the fixed coordinate $y_0= e^{\tilde{y}_0}$:
\be
\text{Area}(\CF_0)= \int _{-\frac{1}{2}}^{\frac{1}{2}} dx
\int _{\sqrt{1-x^2}}^{y_0 }\frac{dy}{y^2}  =\frac{ \pi }{3}-  e^{-\tilde{y}_0}\, = \text{Area}(\CF)-  e^{-\tilde{y}_0}\,.
\ee 
The area $\text{Area}(\CF) - \text{Area}(\CF_0)$ above  the ordinate $\tilde{y}_0$ is therefore exponentially small $ e^{-\tilde{y}_0}$. The horizontal ( $dy =0$) size of the Artin surface is also decreases exponentially in the vertical direction:
\be\label{111}
L_0=2 \int  ds =2 \int \frac{\sqrt{dx^2+dy^2}}{ y}=  \int _{-\frac{1}{2}}^{\frac{1}{2}} {dx\over y_0 } =  e^{-\tilde{y}_0}.
\ee 
The cavity has its eigenmodes and as the energy of the incoming wave $E =p^2  + \frac{1}{4}$ became  close to the eigenmodes of the cavity one should expect a pronounced resonance behaviour of the scattering amplitude \cite{Savvidy:2018ffh}. 
\begin{figure}
 \centering
 \includegraphics[angle=0,width=6cm]{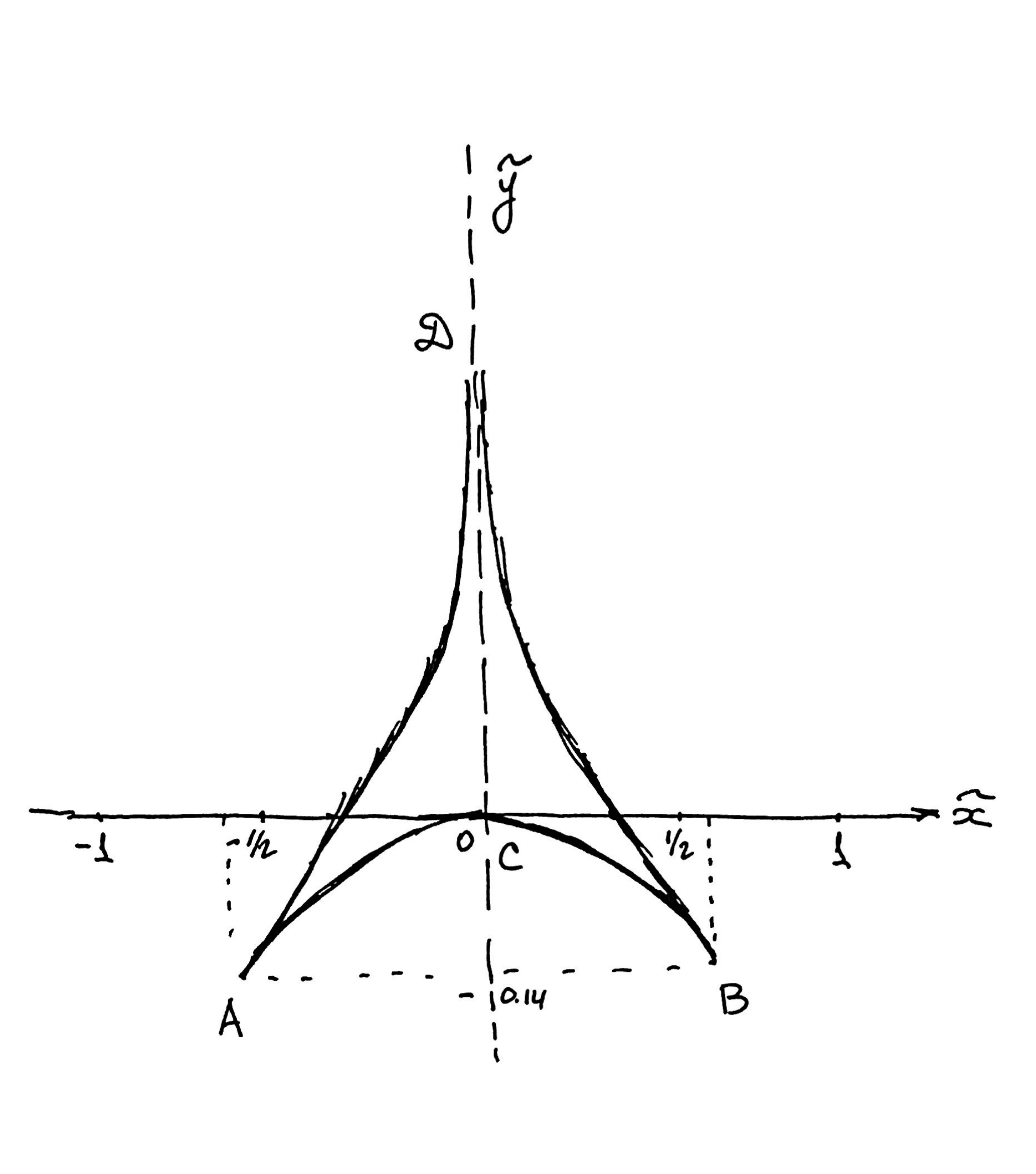}
\caption{ The system can be described as a narrow (\ref{111}) infinitely long waveguide stretched to infinity along the vertical direction  and a cavity resonator attached to it at the bottom $ACB$.}
\label{fig10}
\end{figure}

To trace such behaviour let us consider the analytical continuation of the Maass wave function (\ref{alterwave}) to the complex energy plane $E$. The analytical continuation of the scattering 
amplitudes as a function of the energy $E$ considered as a complex variable allows to establish important spectral properties of the quantum-mechanical system. In particular,  the method of analytic continuation allows to determine the real and complex S-matrix poles. The real  poles on the physical sheet correspond  to the  discrete energy levels and the complex poles on the second sheet below the cut correspond  to the resonances in the quantum-mechanical system  \cite{landauqmech}.  

The asymptotic form of the wave function  can be represented in the following form: 
\be\label{reswave4}
\psi = A(E)\, e^{i p \tilde{y}} + B(E)\, e^{-i p \tilde{y}} , ~~~p = \sqrt{E-1/4}.
\ee
In order to make the functions $A(E) $ and $B(E) $ single-valued one should cut the complex plane 
along the real axis \cite{landauqmech} at the  $E=1/4$. The complex plane with this cut defines  a physical sheet.  To the left from the cut,  at  energies  $E < 1/4$,  the wave function takes the following form: 
\be\label{reswave1}
\psi = A(E)\, e^{- \sqrt{\vert E-1/4 \vert } \tilde{y}} + B(E)\, e^{ \sqrt{\vert E-1/4 \vert} \tilde{y}}, 
\ee
where the exponential factors are real and one of them decreases and the other one increases at $\tilde{y} \rightarrow \infty$.  The  bound states are characterised  by the fact that the corresponding wave function tends to zero at  $\tilde{y} \rightarrow \infty$,  thus for the bound states the second term in (\ref{reswave1}) vanishes $B(E_n)=0$ \cite{landauqmech}. 
If a system is unbounded then its energy spectrum may be continuous and the energy spectrum can be quasi-discrete, consisting of smeared levels of  a width $\Gamma$ \cite{landauqmech}.  These states are also described by the absence incoming waves now in  (\ref{reswave4}) \cite{landauqmech}:     
\be\label{zerosquasi1}
B(E_n - i {\Gamma_n \over 2}) =0
\ee
and define the complex eigenvalues of the form \cite{landauqmech}
\be\label{compeigen1}
\varepsilon_n = E_n - i {\Gamma_n \over 2},
\ee
where $E+n$ and $\Gamma_n$ are both real and positive.  
\begin{figure}
 \centering
 \includegraphics[angle=0,width=8cm]{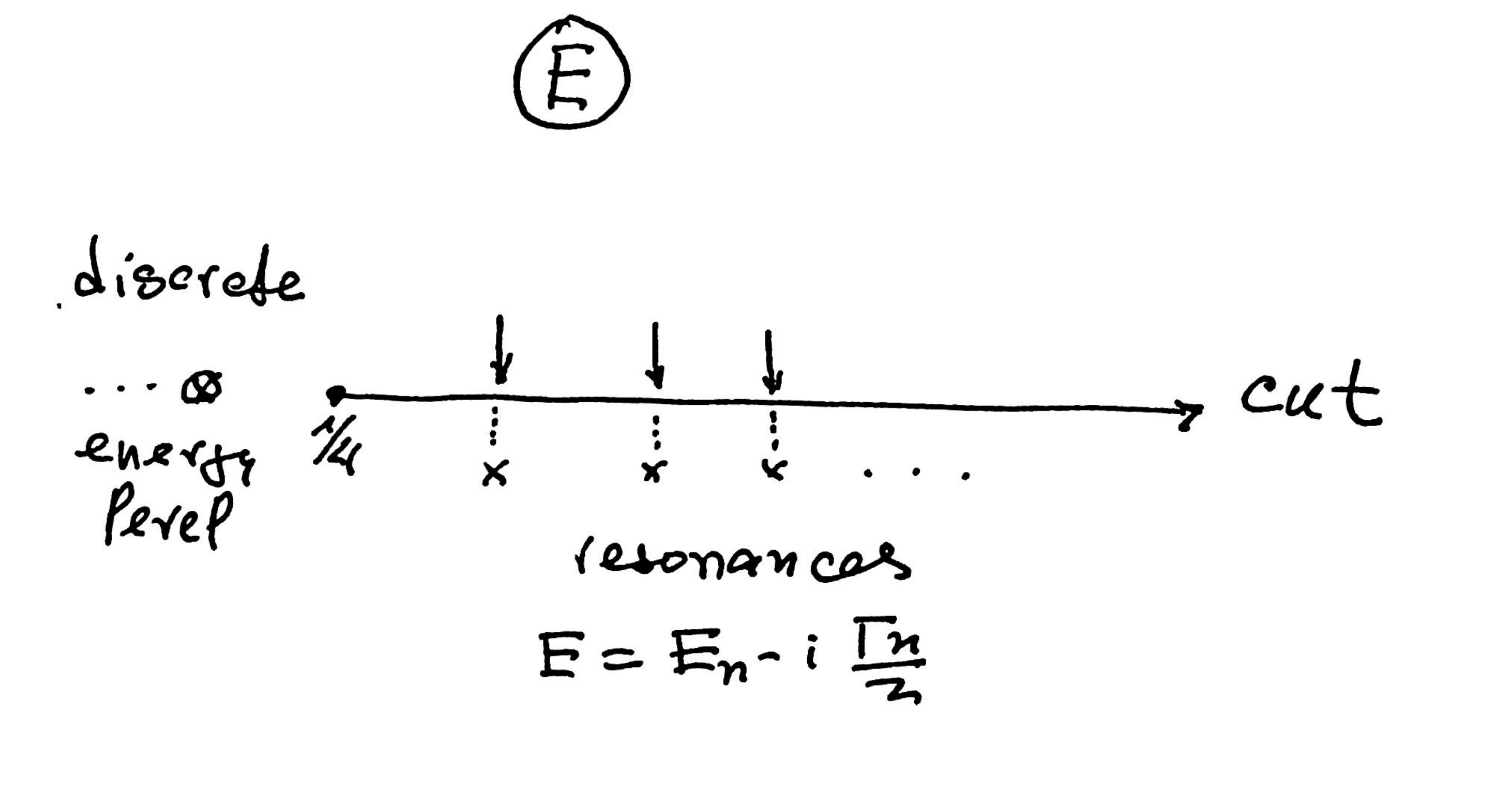}
\caption{ The resonances  $E_n - i {\Gamma_n \over 2}$ are located under the cut on the right hand side of the real axis.}
\label{resonances}
\end{figure}
The complex poles  $E_n - i {\Gamma_n \over 2}$ are located under the cut on the right hand side of the real axis Fig.\ref{resonances}.   To find a position of a resonance one can  expand $B(E)$ near a resonance (\ref{compeigen1}) as $B(E) =(E - E_n + {i \Gamma_n \over 2}) ~b_n +...$ and represent the wave function (\ref{reswave4}) in the form 
\be
\psi~~ \approx~~ b^*_n (E - E_n - {i \Gamma_n \over 2})  e^{i p \tilde{y}} +b_n ~ (E - E_n + {i  \Gamma_n \over 2})  e^{-i p \tilde{y}}.
\ee
The S-matrix  takes the following form \cite{landauqmech}
\be\label{resonphase}
S=e^{2 i \delta} = {E - E_n - i \Gamma_n / 2 \over E - E_n +i \Gamma_n / 2} e^{2 i \delta_n},
\ee
where $e^{2 i \delta_n} =  b^*_n / b_n $.  We can find now the positions of the poles in quantum Artin system. Let us consider the asymptotic  behaviour of the Maass wave function (\ref{alterwave}) at large $\tilde{y}$.  The conditions  (\ref{zerosquasi1}) of the absence of incoming wave due to  (\ref{thata1}) will takes the form \cite{Savvidy:2018ffh}:
\bea 
\theta(\frac{1}{2} -i p)=  {\zeta (1- 2 i p) \Gamma (\frac{1}{2} -i p) \over  \pi ^{\frac{1}{2} -i p} }=0.
\eea  
The solution  can be expressed in terms of the zeros  of the Riemann zeta function
\cite{Riemann}:
\be
\zeta(\frac{1}{2} - i u_n) =0,~~~~  n=1,2,....~~~~u_n > 0.
\ee
Thus one should solve the equation 
\be
1- 2 i p_n = \frac{1}{2} - i u_n~.
\ee 
The location of poles is therefore at the following values of the complex momenta 
\be\label{poles}
p_n = {u_n \over 2} - i \,{1 \over 4} \, ,  ~~~~~  n=1,2,..... 
\ee
and at the complex energies (\ref{energy}) :
\be\label{resonances34} 
E = p^2_n +{1 \over 4}~ =~ ({u_n \over 2} -   \,{1 \over 4}\,i)^2  +{1 \over 4} ~= ~{u^2_n \over 4} + {3\over 16}
- i \, {u_n \over 4}.
\ee
These poles correspond to the resonances (\ref{compeigen1}) \cite{Savvidy:2018ffh}:
\be\label{exactresonance}
E_n = {u^2_n \over 4} + {3\over 16},~~~~~~\Gamma_n = {u_n \over 2} . 
\ee
One can conjecture the following representation of the S-matrix (\ref{Smatrix}):
\be\label{Smatirxphase}
S=e^{2 i\, \delta} =  {\theta(\frac{1}{2} +i p) \over \theta(\frac{1}{2} -i p)} = \sum^{\infty}_{n=1}{E - E_n - i \Gamma_n / 2 \over E - E_n +i \Gamma_n / 2}~ e^{2 i \delta_n}
\ee
with yet unknown phases $\delta_n$. In order to justify the above representation of the S-matrix one can find the location of the poles on the second  sheet by using expansion of the S-matrix
(\ref{Smatrix}) at the "bumps" which occur  along the real axis at energies  
\be \label{approxresonance}
E_n  = {u^2_n \over 4}  +{3\over 16}~.
\ee
The expansion will take the following form:
\bea\label{Smatirxphase1}
S\vert_{E \approx E_n} &  
= &  {\theta(\frac{1}{2} +i \sqrt{E - {1\over 4}} ) \over 
\theta( \frac{1}{2} -i  \sqrt{E - {1\over 4} }) } \vert_{E \approx E_n}~=~{ \theta(\frac{1}{2} +i \sqrt{E_n - {1\over 4}} )  +   \theta^{'}(\frac{1}{2} 
+i \sqrt{E_n - {1\over 4}} ) ~ (E - E_n)
 \over  \theta(\frac{1}{2} -i \sqrt{E_n - {1\over 4}} )  +   \theta^{'}(\frac{1}{2} -i \sqrt{E_n - {1\over 4}} ) ~  (E - E_n)} \nn\\
&=& 
 {    E - E_n + \theta(\frac{1}{2} +i \sqrt{E_n - {1\over 4}} ) /\theta^{'}(\frac{1}{2} 
 +i \sqrt{E_n - {1\over 4}} ) 
 \over    E - E_n + \theta(\frac{1}{2} -i \sqrt{E_n - {1\over 4}} )/\theta^{'}(\frac{1}{2} 
 -i \sqrt{E_n - {1\over 4}} ) } ~~ {\theta^{'}(\frac{1}{2} +i \sqrt{E_n - {1\over 4}} )   \over \theta^{'}(\frac{1}{2} -i \sqrt{E_n - {1\over 4}} )   }  \nn\\ 
 &\equiv &  ~~~
 {    E - E^{'}_n - i \Gamma^{'}_n/2   
 \over    E - E^{'}_n + i \Gamma^{'}_n/2  }~~  e^{2 i \delta^{'}_n}~,\nn\\
 \eea
where
\bea
E^{'}_n - i \Gamma^{'}_n/2 = E_n - {\theta(\frac{1}{2} -i \sqrt{E_n - {1\over 4}} ) 
\over \theta^{'}(\frac{1}{2} -i \sqrt{E_n - {1\over 4}} ) } ,~~~~ e^{2 i \delta^{'}_n}=
 {\theta^{'}(\frac{1}{2} +i \sqrt{E_n - {1\over 4}} )   \over \theta^{'}(\frac{1}{2} 
 -i \sqrt{E_n - {1\over 4}} )   } , 
\eea 
 thus
\bea\label{approxresonance1}
E^{'}_n - E_n = - \Re {\theta(\frac{1}{2} -i \sqrt{E_n - {1\over 4}} ) 
\over \theta^{'}(\frac{1}{2} -i \sqrt{E_n - {1\over 4}} ) } ,~~~~- i  \Gamma^{'}_n/2 = - \Im  {\theta(\frac{1}{2} -i \sqrt{E_n - {1\over 4}} ) 
\over \theta^{'}(\frac{1}{2} -i \sqrt{E_n - {1\over 4}} ) } 
\eea
and all quantities $E^{'}_n$ , $ \Gamma^{'}_n/2$ and $\delta^{'}_n$ are real.  Considering the first ten zeros of the zeta function which are known numerically \cite{Turing,Gourdon} one can calculate the position of the resonances and their widths using the approximation formulas (\ref{approxresonance1}) and get convinced that the energies and the widths of the resonances given by the exact formula (\ref{exactresonance}) and  the one given by the approximation formulas   (\ref{approxresonance1}) are consistent  within the two precent deviation.  

It has been observed that numerical calculation of the discrete energy eigenstates (\ref{wavedisc0}) of the Artin system is unstable and shows slow convergence \cite{hejhal,hejhal1,winkler}. It seems that this can be caused by the presence of the resonances.  If the eigenvalue lies within the continuous energy band of a resonance it should be difficult to distinguish and separate it from the continuous resonance spectrum.   

\section{\it  C-cascade. Entropy and Periodic Trajectories }

\begin{figure}
 \centering
\includegraphics[width=5cm]{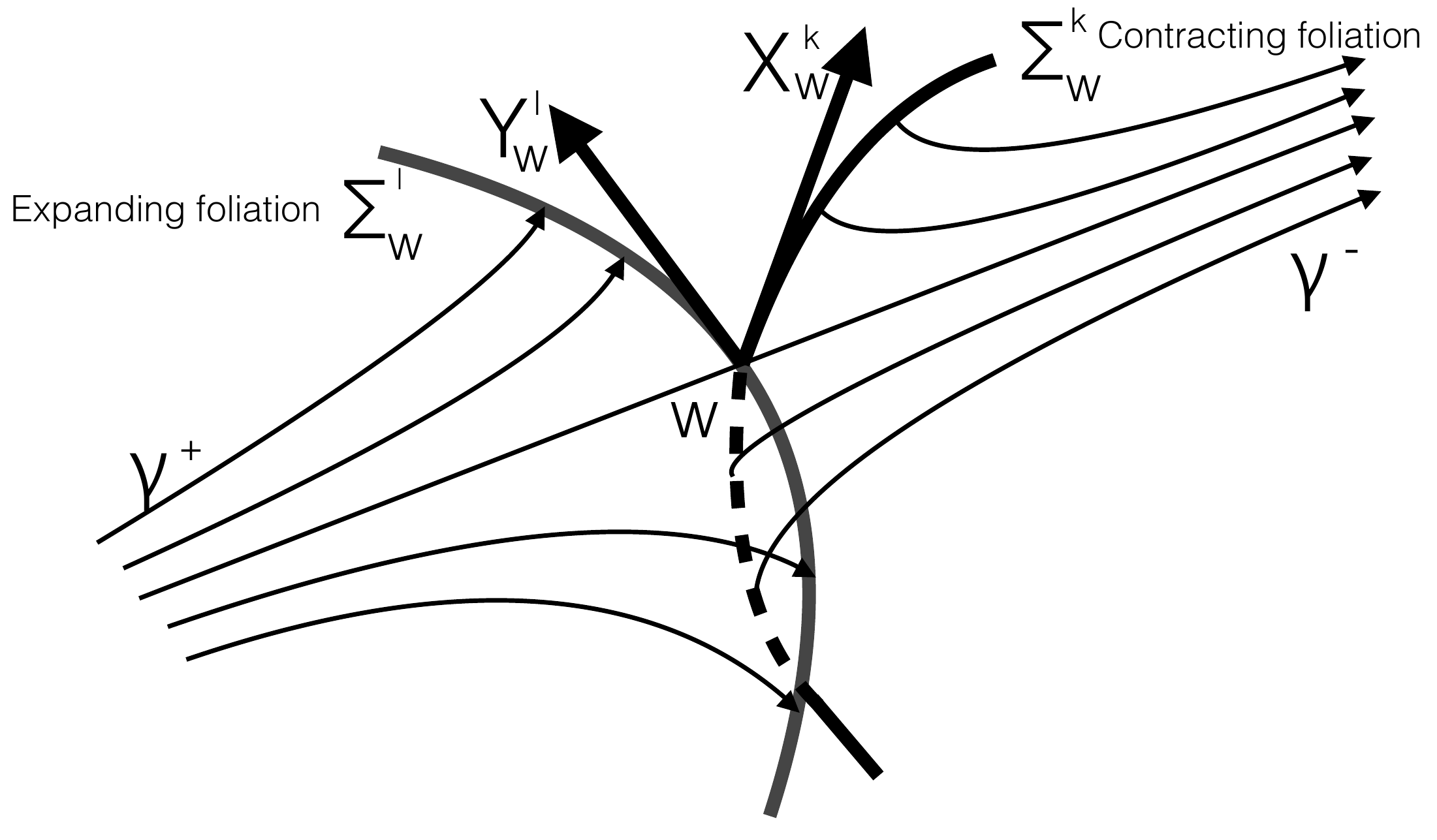}
\centering
\caption{ At each point $w$ of the C-system the tangent space $R^{m}_{w}$  
is  decomposable  into a direct sum of two linear spaces  $Y^l_{w} $ and $X^k_{w} $. 
The expanding and contracting geodesic flows
are $\gamma^+$ and $\gamma^-$. The expanding and 
contracting invariant  foliations  $\Sigma^l_{w} $ and $\Sigma^k_{w} $ 
{\it are transversal to the geodesic flows} and their corresponding tangent spaces 
are  $Y^l_{w} $ and $X^k_{w} $.  } 
\label{fig13}
\end{figure}
In this section we shall turn our attention to the investigation of the second class of the MCDS defined on high dimensional tori with a  discrete in time evolution \cite{anosov}. The systems with discrete time \cite{anosov} is defined as a  {\it cascade}  on the d-dimensional compact phase space $M^{d}$ induced by the diffeomorphisms $T: M^d \rightarrow M^d$.  The iterations
are defined by  a repeated action of the operator  
$\{ T^n, -\infty < n < +\infty  \}$, where $n$ is an  integer number.  
The tangent space at 
the point $x \in M^d$ is  denoted by $R^d_{x}$ and the 
tangent vector bundle by $\CR(M^d)$. 
The diffeomorphism $\{T^n\}$ induces the mapping
of the tangent spaces $\tilde{T}^{n}: R^d_x \rightarrow R^d_{ T^{n}x}$.
The C-condition requires that the tangent space $R^{d}_{x}$ at each point $x$
of the d-dimensional phase space $M^{d}$ of the dynamical system $\{T^{n}\}$ 
should be decomposable  into a 
direct sum of the two linear spaces  $X^{k}_{x}$ and $Y^{l}_{x}$ with the following 
properties \cite{anosov}:
\bea\label{ccondition}
C1.&R^{d}_{x}= X^{k}_{x} \bigoplus Y^{l}_{x} ~~\\
C2.&~~a) \vert \tilde{T}^{n} \xi  \vert  \leq ~ a \vert   \xi \vert e^{-c n}~ for ~n \geq 0 ; ~
\vert \tilde{T}^{n} \xi  \vert  \geq~ b \vert \xi \vert e^{-c n} ~for~ n \leq 0,~~~\xi \in  X^{k}_{x}, \nn\\
&b) \vert \tilde{T}^{n} \eta  \vert  \geq~ b \vert \eta \vert e^{c n} ~~for~ n \geq 0;~
\vert \tilde{T}^{n} \eta  \vert  \leq ~ a \vert   \eta \vert e^{c n}~ for~ n \leq 0,~~~\eta \in Y^{l}_{x},\nn
\eea
where the constants a,b and c are positive and are the same for all $x \in M^d$ and all 
$\xi \in  X^{k}_{x}$, $\eta \in Y^{l}_{x}$.
The length $\vert ...\vert$ of the tangent vectors   $\xi $ and $  \eta $  
is defined by the Riemannian metric on $M^d$.
The linear spaces $X^{k}_{x}$ and $Y^{l}_{x}$ are invariant  with respect to 
the derivative  mapping  $\tilde{T}^{n} X^{k}_{x} = X^{k}_{T^n x}, ~
\tilde{T}^{n} Y^{l}_{x} = Y^{l}_{T^n x}$ and represent the {\it contracting and expanding 
linear spaces} (see Fig.\ref{fig13}).
The C-condition describes the behaviour of all trajectories $\tilde{T}^n \omega$ 
on the tangent vector bundle  $\omega \in R^{d}_{x}$. 
Anosov proved that the vector spaces  $X^{k}_{x}$ and $Y^{l}_{x}$ are continuous 
functions of the coordinate $x$ and that they are the target vector spaces  to 
the foliations $\Sigma^k$ and $\Sigma^l$ which are  the {\it  surfaces transversal to 
the trajectories}  $T^n x$ on $M^d$ (see Fig.\ref{fig13}). 
The contracting and expanding foliations  $\Sigma^k_x$ and $\Sigma^l_x$  are invariant 
with respect to the cascade $T^n$ in the sense that,  under the action of 
these transformations 
a foliation transforms into a foliation  \cite{anosov}. 

A uniform instability of trajectories of MCDS (\ref{ccondition}) leads to the appearance   
of strong statistical properties of the MCDS \cite{leonov}.  It appears that the time average  $
 \bar{f_N}(x) ={1 \over N} \sum^{N-1}_{n=0} f(T^n x)
$
of the function $f(x)$ on phase space $M$  behaves  as a superposition of quantities which are statistically weakly dependent. 
Therefore for the C-systems on a torus  it was  demonstrated that
the fluctuations of the time averages from the phase space integral 
$
\langle f \rangle= \int_{M} f(x) d x 
$
multiplied by $\sqrt{N}$ 
have at large $N \rightarrow \infty$ a Gaussian distribution \cite{leonov}:
\be\label{gauss}
\lim_{N\rightarrow \infty}\mu \bigg\{ x	:\sqrt{N}   \left(  \bar{f_N}(x)  - \langle f \rangle \right) < z    \bigg\}
= {1 \over \sqrt{2 \pi} \sigma_f}\int^{z}_{-\infty} e^{-{y^2 \over 2 \sigma^2_f}} dy.
\ee
The quantity  
$
\sqrt{N}   \Bigg(  \bar{f_N}(x)  -  \langle f \rangle  \Bigg)
$
converges in distribution to the normal random variable with standard deviation  $\sigma_f$  
\be
\sigma^2_f = \sum^{+ \infty}_{n=-\infty} [\langle f(x) f(T^n x) \rangle-
\langle f(x)  \rangle^2 ].
\ee
Let us turn now to the calculation of the corresponding Kolmogorov-Sinai entropy. The most convenient way to calculate the entropy of MCDS  is to integrate over the phase space the logarithm of the volume expansion rate $\lambda(x)$ of a $l$-dimensional infinitesimal cube which is embedded  into the foliation  $\Sigma^{l}_x$. The derivative map
$\tilde{T}$ maps the linear space $Y^{l}_{x}$ into the $Y^{l}_{Tx}$ and if 
the rate of expansion of the volume of the $l$-dimensional cube is  $\lambda(x)$,
then \cite{anosov,sinai3,rokhlin2,sinai4,gines}
\be\label{biuty}
h(T) = \int_{M^d} \ln \lambda(x) d x.
\ee
Here the volume of the $M^d$ is normalised to 1. 

Let us consider the automorphisms of a torus generated by the linear transformation  
\bea\label{cmap}
x_i \rightarrow \sum^{n}_{j=1} T_{ij} x_j,~~~~(mod ~1),
\eea
where the integer matrix $T$ has a determinant equal to one $Det~ T =1$. 
In order for the automorphisms of the torus (\ref{cmap})  {\it to fulfil  the C-condition (\ref{ccondition}) it is necessary and sufficient that the matrix $T$ has no eigenvalues on the unit circle.}  
Thus the  spectrum $\{ \Lambda = {\lambda_1},...,
\lambda_n \}$ of the matrix $T$ should fulfil the following 
two conditions \cite{anosov}: 
\bea\label{mmatrix}
1)&~Det~ T=  {\lambda_1}{\lambda_2}....{\lambda_n}=1\nn\\
2)&\vert {\lambda_i} \vert \neq 1, ~~~~~~~~~~~\forall i.
\eea
Because the determinant of the matrix $T$ is equal to one,
the Liouville's measure $d\mu = dx_1...dx_m$ is invariant under the action of $T$.
The inverse matrix $T^{-1}$ is also an integer matrix because $Det~ T=1$.
Therefore $T$ is an automorphism of  the torus   onto itself. All trajectories with rational coordinates $(x_1,...,x_n)$, and only they, 
are periodic trajectories of the automorphisms of the  torus (\ref{cmap}).
The above conditions (\ref{mmatrix}) on the eigenvalues of the matrix $T$ are  sufficient 
to prove that the system belongs to the class of  Anosov C-systems and therefore has {\it mixing} properties defined 
above  (\ref{mix1})- (\ref{mixnn}). Because the C-systems have {\it mixing of 
all orders} \cite{anosov}  it follows that the C-systems are exhibiting  the decay of the 
 correlation functions of any order. 
 
For the automorphisms  on a torus (\ref{cmap}) the coefficient $\lambda(x)$  does not depends of the phase space coordinates $x$ and is equal to the product of eigenvalues $\{  \lambda_{\beta }  \} $  with modulus  larger than one (\ref{eigenvalues}): 
\be\label{more}
\lambda(w) = \prod^{l}_{\beta=1} \lambda_{\beta}. 
\ee
Thus the entropy of the Anosov automorphisms on a torus (\ref{cmap}), (\ref{mmatrix}) can be calculated and is equal to the sum \cite{anosov,smale,sinai2,margulis,bowen0,bowen,bowen1}:
\be\label{entropyofT}
h(T) = \sum_{\vert \lambda_{\beta} \vert > 1} \ln \vert \lambda_{\beta} \vert.
\ee
\begin{figure}
 \centering
\includegraphics[width=5cm]{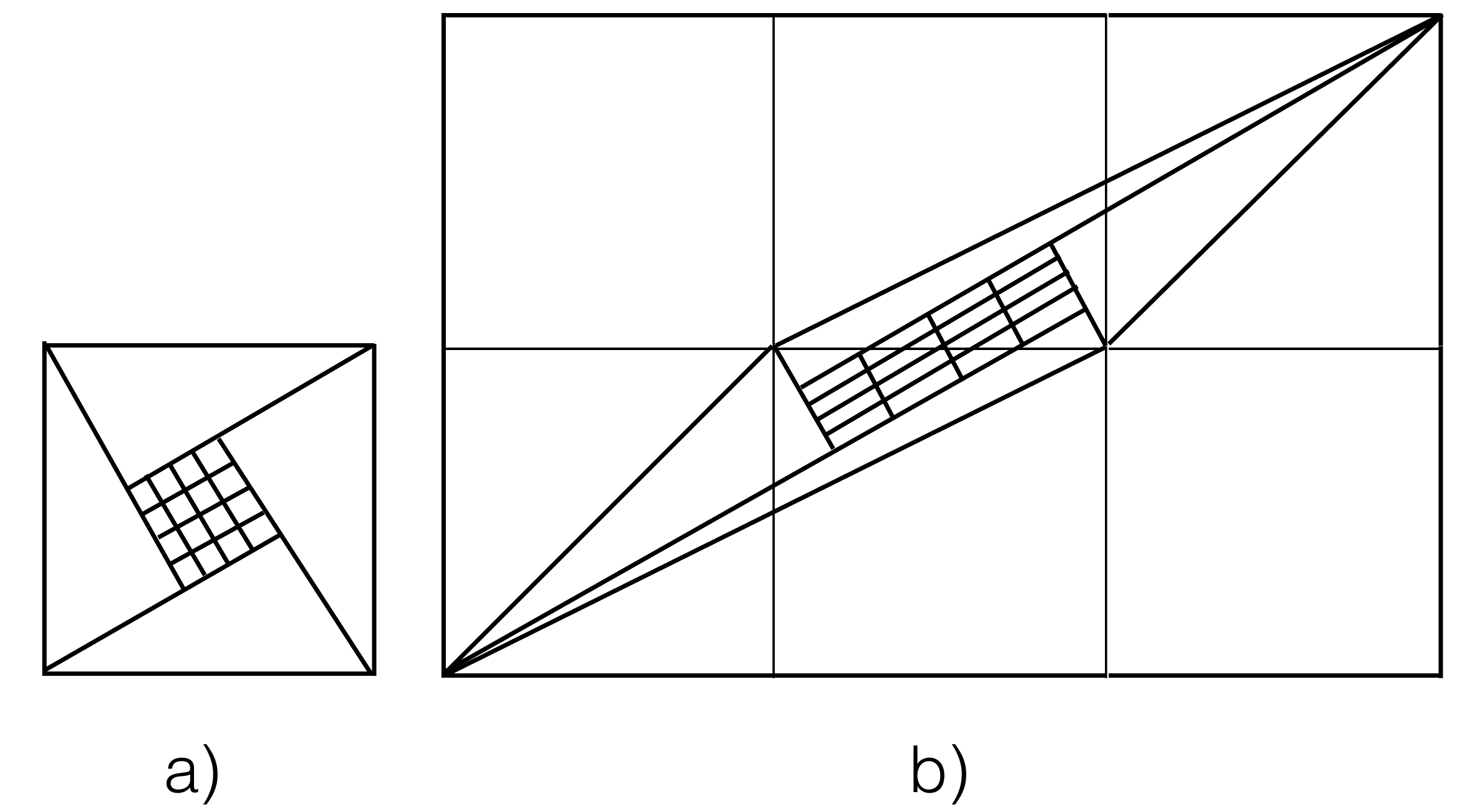}
\centering
\caption{The automorphisms on a two-torus.   
The a) depicts the parallel lines along the eigenvectors  
and b) depicts their positions after the action of the automorphism.} 
\label{fig2}
\end{figure}
{\it Thus the entropy $h(T)$ directly depends on the spectrum  of the operator $T$.  
This fact allows to characterise and compare the chaotic properties of  
dynamical C-systems quantitatively computing and comparing their entropies. }

The entropy defines the variety  and richness of the periodic 
trajectories of the C-systems  \cite{anosov,bowen0,bowen,bowen1,Gorlich:2016fbs}.
The C-systems have a countable set of everywhere dense  
periodic trajectories \cite{anosov}.
The $E^m$ cover of the torus $W^m$ allows 
to translate every set of points on torus into a set of points on Euclidean space $E^m$ and 
the space of functions on torus into the periodic functions on $E^m$.
To every closed curve $\gamma$ on a torus corresponds a curve  $\phi: [0,1] \rightarrow E^m$
for which $\phi(0) = \phi(1)~ mod~1$ and if $\phi(1) - \phi(0)=(p_1,...,p_m)$, then the 
corresponding winding numbers on a torus are $p_i \in Z$.

Let us fix the integer number $N$, then the points on a torus with 
the coordinates having a denominator $N$ form a finite set $\{p_1/N,...,p_m/N \}$. The 
automorphism (\ref{mmatrix}) with integer entries transform this set 
of points into itself, therefore all these points belong to periodic trajectories.
Let  $w=(w_1,...,w_m)$ be a point of a trajectory with the period $n > 1$. Then 
 \be\label{periodictrajectories1}
 T^n w = w + p,
 \ee
 where $p$ is an integer vector. The above equation with 
 respect to $w$ has nonzero determinant, therefore the components 
 of $w$ are {\it rational}. 
 
Thus the periodic trajectories of the period $n$ of the automorphism $T$ are given 
by the solution of the equation (\ref{periodictrajectories1}),
where $p \in Z^m$ is an integer vector and $w=(w_1,...,w_m) \in W^m$.  
 As $p$ varies in $Z^m$ the solutions of the equation (\ref{periodictrajectories1}) determine
 a fundamental domain $D_n$ in the covering Euclidian space $E^m$  of the volume 
 $\mu(D_n)=1/\vert Det (T^n-1) \vert $. Therefore the number of all points $N_n$ on the 
 periodic trajectories of the period $n$  
 is given by  the corresponding inverse volume \cite{smale,sinai2,margulis,bowen0,bowen,Gorlich:2016fbs}: 
 \be\label{numbers}
N_n = \vert Det (T^n-1) \vert = \vert  \prod^{m}_{i=1}(\lambda^n_i -1) \vert .
 \ee
Using the theorem of Bowen \cite{bowen,bowen1} which states that the entropy  
of the automorphism $T$ can be  represented  in terms of $N_n$
defined in   (\ref{numbers}): 
\be\label{bowen}
h(T)= \lim_{n \rightarrow \infty}{1\over n}~ \ln  N_n ~~,  
\ee
one can derive the formula for the entropy (\ref{entropyofT}) for the automorphism $T$
in terms of its eigenvalues: 
\be
h(T)= \lim_{n \rightarrow \infty}{1\over n} \ln (\vert  \prod^{m}_{i=1}(\lambda^n_i -1) \vert) =
\sum_{\vert \lambda_{\beta} \vert > 1} \ln \vert \lambda_{\beta} \vert.
\ee
Let us now define the number of periodic trajectories of the period  $n$  by $\pi(n)$.
Then the number of all points $N_n$ on the 
 periodic trajectories of the period $n$  can be written in the following form:
 \be
 N_n = \sum_{l~ divi~ n  } l ~\pi(l) ,
 \ee
where $l$ divides $n$. Using again the Bowen result (\ref{bowen}) one can get 
\be\label{assimpto}
N_n =\sum_{l~ divi~ n  } l \pi(l) \sim e^{n h(T)}.
\ee
This result can be rephrased as a statement that the number of points 
on the periodic trajectories of the period n exponentially 
grows  with the entropy.

Excluding the periodic trajectories which divide n 
(for example $T^n w=T^{l_2}(T^{l_1}w)$, where $n=l_1 l_2$ and $T^{l_i}w =w$)  
one can get the number of periodic trajectories of period n 
which are not divisible.
For that one should represent  the $\pi(n)$ in the following form:
\be\label{density}
\pi(n) = {1\over n} \big( \sum_{l~ divi~ n  } l ~\pi(l) -
\sum_{l~ divi~ n,~ l <n  } l ~\pi(l) \big)
\ee
and from (\ref{density})  and (\ref{assimpto}) it follows that 
\be\label{dencity}
\pi(n) \sim { e^{n h(T)} \over n}  \big( 1   -
{\sum_{l~ divi~ n,~ l <n  } l ~\pi(l) \over \sum_{l~ divi~ n  } l ~\pi(l)} \big) \sim { e^{n h(T)} \over n}, 
\ee
because  the ratio in the bracket is strictly smaller than one. This result tells that a
system with larger entropy  $\Delta h = h(T_1) - h(T_2) >0$ is more densely populated by the 
periodic trajectories of the same period $n$:
\be
{\pi_1(n) \over \pi_2(n)} \sim e^{n\ \Delta h}.
\ee
The next important result of the Bowen theorem \cite{bowen,bowen1} states that 
\be\label{ation9}
\int_{W^m} f(w) d\mu(w) = \lim_{n \rightarrow \infty}  {1\over N_n} \sum_{ w \in \Gamma_n} f(w),
\ee
where $\Gamma_n$ is a set of all points on the trajectories of period
$n$. 
The total number of points in the set $\Gamma_n$ we defined earlier as $N_n$.

This result has important consequences for the calculation of the 
integrals on the manifold $W^m$, because, as it follows from (\ref{ation9}), the integration 
reduces to the summation 
over all points of periodic trajectories. It is appealing to consider periodic trajectories 
of the period $n$ which is a prime number. Because every infinite subsequence of convergent  
sequence converges to the same limit we can consider in (\ref{ation9}) only terms 
with the prime periods.  In that case $N_n = n \pi(n)$ and the above formula becomes:
\be\label{integral}
\int_{W^m} f(w) d\mu(w) = \lim_{n \rightarrow \infty}  {1\over n \pi(n)} \sum^{\pi(n)}_{j=1}
\sum^{n-1}_{ i=0} f(T^i w_{j}),
\ee
where the summation is over all points of the trajectory $T^i w_{j}$ and  over all 
distinct  trajectories of period n which are enumerated by index $j$. The $w_{j}$ is the initial point of the 
trajectory $j$ \footnote{It appears to be a difficult mathematical problem to decide whether two  vectors 
$w_{1}$ and $w_{2}$ belong to the same or to distinct trajectories.}.  
From the above consideration it follows that the convergence is guaranteed 
if one sums over all trajectories of the same period $n$.    One can conjecture     
that all $\pi(n)$ trajectories at the very large period $n$  contribute  
equally into the sum (\ref{integral}),  
therefore the integral (\ref{integral}) can be reduced to a  sum over fixed trajectory
\be\label{reduce}
  {1\over n } \sum^{n-1}_{ i=0} f(T^i w).
\ee
Thus the knowledge of the spectrum allows to calculate the entropy (\ref{entropyofT}) and the number of periodic trajectories of a period less than $n$ grows exponentially (\ref{dencity}). 

\section{\it  MIXMAX Random Number Generator} 

 It was suggested in 1986 in \cite{yer1986a} to use the MCDS  defined on a torus to generate high quality pseudorandom numbers for Monte-Carlo method. The modern powerful computers open a new era for the application of the Monte-Carlo Method  \cite{metropolis,neuman,neuman1,sobol,yer1986a,Demchik:2010fd,falcion}  for the simulation of physical systems with many degrees of freedom and of 
higher complexity. The Monte-Carlo simulation is an important computational 
technique in many areas of natural sciences, and it has significant application 
in particle and nuclear physics, quantum physics, statistical physics, 
quantum chemistry, material science, among many other multidisciplinary applications. 
At the heart of the Monte-Carlo (MC) simulations are pseudorandom  number generators (RNG).

Usually  pseudo random numbers are generated by deterministic recursive rules
\cite{yer1986a,metropolis,neuman,neuman1,sobol}. 
Such rules produce pseudorandom numbers, and it is a great challenge to design 
pseudo random number generators that produce high quality sequences. 
Although numerous RNGs introduced in the last  decades fulfil most of the 
requirements and are frequently used in simulations, each of them has some 
weak properties which influence the results \cite{pierr} and are less suitable for demanding 
MC simulations which are performed for the high energy experiments at CERN
and other research centres. 
The RNGs are essentially used in high energy experiments at CERN  for the design of the efficient particle detectors and for the statistical analysis of the experimental data. 
\begin{figure}
  \centering
\includegraphics[width=4cm]{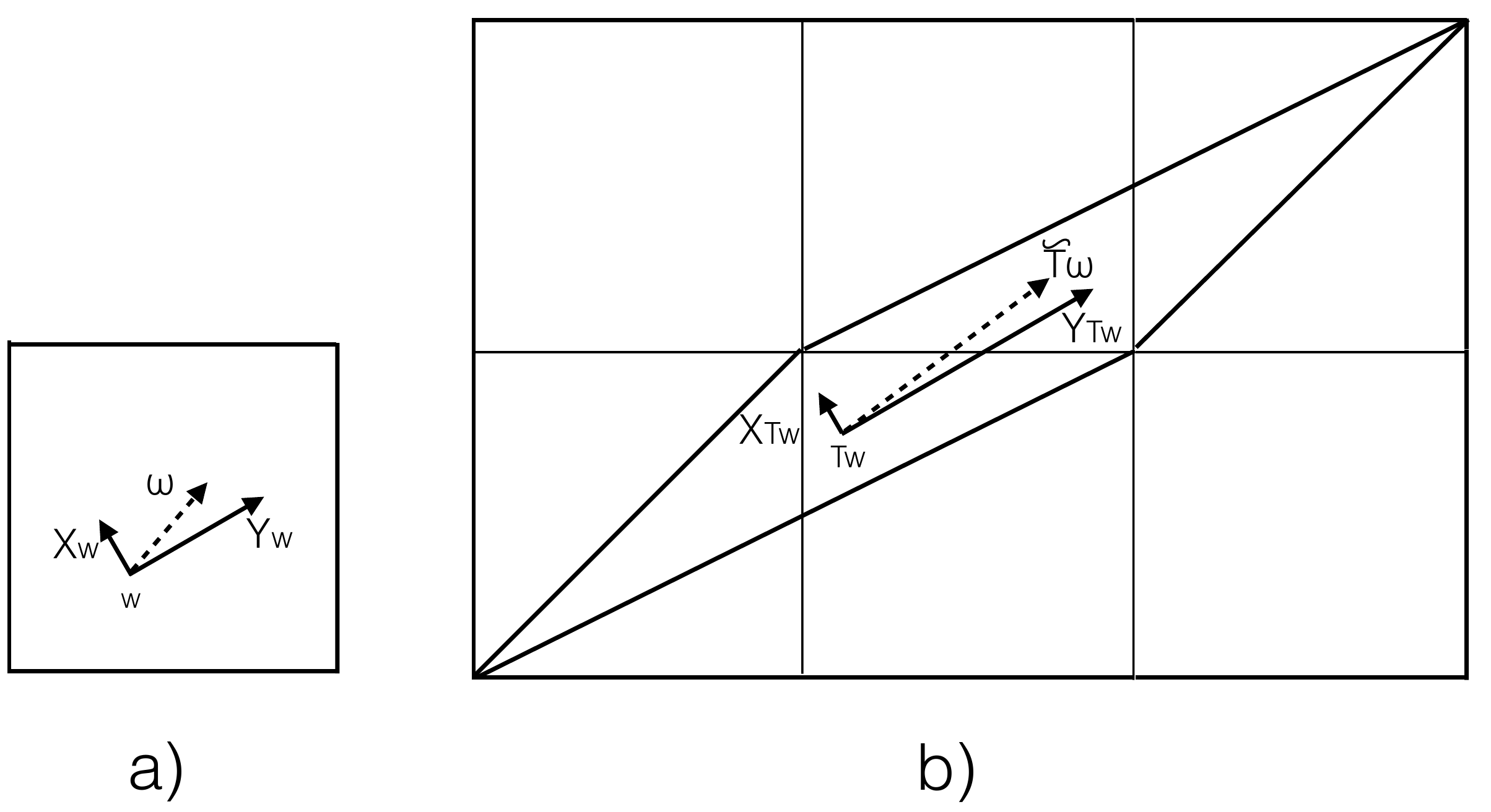}
\centering
\caption{The tangent vector $\omega \in R_{x}$ at  $x \in M^2$  is decomposable  
into the sum $R_{x}= X_{x} \bigoplus Y_{x}$
where the spaces $X_x$ and $Y_x$ are defined by the  
eigenvectors of the $2 \times 2 $ matrix $T(2,0)$ (\ref{eq:matrix}).  
 It is exponentially contracting  the distances on  $ X_{x}  $ and expanding the 
distances on  $ Y_{x} $ (details are given in Appendix A).} 
\label{fig3}
\end{figure}

In order to fulfil these demanding requirements it is necessary to have a solid theoretical 
and mathematical background on which the RNG's are based. RNG  should have a long period, 
be statistically robust, efficient, portable and have a possibility to change and 
adjust the internal characteristics in order to make RNG suitable for concrete 
problems of high complexity. In \cite{yer1986a} it was suggested that  
Anosov C-systems \cite{anosov},  defined on a high dimensional 
torus, are excellent candidates for the pseudo-random number generators.
The C-system chosen in \cite{yer1986a} was the one which realises 
linear automorphism  $T$  defined  in (\ref{cmap}).   For convenience 
in this section the dimension $n$ of the phase space $M$ is denoted by  $N$.
A particular matrix chosen in \cite{yer1986b} was defined for all $N \geq 2$. The operators $T(N,s)$  
are parametrised by the integers  $N$ and $s$ 
\be
\label{eq:matrix}
T(N,s) = 
   \begin{pmatrix} 
      1 & 1 & 1 & 1 & ... &1& 1 \\
      1 & 2 & 1 & 1 & ... &1& 1 \\
      1 & 3 & 2 & 1 & ... &1& 1 \\
      1 & 4 & 3 & 2 &   ... &1& 1 \\
      &&&...&&&\\
      1 & N & N-1 &  N-2 & ... & 3 & 2
   \end{pmatrix}
\ee
Its  entries are all integers  $T_{ij} \in \mathbb{Z}$ and 
 $Det~ T =1$. The spectrum and the value of the Kolmogorov entropy can be calculated. 
 \begin{figure}
 \centering
\includegraphics[width=3cm]{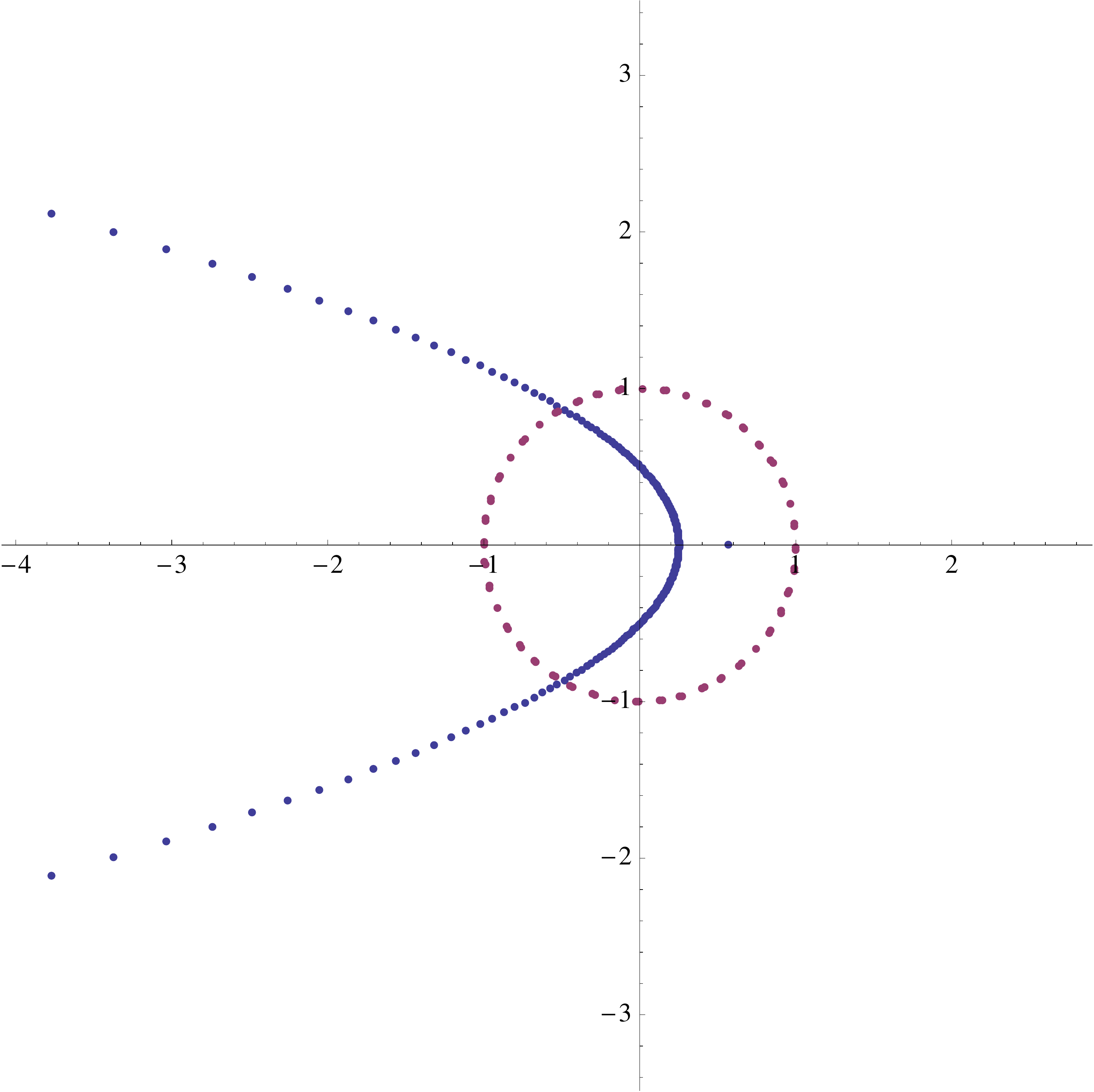}~ 
\includegraphics[width=3cm]{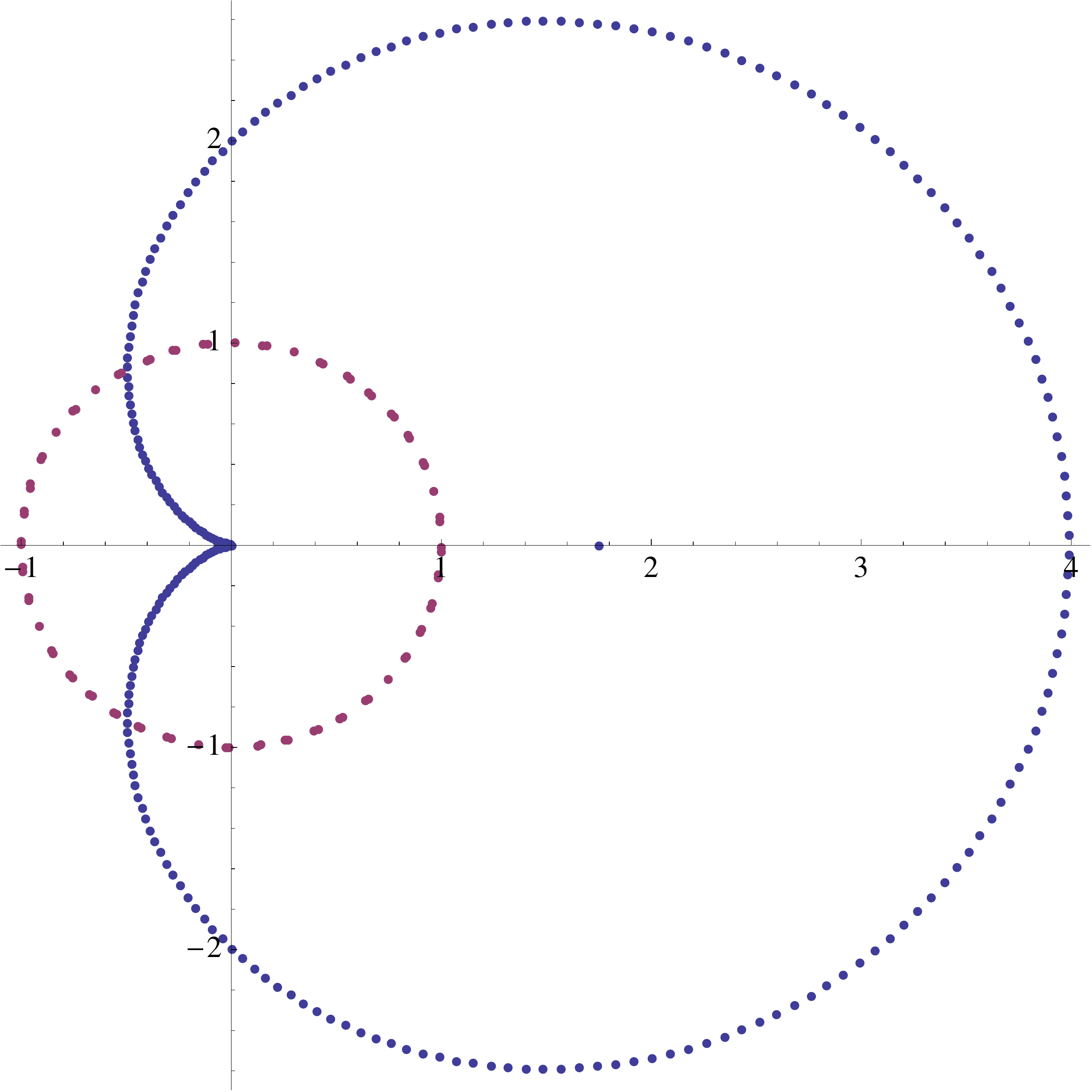}~~~~~~~~~
\centering
\caption{The distribution of the eigenvalue of operator $A=T(N,s)$ and of ots inverse $T^{-1}(N,s)$ in the complex $\lambda $ plane.  All eigenvalues are lying  outside of the unit circle.   The points of the curve that are inside the unit circle represent the eigenvalues with modulus less than one $\{\lambda_{\alpha}\}$ and outside the circle the eigenvalues $\{\lambda_{\beta}\}$ with modulus larger than one. } 
\label{fig4}
\end{figure}
 It is defined recursively, 
since the matrix of size $N+1$ contains in it the matrix of the size $N$.  
In order to generate pseudo-random vectors  $x_n = T^n x$, one should 
 choose  the initial vector $x=(x_{1},...,x_{N})$, called the ``seed",  
 with at least one non-zero component to avoid fixed point of $T$, which is at the origin.
The eigenvalues of the $T$ matrix (\ref{eq:matrix}) are widely dispersed for all $N$,
see Fig.\ref{fig4}
from reference \cite{konstantin}. The spectrum  is "multi-scale", with trajectories exhibiting exponential instabilities at different scales \cite{yer1986a}. The spectrum of the operator $T(N,s)$
has two real eigenvalues for even $N$ and three for odd $N$, all the rest
of the eigenvalues are complex and lying on leaf-shaped curves.
It is seen that the spectrum tends to a universal limiting form as
$N$ tends  to infinity, and the complex eigenvalues  $ 1/\lambda$
(of the inverse operator) lie asymptotically on the cardioid  curve Fig.\ref{fig4} which has the representation
\be 
r(\phi) = 4 \cos^2(\phi/2)
\label{eq:curve}
\ee
in the polar coordinates $ \lambda = r \exp(i\phi)$. From the above 
analytical expression for eigenvalues it follows that the eigenvalues satisfying the 
condition $ 0 <  \vert \lambda_{\phi} \vert   < 1$  are in the range $-2\pi/3 <  \phi < 2\pi/3$ 
and the ones satisfying the 
condition $ 1 <  \vert \lambda_{\phi} \vert $ are in the interval $2\pi/3 <  \phi < 4\pi/3$.
One can conjecture that there exists a limiting infinite-dimensional dynamical system with continuous space coordinate and discrete time with the above spectrum.  
The entropy of the C-K system $T(N,s)$ can now be calculated for
large values of $N$  as an integral over eigenvalues (\ref{eq:curve}):
\be\label{linear}
h(T)= \sum_{\alpha   } \ln \vert {1 \over \lambda_{\alpha} }\vert = \sum_{-2\pi/3 <  \phi_i < 2\pi/3} \ln (4\cos^2(\phi_i/2)~ \rightarrow  ~N \int^{2\pi/3}_{-2\pi/3} \ln (4\cos^2(\phi/2){d\phi \over 2\pi}=
{2\over \pi}~ N
\ee
and to confirm  that the entropy  {\it increases linearly with the dimension $N$ of the operator} $T(N,s)$.

The period of the trajectories of the system $T(N,s)$ was found in \cite{konstantin} and is characterised by a prime number $p$\footnote{The general theory of Galois field and the periods
of its elements can be found in \cite{mixmaxGalois,lnbook,nied,niki}.} .
In  \cite{konstantin} the necessary and sufficient criterion were formulated for the sequence to be of the maximal possible period:
\be\label{period1}
\tau={p^N-1 \over p-1} \sim e^{N \ln p},~~~N \gg 1.
\ee
It follows then that the period of the trajectories exponentially increases with
the size of the  operator $T(N,s)$. 

{\it Computer Implementation.} In a typical computer implementation of the automorphism \eqref{eq:matrix} the initial vector will have rational components $u_i=a_i/p$, where $a_i$ and $p$ are natural numbers.  Therefore it is convenient to represent $u_i$ by its numerator $a_i$ in computer memory and define the iteration in terms of $a_i$ \cite{mixmaxGalois}:
\be
\label{eq:recP}
a_i \rightarrow\sum_{j=1}^N T_{ij} \, a_j ~\textrm{mod}~ p .
\ee
If the denominator p is taken to be a prime number \cite{mixmaxGalois}, 
then the recursion is realised on extended 
Galois field $GF[p^N]$  \cite{niki,nied} and 
allows to find the period of the trajectories in terms of p and the properties of the 
characteristic polynomial $P(x)$ of the matrix T \cite{mixmaxGalois}. If 
the characteristic polynomial $P(x)$ of matrix $T$ is primitive in the 
extended Galois  field $GF[p^N]$, then
\cite{mixmaxGalois,nied,lnbook}
\be\label{period}
 T^q = p_0~ \mathbb{I}~~\textrm{ where}~~  q=\frac{p^N-1} {p-1} ~,
\ee
where $p_0$ is a free term of the  polynomial $ P(x)$ and is a {\it primitive element} of $GF[p]$.
Since our matrix T has $p_0=Det T= 1$, the polynomial $ P(x)$ of T cannot be primitive. 
The solution suggested  in \cite{konstantin} is to define the necessary and 
sufficient conditions for the period $q$  
to attain its maximum  are the following:
\begin{enumerate}
\item[\bf{1.}] $T^q = \mathbb{I} ~(mod~ p) $,~~~where $q=\frac{p^N-1} {p-1}$
\item[\bf{2.}] $T^{q/r} \neq \mathbb{I} ~(mod~ p)$,~~~~ for any r which is a prime divisor of q .
\end{enumerate}
The first condition is equivalent to the requirement  that the characteristic polynomial is irreducible. 
The second condition can be checked if the integer factorisation of $q$ is available \cite{konstantin}, then
the period of the sequence is equal to (\ref{period}) and is independent of the seed. 
There are precisely $p-1$ distinct  trajectories which together fill up all states of the $GF[p^N]$ lattice: 
\be
 q~ (p-1) = p^N-1.
\ee
In \cite{konstantin} the actual value of p was  taken as $p=2^{61}-1$, 
the largest Mersenne number that fits into an 
unsigned integer on current 64-bit computer architectures.  For the matrix of the 
size $N=256$ the period in that case is  $q \approx 10^{4600}$.
The algorithm  which allows the efficient implementation of the generator 
in actual computer hardware, reducing the matrix multiplication to the O(N) operations
was found in \cite{konstantin}.
The other advantage of this implementation is that it allows to make  "jumps" into 
any point on a periodic trajectory 
without calculating all previous coordinates on a trajectory, which typically has a 
very large  period $q \approx 10^{4600}$.
The MIXMAX generators were integrated into the concurrent and distributed MC toolkit Geant4 \cite{Geant4}, the foundation library CLHEP \cite{CLHEP} and data analysis framework ROOT \cite{root}. These software tools have wide applications in High Energy Physics at CERN, in CMS experiment \cite{CMSrunII,CMS}, at SLAC, FNAL and KEK National Laboratories and are part of the CERN's active Technology Transfer policy. The generator is available in the PYTHIA event generator \cite{PYTHIA}. The MIXMAX code can be downloaded from the GSL-GNU Scientific Library \cite{GSL}.

\section{\it  Turning C-cascade into C-flow}

\begin{figure}
 \centering
\includegraphics[width=4cm]{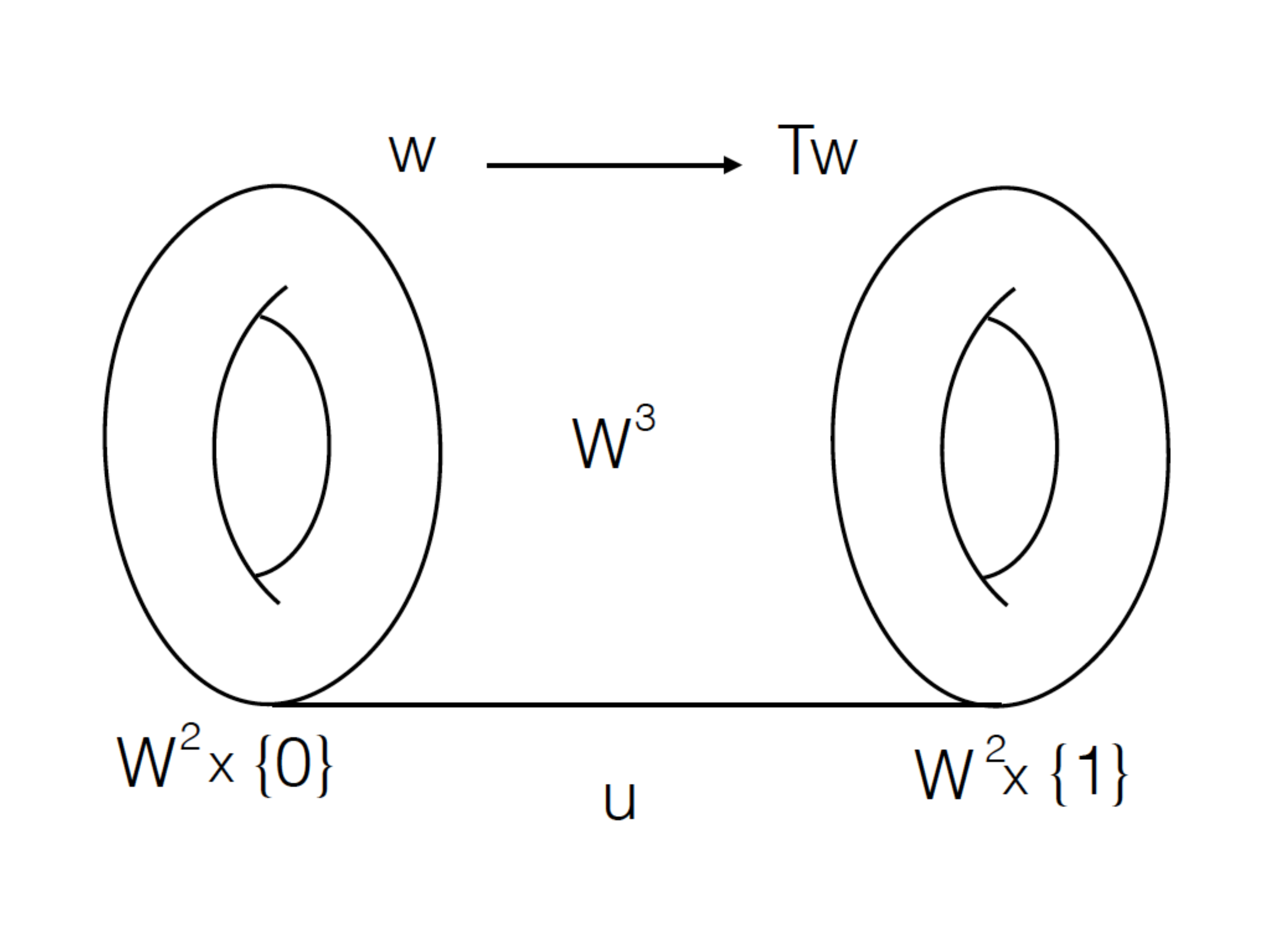}
\caption{The  identification of the  $W^2 \times \{0\}$ with $W^2 \times \{1\}$ by the 
formula $(w,1) \equiv  (Tw,0)$ of a cylinder $W^2 \times [0,1]$, where $[0,1]= 
\{  u~ \vert ~ 0 \leq u \leq 1 \}$.
The resulting compact manifold $W^{3}$ has a bundle structure 
with the base $S^1$ and fibres of the type $W^2$. The manifold $W^{3}$ 
has the local coordinates $\tilde{w}=(w^1,w^2,u)$ . 
} 
\label{fig14}
\end{figure}
In \cite{anosov} Anosov demonstrated how any C-cascade on a torus can be
embedded into a certain  C-flow. The embedding was defined by the identification 
(\ref{identification}) and the corresponding  C-flow on a smooth Riemannian manifold 
$W^{m+1}$ with the metric (\ref{metric}) was defined by the equations (\ref{velo}).
We are interested here to analyse the {\it geodesic flow} on the same Riemannian manifold $W^{m+1}$. 
The geodesic flow has different dynamics (\ref{geodesicflow} ) and  as we shall demonstrate below 
has very interesting hyperbolic components different from (\ref{velo}). 

Let us consider a C-cascade on a torus $W^m$ and increase its dimension m  by one unit
constructing a cylinder $W^m \times [0,1]$, where $[0,1]= \{  u~ \vert ~0 \leq u \leq 1 \}$,
and identifying $W^m \times \{0\}$ with $W^m \times \{1\}$ by the formula:
\be\label{identification}
(w,1) \equiv  (Tw,0).
\ee
Here T is diffeomorphism:
\bea\label{cmap1}
 w^i \rightarrow \sum T_{i,j} w^j,~~~~(mod ~1).
\eea
The resulting compact Riemannian manifold $W^{m+1}$ has a bundle structure
with the base $S^1$ and fibres of the type $W^m$. The manifold $W^{m+1}$
has the local coordinates $\tilde{w}=(w^1,...,w^m,u)$ shown in Fig.\ref{fig14}.
The C-flow $T^t$ on the manifold $W^{m+1}$ is defined by the equations \cite{anosov}
\be\label{velo}
{d  w^1 \over d t}=0~, ....,~ {d  w^m  \over d t} = 0,~ {d  u \over d t}=1.
\ee
For this flow the tangent space $R^{m+1}_{\tilde{w}}$ can be represented as
a direct sum of three subspaces:  contracting  and  expanding
 linear spaces  $X^k_{\tilde{w}} $,~$ Y^l_{\tilde{w}} $   and $ Z_{\tilde{w}}$:
 \be
R^{m+1}_{\tilde{w}} = X^k_{\tilde{w}}  \oplus  Y^l_{\tilde{w}}  \oplus  Z_{\tilde{w}}.
\ee
The linear space  $X^k_{\tilde{w}} $ is tangent to the fibre  $W^m \times u$ and is parallel
to the eigenvectors corresponding to the eigenvalues which are lying inside the unit circle  $0 <  \vert \lambda_{\alpha} \vert   < 1$ and $Y^l_{\tilde{w}} $ is tangent to the fibre  $W^m \times u$ and is parallel to the eigenvectors  corresponding to the eigenvalues
which are lying outside of the unit circle $1 <\vert \lambda_{\beta}\vert$. $ Z_{\tilde{w}}$ is collinear to the phase space velocity (\ref{velo}). Under the derivative mapping of the (\ref{velo}) the vectors (\ref{eigenvectros}) from $X^k_{\tilde{w}} $ and  $Y^l_{\tilde{w}} $ are contracting  and expanding:
\be
\vert \tilde{T}^{t} v_1 \vert =  \lambda_2^{ t}~ \vert v_1 \vert,~~~~
\vert \tilde{T}^{t} v_2 \vert =  \lambda_1^{ t}~ \vert v_2 \vert.
\ee
This identification of contracting and expanding spaces proves  that (\ref{velo})
indeed defines a C-flow \cite{anosov}.

It is also interesting to analyse the {\it geodesic flow} on a  Riemannian manifold
$W^{m+1}$. The equations for the geodesic flow on $W^{m+1}$
\be
{d^2 \tilde{w}^{\mu} \over d t^2} +\Gamma^{\mu}_{\nu \rho}
{d  \tilde{w}^{\nu} \over d t } {d  \tilde{w}^{\rho} \over d t } =0
\ee
are different from the flow equations
defined by the equations (\ref{velo}) and our goal is to learn if the geodesic flow
has also the properties of the C-flow. The answer to this  question is not obvious and
requires investigation of the curvature structure of the manifold $W^{m+1}$. If all sectional
curvatures  are negative then geodesic flow defines a C-flow \cite{anosov}. For simplicity 
let us consider the automorphisms of  a two-dimensional torus which is defined by the $2 \times 2 $ matrix $T(2,0)$ (\ref{eq:matrix}).
The metric on the corresponding manifold $W^3$ can be defined as \cite{arnoldavez}
\be\label{metric}
ds^2 = e^{2u} [\lambda_1 d w^1 + (1-\lambda_1) d w^2]^2 +
e^{2u} [\lambda_2 d w^1 + (1-\lambda_2) d w^2]^2 +du^2 =\nn\\
 g_{\mu\nu} d\tilde{w}^{\mu} d\tilde{w}^{\nu},
\ee
where $0 < \lambda_2  < 1 <  \lambda_1$ are eigenvalues of the matrix $T(2,0)$ and
fulfil the relations $\lambda_1 \lambda_2 =1,\lambda_1+ \lambda_2=3$.
The metric  is invariant  under the transformation
\be\label{trans}
w^1 = 2 w^{'1} - w^{'2},~~~ w^2 = -w^{'1}_1 + w^{'2}, ~~~u  = u^{'}-1
\ee
and is therefore consistent with the identification (\ref{identification}). The metric
tensor has the form
\be\label{metric}
 g_{\mu\nu}(u)= \begin{pmatrix}
\lambda_1^{2 + 2 u} + \lambda_2^{2 + 2 u} & (1 - \lambda_1) \lambda_1^{1 + 2 u} + (1 - \lambda_2) \lambda_2^{1 + 2 u}& 0 \\
(1 - \lambda_1)\lambda_1^{1 + 2 u} + (1 - \lambda_2) \lambda_2^{1 + 2 u}&(1 - \lambda_1)^2 \lambda_1^{2 u} + (1 - \lambda_2)^2 \lambda_2^{2 u}&0\\
0&0&1\\
 \end{pmatrix}
 \ee
and the corresponding geodesic equations take the following form:
 \bea\label{geodesicflow}
&  \ddot{w}^1 + 2{(\lambda_1-1) \ln\lambda_1 \over \lambda_1+1} \dot{w^1}\dot{u}
-4 {(\lambda_1-1) \ln\lambda_1 \over \lambda_1+1} \dot{w^2}\dot{u}  =0
\nn\\
&  \ddot{w}^2 - 2{(\lambda_1-1) \ln\lambda_1 \over \lambda_1+1} \dot{w^2}\dot{u}
  - 4 {(\lambda_1-1) \ln\lambda_1 \over \lambda_1+1} \dot{w^1}\dot{u} =0\\
& \ddot{u} + {(1-\lambda^{4u+4}_1) \ln\lambda_1 \over \lambda^{2u+2}_1} \dot{w^1}\dot{w^1}
 + 2{(1+\lambda^{4u+3}_1)(\lambda_1-1) \ln\lambda_1 \over \lambda^{2u+2}_1} \dot{w^1}\dot{w^2}+\nn\\
& +  {(1-\lambda^{4u+2}_1)(\lambda_1-1)^2 \ln\lambda_1 \over \lambda^{2u+2}_1} \dot{w^2}\dot{w^2}=0.\nn
\eea
One can get convinced that these equations are invariant under the transformation (\ref{trans}).
 In order to study a stability of the geodesic flow one has to compute the sectional curvatures.
We shall choose the orthogonal frame in the directions of the linear spaces  $X^1_{\tilde{w}} , Y^1_{\tilde{w}} $   and $ Z_{\tilde{w}}$. The corresponding vectors are:
\be\label{eigenvectros}
v_1 = (\lambda_1-1, \lambda_1,0),~~~v_2=(\lambda_2 -1, \lambda_2,0),~~~ v_3=(0,0,1)
\ee
and in the metric (\ref{metric}) they have the lengths:
\be
\vert v_1  \vert^2= (\lambda_1 - \lambda_2)^2 \lambda_2^{2 u},~~~~
\vert v_2  \vert^2= (\lambda_1 - \lambda_2)^2 \lambda_1^{2 u},~~~~
\vert v_3 \vert^2 = 1.
\ee
The corresponding sectional curvatures can be computed and the following values:
\bea
K_{12} = {R_{\mu\nu\lambda\rho} v^{\mu}_1 v^{\nu}_2  v^{\lambda}_1 v^{\rho}_2 \over
\vert v_1 \wedge v_2 \vert^2}=   \ln^2 \lambda_1
\nn\\
K_{13} = {R_{\mu\nu\lambda\rho} v^{\mu}_1 v^{\nu}_3  v^{\lambda}_1 v^{\rho}_3 \over
\vert v_1 \wedge v_3\vert^2}= -  \ln^2 \lambda_2
\\
K_{23} = {R_{\mu\nu\lambda\rho} v^{\mu}_2 v^{\nu}_3  v^{\lambda}_2 v^{\rho}_3 \over
\vert v_2 \wedge v_3 \vert^2}= -  \ln^2 \lambda_1. \nn
\eea
It follows from the above equations that the geodesic
flow is exponentially unstable on the planes (1,3) and (2,3)
and is stable in the plane (1,2). This behaviour is dual to the flow (\ref{velo}) which
is unstable in (1,2) plane and is stable in (1,3) and (2,3) planes.
 The scalar curvature is
\be
R= R_{\mu\nu\lambda\rho} g^{\mu\lambda}g ^{\nu\rho} = 2(K_{12} +K_{13}+K_{23}) =   - 2 
\ln^2 \lambda_1 = -2 h(T)^2 ,
\ee
where $h(T)$ is the entropy of the automorphism $T(2,0)$.

\section{\it  Infinite Dimensional Limit of the C-cascade}

We are interested to consider the infinite-dimensional limit $N\rightarrow \infty $ of the system  (\ref{cmap}) when the operator $T(N)$ is given by the $N\times N$ matrix with all integer entries   $T_{ij} \in \mathbb{Z}$ and has the form given in (\ref{eq:matrix}) \cite{yer1986a,konstantin,Savvidy:2015jva}:
 It has the determinant equal to one and its spectrum has the form  \cite{Savvidy:2015jva} (see also (\ref{eq:curve}))
\bea\label{eq:curve1}
&\lambda_j = 1 -2 \exp(i\, \pi j/N) + \exp(2i\, \pi j/N)= 
-4 \sin^2({\pi j\over 2 N})  \exp(i\, \pi j/N)  \\
& ~ j=-N, -N+2,...N-2, N ~,\nn
\eea
shown  in Fig.\ref{fig4}. The kernel  $T(N) u_0=0$ of the operator $T(N)$ consists of only one vector with zero components $u_0= (0,...,0)$. The eigenvalues  fulfil the C-condition (\ref{mmatrix}) and the entropy of the system  can be calculated for the large values of $N$  as a sum over eigenvalues:
\be\label{linear}
h(A)= \sum_{\beta   } \ln \vert  \lambda_{\beta} \vert = \sum_{-2\pi/3 <  \phi_j < 2\pi/3} \ln (4\cos^2(\phi_j/2)~ \rightarrow  ~N \int^{2\pi/3}_{-2\pi/3} \ln (4\cos^2(\phi/2){d\phi \over 2\pi}=
{2\over \pi}~ N,
\ee
where the eigenvalues are given in (\ref{eq:curve}) and  $\phi_j = \pi {j\over N} $. The entropy increases linearly with the dimension $N$ of the operator $T(N)$. We note that the special form of the matrix $T(N)$ in \eqref{eq:matrix} has highly desirable property of having a widely spread, nearly continuum spectrum of eigenvalues (\ref{eq:curve}) shown in Fig.\ref{fig4}) and indicating that the exponential mixing takes place in many scales \cite{yer1986a}.

 We are interested in defining and investigating  the system  that appears in the infinite-dimensional limit of $T(N)$ when $N \rightarrow \infty$. The size of state vector $u= (u_1,...,u_N)$ tends to infinity and it seems natural to expect that it can be represented by a continuous function $\psi$ defined in an infinite-dimensional space $\CH=\{\psi(u)\}$, possibly a Hilbert space, and the operator $T(N)$ will reduce to a differential operator $T$ acting in $\CH$. The existence of such a limit would mean that $T$ defines a "measure preserving"  transformation of continuous functions $  T^n \psi = \psi^{(n)} $ in $\CH$ that will have maximally strong chaotic properties. Our intension is to define this limiting system, to explore its properties and possible applications in Monte Carlo method.  It seems that the investigation of infinite-dimensional "fully chaotic" transformation of continuous functions may also help to understand better a chaotic/turbulent motion of fluids.  As it was demonstrated by Arnold \cite{turbul,Arakelian:1989},  the solutions of the partial differential equation describing the evolution  of the hydrodynamical flow of incompressible ideal fluid can be considered as a continuous measure preserving transformation of fluid velocity and the evolution is partially chaotic, that is, the flow is exponentially unstable in some directions and is stable in other directions (the details will be discussed in the last paragraph).  

\begin{figure}
\begin{center}
\includegraphics[width=3cm]{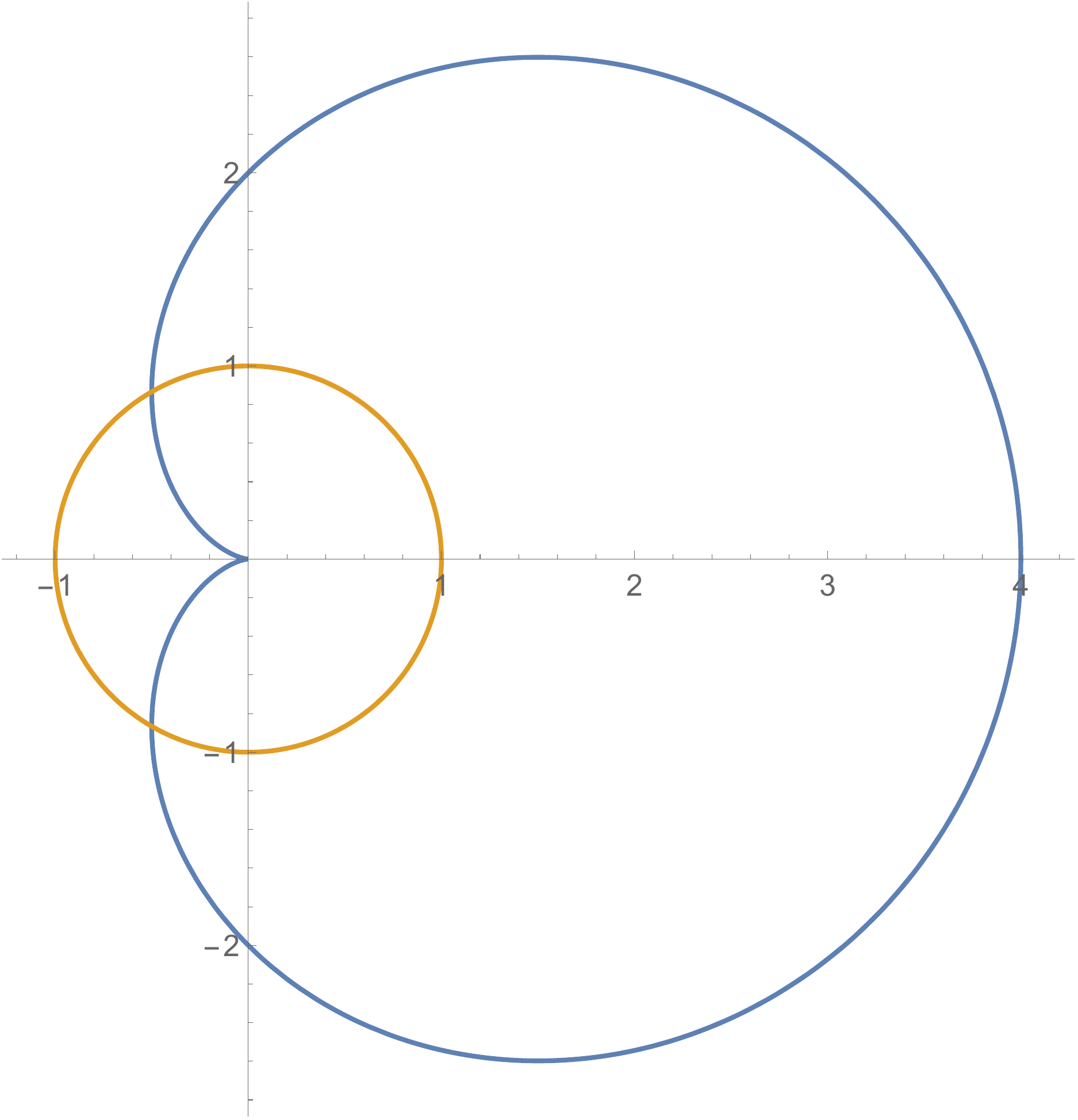}
\caption{
The distribution of the eigenvalues of the infinite-dimensional operator $A$ in the complex $\lambda $ plane. The points of the curve that are inside the unit circle represent the eigenvalues with modulus less than one $\{\lambda_{\alpha}\}$ and outside the circle the eigenvalues $\{\lambda_{\beta}\}$ with modulus larger than one.
}
\label{fig22}
\end{center}
\end{figure}

In many areas of mathematics  the consideration of infinite-dimensional limits is an ambiguous procedure, and a priory there is no guarantee that a sensible limit of a finite-dimensional structure exists. In our case the dimension of the N-dimensional torus $S^1 \otimes....\otimes S^1$  tends to infinity and it is unclear what type of phase space should be taken in the limit.  The hint that a sensible limit may exist comes from the fact that as $N \rightarrow \infty$ the eigenvalues (\ref{eq:curve}) fill out the cardioid curve more and more dense without producing any deformation of the cardioid curve that can be seen in figure Fig.\ref{fig22} \cite{Savvidy:2015jva}. The other important hint is that the inverse matrix $T(N)^{-1}$ is reminiscent to the matrix that represents the discrete version of the second-order differential operator \cite{Savvidy:2015jva}. If one supposes that the state vector becomes a function $\psi(x)$ with its argument on a real line $x \in R^1$, then it seems natural to look for a differential operator of the second order and the one that will reproduce the eigenvalue spectrum distributed on cardioid  curve.  Having in mind the above consideration let us consider the differential operator
\be\label{limitings}
T= 1 - 2 \exp{\Big({d \over d x}\Big)} + \exp{ \Big(2 {d \over d x}\Big)}= {d^2 \over d x^2} +{d^3 \over d x^3} +{7\over 12}{d^4 \over d x^4} +...
\ee
 acting in the Hilbert space of functions $\CH=\{\psi(x)\}$ defined on the interval $x \in [-\infty, +\infty]$. The series expansion of the operator has indeed a second-order differential operator and also infinitely many high derivative terms $$T= {d^2 \over d x^2} +{d^3 \over d x^3} +{7\over 12}{d^4 \over d x^4} +...$$ Its spectral characteristics are defined by the eigenvalue equation 
\be 
T \psi(x) = \lambda \psi(x).
\ee
Searching the eigenfunctions in the form of plane waves 
\be\label{eigenfun}
\psi_a(x)= e^{i a x}
\ee
one can find that the spectrum represents a continuous cardioid curve on the complex plane
\be\label{eigenvalues12}
\lambda(a) = 1-2 e^{i a} + e^{2 i a}= 4 e^{i (a+\pi)} \sin^2{\Big({a\over 2}\Big)}.
\ee
It is similar to the discrete spectrum (\ref{eq:curve}) and has the periodic structure 
\be\label{period}
\lambda(a+2\pi k) =\lambda(a), ~~k=0,\pm 1, \pm2, ....
\ee
The  spectrum is continuous and the eigenvalues are distributed in the complex plane representing the cardioid  curve shown in Fig.\ref{fig2}. As the real momentum parameter $a$ varies on the real line interval  the eigenvalues run around the cardioid infinitely many time (\ref{period}). The eigenvalues of the operator $A$ can be divided into two sets   $\{ \lambda_{\alpha}  \} $ and $\{  \lambda_{\beta }  \} $
with modulus smaller and larger than one:
\bea\label{eigenvalues}
0 <  \vert \lambda_{\alpha} \vert   < 1,~~
1 <  \vert \lambda_{\beta} \vert   . 
\eea
The eigenvalues $\lambda_{\alpha}$ and  $\lambda_{\beta}$ can be found  using (\ref{eigenvalues12}):
\bea\label{lessone}
\lambda_{\alpha} &=& 4 e^{i (a+\pi)} \sin^2{({a\over 2})} ,~~\text{when}~~ 
- {\pi \over 3} +2\pi k   < a < + {\pi \over 3} +2\pi k , ~~~~~\nn\\
\label{largeone}
\lambda_{\beta} &=& 4 e^{i (a+\pi)} \sin^2{({a\over 2})} ,~~\text{when}~~   + {\pi \over 3} +2\pi k   < a < + {5 \pi \over 3} +2\pi k,
\eea
where $k=0,\pm 1, \pm2, ....$.This structure of the spectrum repeats itself with the period $2\pi$.
There are two eigenvalues $\lambda =1$ corresponding to  $a=\pm \pi/3$ where the cardioid intersects a unit circle shown  in Fig.\ref{fig2}.

In order to establish the fact that the operator $T$ is defining a measure-preserving transformation one should calculate the determinant of the operator $T$. The measure-preserving transformations of the phase space is a characteristic property of Hamiltonian systems that is expressed in terms of the Liouville's theorem and represent a large class of dynamical systems that are considered in ergodic theory \cite{Gibbs,hopf1,krilov,arnoldavez,kornfeld}.  Using the fact that $\ln Det T = Tr \ln T $ we will have 
\be\label{det}
\ln Det A_x= \sum^{\infty}_{k=-\infty} \int^{+\pi +2\pi k }_{-\pi +2\pi k} \ln[  e^{i (a+\pi)}  4 \sin^2{\Big({a\over 2}\Big)} ]{d a \over 2 \pi}=0,
\ee 
that is, the determinant is equal to one $Det T=1$ and the operator $T$ is defining  a measure-preserving transformation. One can define now the homomorphism of the Hilbert space  $\CH=\{\psi(x)\}$ in terms of the operator $T$  as 
\be\label{direct}
\psi^{(1)}(x) = T \psi(x)
\ee
and the dynamical system on the infinite dimensional phase space $\CH$ as: 
\be\label{auth}
\psi^{(n)}(x) = T^n  \psi(x)~~~~~\textrm{mod}~ 1 ,~~~~~~~~n=0,1,2,....
\ee
 The homomorphism  (\ref{auth}) is defined by mod 1 operation meaning that the functions are  wrapping around an infinitely long cylinder $R^1 \otimes S^1  $\footnote{One can imagine this cylinder as appearing in the "decompactification" of a 2-torus $S^1 \otimes S^1$, when cutting a 2-torus with a non-contractible circle $S^1$ and stretching the open boundary circles to the infinities. }. The  function $\psi(x)$ maps $R^1$ to $S^1$ ($\psi: R^1 \rightarrow S^1$).

The determinant of the operator $T$ is equal to one and the eigenvalues are distributed inside and outside of the unit circle, and we have an example of infinite-dimensional hyperbolic C-K system of the type (\ref{mmatrix}). To get convinced that the operator $T$ represents a hyperbolic C-K system one should establish the existence of exponentially expanding and contracting foliations \cite{anosov}. Let us consider  the evolution of the infinitesimal perturbation $\psi \rightarrow \psi + \delta \psi $ under the  action of $T$ operator in analogy with the geodesic deviation equation: 
\be\label{primarysys}
\delta\psi^{(n)}(x) = T^n  \delta\psi(x).
\ee
The  deviation $\delta L$ can be evaluated by using the standard inner product in Hilbert space and the mean value theorem\footnote{ If   $f(x)$  is a continuous function and g(x)  is a nonnegative integrable function on $[a, b]$, then there exists some number $\bar{x} \in (a, b) $, such that $\int^{b}_{a}  f(x) g(x) dx =f(\bar{x}) \int^{b}_{a} g(x) dx $.  }:
\bea
\delta L_n &=&   \langle  \delta\psi^+\vert T^n  \delta\psi \rangle  =\int^{+\infty}_{-\infty} \delta\psi^+(x) T^n  \delta\psi(x) dx  \\
&=& \int^{+\infty}_{-\infty} \int_{\Delta a}
 {da^{'} \over 2 \pi}\int_{\Delta a} {da \over 2 \pi}~ e^{- i a^{'} x} \delta\phi^+(a^{'}) ~  \Big[4  \sin^2{({a\over 2})} \Big]^n ~ e^{in (a+\pi)} e^{ i a x }   \delta\phi(a)~  
dx \nn\\
&=&  {1\over 2\pi }\int_{\Delta a} da  ~    e^{in (a+\pi)} ~ \Big[4  \sin^2{({a\over 2})} \Big]^n   ~\vert \delta\phi(a) \vert^2   ~ =  ~e^{in (\bar{a}+\pi)}  e^{ n \ln{[4  \sin^2{({\bar{a}\over 2})}]} }~  \overline{\vert \delta\phi^2 \vert}, \nn
\eea 
where $\bar{a}$ is a  number in the interval $\Delta a \subset ( \pi/3,5\pi/3 ) $ and 
$ \overline{\vert \delta\phi^2 \vert}  =   {1 \over 2 \pi} \int_{\Delta a}  \vert \delta\phi \vert^2  da $.
The absolute value of the deviation  is growing exponentially with the iteration time  $n$  as 
\be
\vert \delta L_n \vert \sim  \overline{\vert \delta\phi^2 \vert} ~e^{ n \ln{[4  \sin^2{({\bar{a}\over 2})}]} } .
\ee 
In the above perturbation the Fourier  spectrum of the  function  $\delta\psi(x)$ is localised in the region of the spectrum $  \Delta a $,   where the eigenvalues $\lambda_{\beta}$ are larger than one, $\Delta a \subset ( \pi/3,5\pi/3 )~ $,  and are defined in (\ref{largeone}).  The integration over $a$ and $a^{'}$ was specified to be in that region $  \Delta a $  and the perturbation function had the following form:
\be
 \delta\psi(x) =   \int_{ \Delta a}  \delta\phi(a)    e^{i a x}     {da\over 2 \pi}.
\ee
In a similar way one can get convinced that the exponential contraction takes place when the Fourier spectrum of the perturbation function is localised in the part of the spectrum $\lambda_{\alpha}$, where the eigenvalues are less than one $\Delta a \subset ( -\pi/3,\pi/3 )$ in (\ref{lessone}). The existence of exponentially expanding and contracting foliations is a sufficient condition for a dynamical system to expose strong statistical/chaotic  properties and, in particular, to have nonzero Kolmogorov-Sinai entropy  and to be classified as a  hyperbolic C-K system. In physical terms this means that the Fourier amplitudes $ \phi(a)$ of the initial state vector $\psi(x)$ are stretched and compressed depending on whether the value of $a$  is in the interval $a \in ( {\pi \over 3} +2\pi k ,  {5 \pi \over 3}+2\pi k) $ or in the interval   $a \in ( -{\pi \over 3}+2\pi k,{\pi \over 3} +2\pi k) $, where $~k=0,\pm 1, \pm2, ....$.

The above consideration allows to  calculate the Kolmogorov entropy of the system per unit of the iteration time $n$ as it is was defined by Kolmogorov \cite{kolmo1}.  The new aspect that appears  in this infinite-dimensional system case is that the spectrum (\ref{eigenvalues12}) is continuous and repeats itself infinitely many times. For that reason the standard Kolmogorov definition of the entropy per unit iteration time is equal to infinity.  That can be observed also from the equation (\ref{linear}) when $N \rightarrow \infty$. In this circumstances  one can propose to calculate the entropy per unit period  (\ref{period}) of the spectrum  (\ref{eigenvalues12}):
\be\label{entropyofAA}
h(T)=\sum_{\alpha   } \ln {1\over \vert \lambda_{\alpha} \vert}=   -\int^{\pi/3}_{-\pi/3} \ln [4\sin^2({a\over 2})]{da \over 2\pi}=
2 i  [Li_2(e^{i5\pi/3}) -  Li_2(e^{i\pi/3})] \sim {2\over \pi} ,
\ee
 where we used the fact that   $ \prod_{\alpha} \lambda_{\alpha}   \prod_{\beta} \lambda_{\beta} =1$ and $Li_n(z)$ is the polylogarithm function. This result is understandable in the sense that the finite-dimensional matrix  system (\ref{eq:matrix}) considered above  had the entropy $\sim {2\over \pi} N$,  where $N$ is the dimension of the matrix operator.  As far as the operator $T$  can be considered  as the infinite dimensional limit $N\rightarrow \infty$ of (\ref{eq:matrix}), the standard Kolmogorov entropy of the system (\ref{auth}) tends to infinity but its entropy "per spectral period" is finite.  
  
The inverse operator $G$ is defined by the  equation 
\be\label{inverse}
T ~G(x-y) =  \sum^{\infty}_{k=-\infty} \delta(x-k)
\ee
and has the following solution: 
\be\label{fixed}
G(x-y) =  {1\over 2} (x-y)^2 \sum^{\infty}_{k=-\infty} e^{2 \pi k (x-y)}.
\ee
The general solution of (\ref{inverse}) can be expressed as a sum the fixed solution (\ref{fixed}) and an arbitrary element of the kernel $\CK$.  The kernel subspace $\CK$ of the operator $T$ is defined by the equation $T \psi_0(x)=0$ and has the following solution:
\be\label{kernel}
 \psi_0(x) = (c_1 x +c_2) \sum^{\infty}_{k=-\infty} a_k e^{2 \pi k i x},
\ee
where $c_1,c_2, a_k$ are arbitrary constants. The inverse transformation $\psi^{(-n)}   = G^n \psi,~ n=0,1,2,...$ is therefore defined modulo kernel (\ref{kernel}).  We will defined it in its most simple form (\ref{fixed}):
\be\label{inversetime}
\psi^{(-1)}  = \int G(x-y) \psi(y) dy = \sum^{\infty}_{k=-\infty} k^2 ~\psi(x-k)~~~~~\textrm{mod}~ 1.
\ee
Thus the evolution of the system is given in both "time directions" by (\ref{inversetime})  and  (\ref{direct}).  
 \begin{figure}
 \begin{center}
\includegraphics[width=5cm]{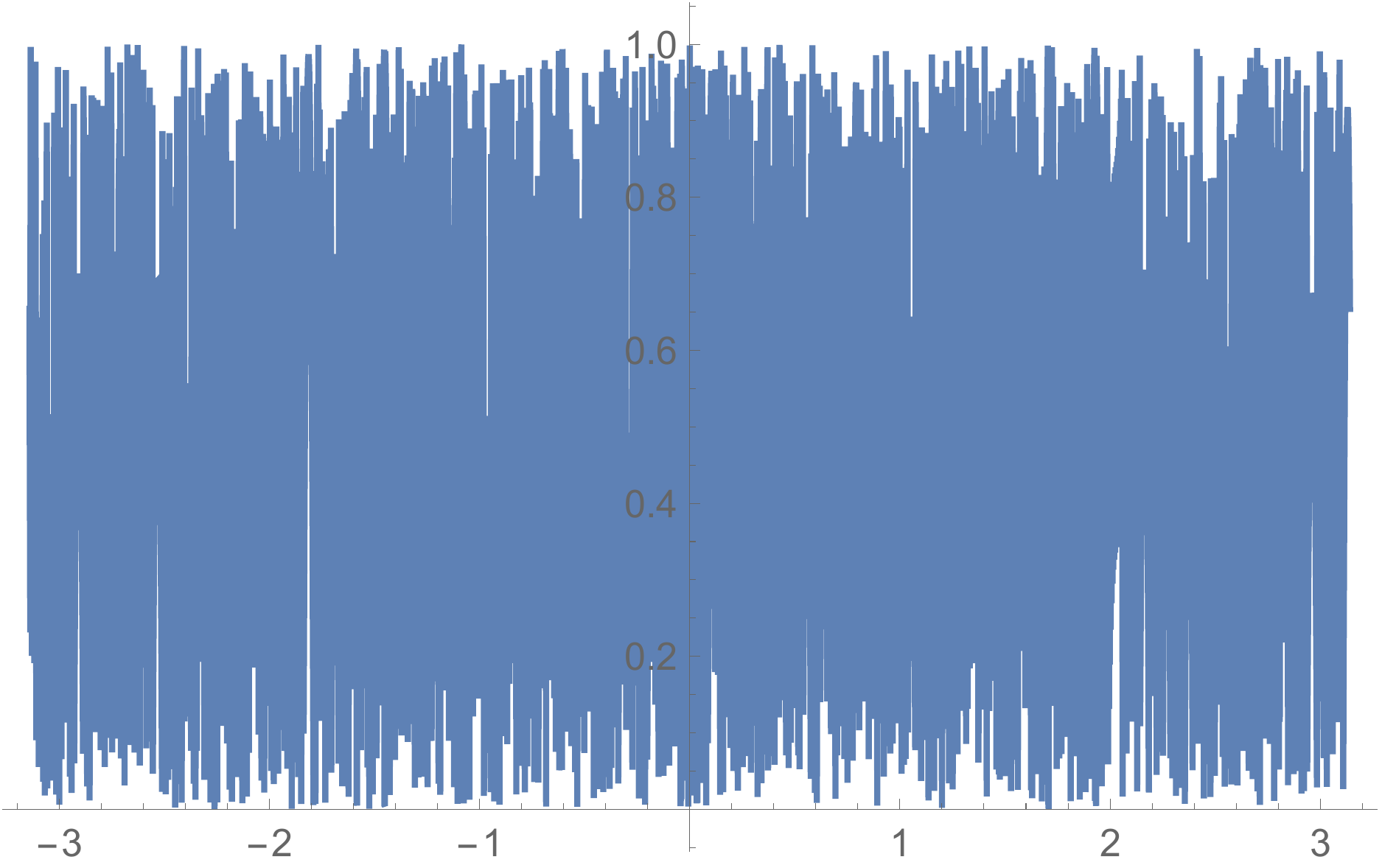}
\end{center}
\caption{\label{fig33}
The figure demonstrates the result of a triple iteration (\ref{auth}) $\psi^{(3)}(x)= T^3 \psi(x)$ of the smooth function $\psi(x) = \sin(x) +\sin(2x) - \cos(3x) +\sin(4x)$.}
\end{figure}
Because the determinant of the operator $T$ is equal to one on a quotient space $\CH/\CK$ of functions  on a cylinder $R^1 \otimes S^1 $, where $\CK$ is a kernel (\ref{kernel}),  it is natural to think that the operator  $T$ defines a measure preserving  transformation.  In this circumstances one can try to define a measure  that is  invariant with respect to the transformations generated by $T$.  The volume element $ \CV$ in the quotient space $\CH/\CK$ can be defined by using the Wiener-Feynman functional integral:
 \be\label{measure}
 \CV_A= \int_{C} F[\psi] \CD \psi(x),
 \ee 
where $A $ is a subset in $\CH/\CK$ and the functional $F[T \psi]=F[\psi]$  should be  invariant under the action of the transformations generated by the operator $T$. The measure $\CD \psi(x)$ is invariant because the determinant of the corresponding Jacobian operator is equal  to one (\ref{det}). The invariant functional $F[\psi]$ can be appropriately chosen.

\section{\it  Fluid Dynamics and Stability of Atmosphere} 

It is interesting to know if the similar systems were investigated in the past? The solutions of the partial differential equation describing the evolution $t \rightarrow g_t(x)$ of the hydrodynamical flow of incompressible ideal fluid filled in a two dimensional torus $x \in \CT^2$ can be considered as a continuous area preserving diffeomorphims $SDiff(\CT^2)$ of a torus $\CT^2 \rightarrow \CT^2$.  In Arnold approach \cite{turbul} the ideal fluid flow is described by the geodesics $g_t(x) \in G$ on the  diffeomerphism group $G=SDiff(\CT^2)$ with the velocity $v_t(x)= \dot{g}_t(x) g^{-1}_t(x)$ belonging  to the corresponding algebra \textfrak{g}=sdiff$(\CT^2)$  of divergence free vector fields. The Riemannian metric on the group $G$ is induced from the metric on a torus \cite{turbul} and the stability of the geodesic flows on  the group $G$  can be analysed by investigating the behaviour of the corresponding sectional curvatures $K(v,\delta v)$ \cite{Milnor,turbul,anosov,Arakelian:1989,Arakelian:1988gm,Lukatzki:1981,Smolentsev,Yoshida,Dowker:1990tb}.  It was found that the  flows that are defined by a parallel velocity field on $\CT^2$ are unstable because the sectional curvatures are negative and the flow is exponentially unstable.  In other directions the sectional curvatures are positive and the flows are stable.   
The extension to high-dimensional torus $\CT^N$ can be found in \cite{Lukatzki:1981,Smolentsev}.

This shows that it is not possible to reasonably predict the weather beyond a certain period if one assume that the Earth has torus topology and its atmosphere is a two-dimensional incompressible fluid. A similar  stability analysis was performed  for the hydrodynamical flow on a two-dimensional sphere  $\CS^2$ in \cite{Arakelian:1989,Smolentsev,Smolentsev,Yoshida,Dowker:1990tb} as it is important to use the more realistic assumption that the surface of the Earth is a sphere. 
 In all these  cases  the flow is exponentially unstable in some directions and is stable in some other directions, resulting in the limitation of predictability of the hydrodynamical flow and  leading to the principal difficulties of a long-term weather forecasting.  

Comparing these systems with the system considered above one can observe that here
 we have discrete in time transformations of the phase space and, secondly, the system (\ref{limitings}),  (\ref{direct}), (\ref{inversetime}) shows up  exponential instability of its geodesics in the full quotient phase space $\CH/\CK$. This full phase space chaotic behaviour can find application in Monte Carlo method, statistical physics  and  most probably in digital signal processing  in communication systems. 

The main idea behind of this approach is that the motion of an abstract "rigid body" rotating in high-dimensional Euclidean  space $\vec{r} \in E$ that is invariant under the isometry group $g(\xi) \in G$ can be described in terms of geodesic flow on a corresponding group manifold  \cite{turbul,Milnor}.  The stationary frame coordinates  of the "rigid body" are defined as $\vec{r}= g_t \vec{Q}$,  where $g_t = g(\xi(t))$ is a time-dependent  element of the matrix group $G$ and $ \vec{Q} $ are the frame coordinates rigidly fixed to the rotating "body"  $\dot{\vec{Q}}=0$. Thus 
 \be\label{velocity}
 \dot{\vec{r}} = \dot{g}_t \vec{Q}=  \dot{g}_t   g^{-1}_t ~ g_t\vec{Q} =  \hat{\omega}_s \vec{r}.
 \ee
The matrix of angular velocity in stationary frame
 \be\label{rangvel}
 \hat{\omega}_s= \dot{g}_t g^{-1}_t 
 \ee
is a right-invariant one-form ( $d ( g g_0)( g g_0 )^{-1}   =  d  g g^{-1}$, where $g_0$ is a fixed element of the group $G$). The matrix of angular velocity in rotating frame  
 \be\label{langvel}
 \hat{\Omega}_c= g^{-1}_t \dot{g}_t 
 \ee
is left invariant  because $(g_0 g)^{-1} d (g_0 g)   =  g^{-1} d  g  $. It follows that\footnote{The general relation between operators in stationary and rotating frames is $A_s = g_t  A_c g^{-1}_t $ . }
\be\label{angtrans}
 \hat{\omega}_s= \dot{g}_t g^{-1}_t = g_t  g^{-1}_t \dot{g}_t g^{-1}_t  = g_t  \hat{\Omega}_c g^{-1}_t .
 \ee
The kinetic energy is defined as a sum of the kinetic energies of all "parts" of the rotating "body" through the velocities  (\ref{velocity}):  
\be\label{kinener}
 T= {1\over 2}\ \sum_a m_a \dot{\vec{r}}_a \dot{\vec{r}}_a   =  {1\over 2}\ Tr (  \hat{I}_s  \hat{\omega}_s ~\hat{\omega}^+_s)  = -{1\over 2}\ Tr (  \hat{I}_s  \hat{\omega}_s~ \hat{\omega}_s),
\ee 
 where one should use the relation  $\vec{r}= g_t \vec{Q}$, and therefore
 \be\label{inertiatens}
 \hat{I}_s = \sum_a m_a r_a  r^+_a = g_t  \hat{I} g^{-1}_t  , ~~~~\hat{I}  = \sum_a m_a Q_a  Q^+_a.
 \ee
 The matrix  $ \hat{I}$ is a symmetric positive definite constant  matrix that determines the "moment of inertia"  in the frame rigidly fixed to the rotating "body".  The matrix of angular momentum in stationary frame is
\be\label{mominet}
\hat{m}_s=    \hat{I}_s \  \hat{\omega}_s 
\ee
and the corresponding angular momentum in rotating frame can be defined by projection of $\hat{m}_s$ into the rotating frame:
\be\label{transfor}
 \hat{m}_s= g_t \hat{M}_c g^{-1}_t , 
 \ee
thus\footnote{The square of angular momentum is conserved: 
$Tr (\hat{m}^2_s ) = Tr [(g_t \hat{M}_c g^{-1}_t)^2] = Tr (\hat{M}^2_c).$}
\be
\hat{M}_c=   \hat{I} \  \hat{\Omega}_c ,
\ee 
where one should use the relations  (\ref{mominet}), (\ref{inertiatens}) and  (\ref{rangvel}), (\ref{langvel}).
In terms of rotating frame coordinates the kinetic energy (\ref{kinener})  will take the form  
\be\label{kinener1}
 T= - {1\over 2}\ Tr (  \hat{I}_s  \hat{\omega}_s ~\hat{\omega}_s)= -{1\over 2}\  Tr (  \hat{m}_s ~  \hat{\omega}_s ) = -{1\over 2}\  Tr (\hat{M}_c ~\hat{\Omega}_c) =  -{1\over 2} \ Tr (  \hat{I} \hat{\Omega}_c  ~\hat{\Omega}_c),
\ee
where we used the relations (\ref{transfor})  and (\ref{angtrans}).  As it follows from (\ref{kinener1}) and (\ref{Riemann-metric}), in mathematical terms the matrix  $\hat{I}$  defines  the alternative Euclidean structure on the group algebra  $\langle a,b\rangle_I= Tr(T_a \hat{I }T_b)$, where $T^a$ are the generators of the algebra \textfrak{g} and $\hat{\Omega}_c= \sum_aT_a \Omega^{a}$.

 Using the relation (\ref{transfor}) and the conservation of the angular momentum in stationary frame  $\dot{\hat{m}}_s=0$ one can get the generalised Euler equation  
\be
\dot{g}_t \hat{M}_c g^{-1}_t  + g_t \dot{\hat{M}}_c g^{-1}_t - g_t \hat{M}_c g^{-1}_t \dot{g}_t  g^{-1}_t =0,\nn
\ee
wgich can be represented in the following standard form: 
\be
\dot{\hat{M}}_c +    \hat{\Omega}_c  \hat{M}_c  -    \hat{M}_c  \hat{\Omega}_c =0,\nn
\ee
or equivalently as
\be
{d \hat{M}_c \over dt }= [\hat{M}_c ; \hat{\Omega}_c].
\ee 
In terms  of  angular velocity it takes the following form:
\be
  \hat{I} ~ {d \hat{\Omega}_c \over dt }= [   \hat{I} \hat{\Omega}_c ; \hat{\Omega}_c]~,~~~
~~~~
   {d \hat{\Omega}_c \over dt }=  \hat{I}^{-1}~ [   \hat{I} \hat{\Omega}_c ; \hat{\Omega}_c] = \Gamma ( \hat{\Omega}_c , \hat{\Omega}_c).
\ee 
The last Euler equation can be represented in the form of geodesic equation 
\be
 {d \Omega^{a} \over dt } + \Gamma^{a}_{bd}\  \Omega^{b} \ \Omega^{d} =0,
\ee
where $\Gamma^{a}_{bd}$  are the Christopher symbols of the metric (\ref{Riemann-metric}). The kinetic energy defines the left invariant metric on the group: 
\be
ds^2 =  Tr (  \hat{I}  \hat{\Omega}_c ~ \hat{\Omega}_c) dt^2=    Tr (  \hat{I} ~g^{-1} d  g \  g^{-1} d  g    )=  
Tr (  \hat{I} ~g^{-1}  {\partial g \over \partial \xi^a}  ~ g^{-1}  {\partial g \over \partial \xi^b}   ) ~d\xi^a d\xi^b,
\ee 
where the $\xi^a$ are parameters of the Lie group $G$ and 
\be\label{Riemann-metric}
ds^2 = g_{ab} ~d\xi^a d\xi^b,~~~~    g_{ab}= Tr (  g^{-1}  {\partial g \over \partial \xi^b}  ~ \hat{I} ~g^{-1}  {\partial g \over \partial \xi^a}  ~  ).
\ee
The calculation of the components of the Riemann tensor and of the sectional curvatures  
\be
K(\xi,\eta) = { R_{abcd} \xi^a \eta^b  \xi^c \eta^d \over  \vert \xi \wedge  \eta   \vert^2 }
\ee
allows to investigate the stability of the geodesic flows  even in the cases when $G$ is the infinite-dimensional group of diffeomorphims $SDiff(\CM)$ that describes the flow of incompressible ideal fluid filled in a   manifold $\CM$ \cite{Milnor,turbul,anosov,Arakelian:1989,Arakelian:1988gm,Lukatzki:1981,Smolentsev,Yoshida,Dowker:1990tb}.

\section{\it Acknowledgments}
This review article is based on the lectures presented at the International Bogolyubov Conference "Problems of Theoretical and Mathematical Physics"  at the Steklov Mathematical Institute, as well as at the  CERN Theory Department and A. Alikhanian National Laboratory in Yerevan.  I would like to thank Luis Alvarez-Gaume for stimulating discussions, for kind hospitality at Simons Center for Geometry and Physics and providing to the author the references \cite{hejhal}, \cite{hejhal1} and \cite{hejhal2}. I would like to thank  H.Babujyan, R.Poghosyan and K.Savvidy for collaboration and enlightening  discussions. I would like to thank R. Kirschner,  J.Zahn and M.Bordag for kind hospitality in the Institute of Theoretical Physics of the Leipzig University were this work was completed.  This work was supported by the Alexander von Humboldt Foundation GRC 1024638 HFST.

\vfill

\end{document}